\documentclass[twocolumn,english,superscriptaddress,floatfix]{revtex4}
\usepackage[T1]{fontenc}
\usepackage[latin9]{inputenc}
\usepackage{amsbsy}
\usepackage{graphicx}
\usepackage{amssymb}

\makeatletter

\usepackage{amsbsy}

\makeatletter

\usepackage{amsbsy}

\usepackage{amsbsy}

\usepackage{amsbsy}

\usepackage{amsbsy}

\usepackage{amsbsy}

\usepackage{amsbsy}

\usepackage{amsbsy}

\usepackage{amsbsy}

\usepackage{esint}

\usepackage{bm}

\usepackage{babel}

\usepackage{babel}

\usepackage{babel}

\usepackage{babel}

\usepackage{babel}

\usepackage{babel}

\usepackage{babel}

\usepackage{babel}

\makeatletter
\makeatletter
\makeatletter
\makeatletter
\makeatletter
\makeatletter
\makeatletter
\makeatother
\makeatother
\makeatother
\makeatother
\makeatother
\makeatother
\makeatother

\makeatother

\usepackage{babel}

\makeatother

\usepackage{babel}

\begin{document}

\title{Preemptive nematic order, pseudogap, and orbital order in the iron
pnictides}

\author{R. M. Fernandes}

\affiliation{Department of Physics, Columbia University, New York, New York 10027,
USA}

\affiliation{Theoretical Division, Los Alamos National Laboratory, Los Alamos,
NM, 87545, USA}

\author{A. V. Chubukov}

\affiliation{Department of Physics, University of Wisconsin-Madison, Madison,
Wisconsin 53706, USA}

\author{J. Knolle}

\affiliation{Max-Planck-Institut für Physik komplexer Systeme, D-01187 Dresden,
Germany}

\author{I. Eremin}

\affiliation{Institut für Theoretische Physik III, Ruhr-Universität Bochum, D-44801
Bochum, Germany}

\author{J. Schmalian}

\affiliation{Institut für Theorie der Kondensierten Materie, Karlsruher Institut
für Technologie, D-76131 Karlsruhe, Germany}

\date{\today }

\begin{abstract}
Starting from a microscopic itinerant model, we derive and analyze
the effective low-energy model for collective magnetic excitations
in the iron pnictides. We show that the stripe magnetic order is generally
preempted by an Ising-nematic order which breaks $C_{4}$ lattice
symmetry but preserves $O(3)$ spin-rotational symmetry. This leads
to a rich phase diagram as function of doping, pressure, and elastic
moduli, displaying split magnetic and nematic tri-critical points.
The nematic transition may instantly bring the system to the verge
of a magnetic transition, or it may occur first, being followed by
a magnetic transition at a lower temperature. In the latter case,
the preemptive nematic transition is accompanied by either a jump
or a rapid increase of the magnetic correlation length, triggering
a pseudogap behavior associated with magnetic precursors. Furthermore,
due to the distinct orbital character of each Fermi pocket, the nematic
transition also induces orbital order. We compare our results to various
experiments, showing that they correctly address the changes in the
character of the magneto-structural transition across the phase diagrams
of different compounds, as well as the relationship between the orthorhombic
and magnetic order parameters. 
\end{abstract}


\maketitle

\section{Introduction}

The current interest in iron-based superconductors (FeSCs) lies not
only in the superconducting pairing mechanism, but also in the peculiar
normal-state properties of these materials (for reviews, see \cite{reviews}).
Of particular interest is the fact that in weakly doped iron pnictides
the stripe spin-density wave order at $T<T_{N}$, with ordering vectors
$(0,\pi)$ or $(\pi,0)$ in the 1-Fe Brillouin zone \cite{Yildirim08,Xiang08},
is often preceded by a {}``nematic'' phase with broken $C_{4}$
tetragonal symmetry but unbroken $O(3)$ rotational symmetry. The
emergence of such a phase is not only manifested by a tetragonal to
orthorhombic transition at $T_{s}\geq T_{N}$, but also by the onset
of significant anisotropies in several quantities \cite{Fisher11},
such as dc resistivity \cite{Chu10,Tanatar10}, optical conductivity
\cite{Duzsa11,Uchida11}, local density of states \cite{Davis10},
orbital occupancy \cite{Shen11}, uniform susceptibility \cite{Matsuda11},
and the vortex core in the mixed superconducting state \cite{Song11}.
The fact that the spin-density-wave and structural transition lines
follow each other across all the phase diagrams of 1111 materials
($R\mathrm{FeAsO}$, with rare-earth $R$) and 122 materials ($Ae\mathrm{Fe_{2}As_{2}},$with
alkaline-earth $Ae$), even inside the superconducting dome \cite{FernandesPRB10,Nandi09},
prompted researchers to propose that these phases are intimately connected.
The interplay between magnetic and structural transitions in FeSCs
is also quite rich: while in 1111 materials the two transitions are
second-order and split ($T_{s}>T_{N}$), in most of the 122 materials
they seem to occur simultaneously or near-simultaneously at small
dopings, but clearly split above some critical doping - $x\approx0.022$
in $\mathrm{Ba\left(Fe_{1-\mathit{x}}Co\mathit{_{x}}\right)_{2}As_{2}}$,
see \cite{Kim11,Birgeneau11}, and $x\approx0.039$ in $\mathrm{Ca\left(Fe_{1-\mathit{x}}Co\mathit{_{x}}\right)_{2}As_{2}}$,
see \cite{Prokes11}.

Early theoretical proposals explored two alternative scenarios for
the nematic order. One scenario, which borrows concepts from the manganites,
is that the driving force is orbital ordering - it induces the structural
transition and triggers the magnetic transition at a lower temperature
by renormalizing the exchange constants \cite{kruger09,RRPSingh09,Phillips10,w_ku10,bascones,maria,chen10,Craco10,Lin11,Nevidomskyy11}.
An alternative, magnetic scenario, which borrows concepts from the
studies of the $J_{1}-J_{2}$ model of localized spins \cite{chandra},
is build upon the fact that the order parameter manifold for the stripe
magnetic order is $O(3)\times Z_{2}$, with the $Z_{2}$ Ising symmetry
differentiating between the two possible ordering vectors $(0,\pi)$
and $(\pi,0)$. This scenario explores the possibility that the $Z_{2}$
Ising variable orders before the $O(3)$ rotational symmetry is broken,
leading to an intermediate phase with a broken tetragonal symmetry
but no long-range magnetic order. The Ising order makes the Fermi
pockets at $(0,\pi)$ and $(\pi,0)$ non-equivalent and induces orbital
order, since the two electron pockets are constituted of different
orbitals \cite{Graser09,Zhang09}. Furthermore, the breaking of the
$Z_{2}$ symmetry also makes the $x$ and $y$ directions inside the
unit cell inequivalent, inducing a structural instability via the
bi-linear coupling between the $Z_{2}$ order parameter and the orthorhombic
distortion (see below). For this reason, the state with broken Ising
symmetry is often called an Ising-nematic phase. The magnetic scenario
was first applied to localized or nearly-localized spin models for
Fe-pnictides \cite{Si08,Fang08,Xu08}, but was later extended to itinerant
systems in a phenomenological way \cite{Qi09,FernandesPRL10}.

\begin{figure}
\begin{centering}
\includegraphics[width=1\columnwidth]{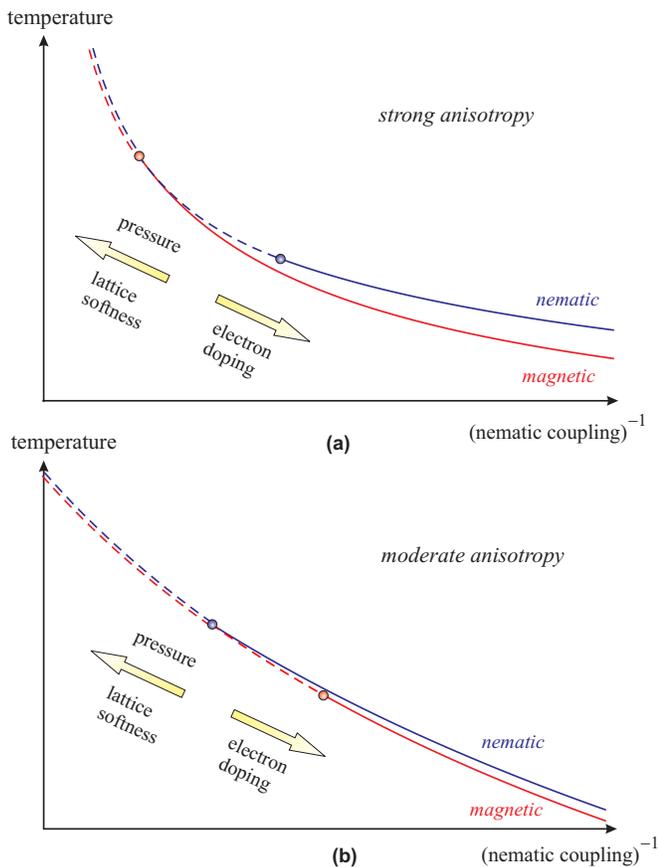} 
\par\end{centering}

\caption{(Color online) Two characteristic phase diagrams of the system behavior
as function of temperature and nematic coupling. Panel (a) is for
strongly anisotropic, quasi-two dimensional systems, while panel (b)
is for moderately anisotropic, less quasi-two dimensional systems.
Red (light gray) and blue (dark gray) curves represent magnetic and
Ising-nematic transitions, respectively. The arrows show how the inverse
nematic coupling changes as function of doping, pressure, and lattice
stiffness. Here and in the other figures, a solid (dashed) line denotes
a second-order (first-order) transition, and a double-dashed line
indicates a simultaneous first-order transition. The two solid points
mark the positions of the nematic and magnetic tri-critical points.
The difference between the two phase diagrams is in the system behavior
at intermediate nematic coupling. In (a), the system displays a first-order
nematic transition followed by a second-order magnetic transition
at lower temperatures. In (b), the system first undergoes a second-order
nematic transition, and then, at smaller temperatures, a meta-nematic
transition which triggers a first-order magnetic transition. At the
meta-nematic transition, the nematic order parameter jumps between
two finite values. \label{fig_exp_phase_diagram} }

\end{figure}

Since orbital order and Ising-nematic order break the same symmetry,
one order generates the other, making the experimental distinction
between the two scenarios rather subtle. The same is true if one wants
to distinguish between itinerant and localized magnetic models, which
we will address as well. In both situations, the only real way to
distinguish between different approaches is by explicit calculations
followed by comparison to the available experimental results. In this
regard, the richness of the phase diagrams of the iron pnictides is
an important cornerstone, because one has to explain not only the
very existence of the nematic phase, but also how this transition
changes as a function of doping, pressure, and material. In addition,
one has to address the feedback effect from the nematic order on fermions.
In particular, there is growing volume of experimental evidence of
pseudogap-like electronic behavior in the iron pnictides, which in
some cases seems to emerge at the same temperature at which nematic
order sets in \cite{He_NaFeAs,Xu_pseudogap,Tanatar_pseudogap,Basov_pseudogap}.

In this paper, we argue that the changes in the character of the nematic
transition with doping, pressure, and alkaline-earth substitutions,
as well as the development of the pseudogap and other experimentally
detectable features, can be understood within a magnetic scenario
for an itinerant fermionic model. We depart from a multi-band model
of interacting fermions and \emph{derive} (instead of assuming) the
Ginzburg-Landau (GL) effective action for the two low-energy collective
$O(3)$ magnetic degrees of freedom $\boldsymbol{\Delta}_{X}$ and
$\boldsymbol{\Delta}_{Y}$ associated with the ordering vectors $(\pi,0)$
and $(0,\pi)$, respectively. This action contains, besides other
terms, the term $g(\boldsymbol{\Delta}_{X}^{2}-\boldsymbol{\Delta}_{Y}^{2})^{2}$
which describes fluctuations in the Ising-nematic channel. We extend
the original $O(3)$ model to $N$ field components and study the
limit $N=\infty$ within mean-field theory, and arbitrary $N$ by
the renormalization group (RG) technique. We find different types
of system behavior depending on the strength of the dimensionless
parameter $\alpha\propto g^{-1}$ and on the degree of magnetic anisotropy.
In all cases, however, we find two distinct multi-critical points
in the phase diagram, namely a \emph{magnetic tri-critical point}
and a \emph{nematic tri-critical point}.

In the case of strongly anisotropic, quasi-2D systems, the nematic
tri-critical point precedes the magnetic one, resulting in three types
of behavior - see Fig. \ref{fig_exp_phase_diagram}(a). (i) For small
$\alpha$ (largest nematic coupling $g$) we find a strong first-order
Ising-nematic transition at $T_{s}$, accompanied by a discontinuous
jump of the magnetic correlation length $\xi$ and a jump of the magnetic
order parameter to a finite value. In this case, the stripe magnetic
order emerges simultaneously with the Ising-nematic order, via a first-order
magnetic transition ($T_{N}=T_{s}$), but because $\xi$ jumps, there
are no critical magnetic fluctuations above the transition. (ii) For
intermediate $\alpha$ we find a first-order preemptive Ising-nematic
transition accompanied by a discontinuous increase of the magnetic
correlation length to a larger but still finite value, such that the
stripe magnetic order does not develop at $T_{s}$ and emerges only
at a smaller $T$. In this case, the magnetic transition is second-order.
(iii) For large $\alpha$ (small nematic coupling $g$) we find a
second-order preemptive Ising-nematic transition followed by a second-order
magnetic stripe transition at a smaller $T$. Near the nematic tri-critical
point we find that the Ising-nematic order emerges very rapidly such
that, in practice, the second-order Ising transition is hard to distinguish
from a first-order transition. The behavior of the nematic and magnetic
order parameters for the phase diagram in Fig. \ref{fig_exp_phase_diagram}(a)
is shown in Fig. \ref{fig_order_parameters}(d), (c), and (b) for
regimes (i), (ii), and (iii), respectively.

In the case of moderately anisotropic, less quasi-2D systems, the
positions of the nematic and magnetic tri-critical points can be reversed,
resulting in a new behavior in the regime of intermediate $\alpha$,
see Fig.\ref{fig_exp_phase_diagram}(b). The regimes (i) and (iii)
are still present, when both the nematic and magnetic transitions
are first-order and second-order, respectively. However, in regime
(ii) the upper Ising-nematic transition is second-order and at a lower
$T$ there is a meta-nematic transition (i.e. the nematic order parameter
undergoes a finite jump from one non-zero value to the other), which
induces a first-order magnetic transition. The behavior of the nematic
and magnetic order parameters for the phase diagram in Fig. \ref{fig_exp_phase_diagram}(b)
is shown in Fig. \ref{fig_metanematic}(c), (b), and (a) for regimes
(i), (ii), and (iii), respectively.

As shown in Figs. \ref{fig_exp_phase_diagram}, we find from our microscopic
model that $\alpha$ decreases with pressure but increases with increasing
lattice stiffness and electron doping, i.e. larger electron doping
results in a larger value of $\alpha$. Our results then predict that
the Ising-nematic and magnetic transitions split upon electron doping,
but tend to remain simultaneous and first-order upon pressure, in
agreement with the experimental data for FeSCs. We also show that
the nematic order parameter couples linearly to the orbital polarization
and lattice distortion, and hence nematic order generates orbital
and structural order. We argue, however, that the sign of the orbital
order may differ between hole and electron-doped materials.

We also argue that, for the phase diagram in Fig. \ref{fig_exp_phase_diagram}a,
the discontinuous (or nearly-discontinuous) increase of the magnetic
correlation length at $T_{s}$ in the regimes (ii) and (iii) greatly
increases the strength of thermal magnetic fluctuations. Once enhanced,
these fluctuations account for spectral weight redistribution (magnetic
precursors), what gives rise to pseudogap behavior in the fermionic
spectral function and other observables. We present more detailed
comparison with the data later in the paper and also compare our results
with earlier studies of Ising-nematic order.

\begin{figure}
\begin{centering}
\includegraphics[width=1\columnwidth]{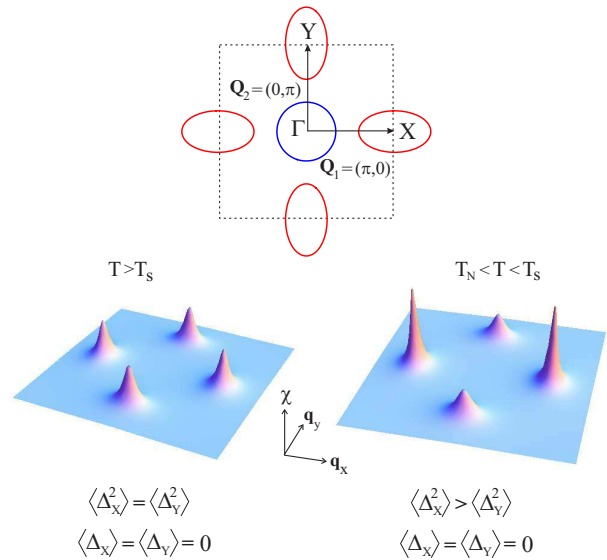} 
\par\end{centering}

\caption{(Color online) (\emph{upper panel}) The band-structure with a circular
hole pocket at $\Gamma$ and two electron pockets at $X$ and $Y$.
The Brillouin zone contains one Fe atom. (\emph{lower panels}) Static
magnetic susceptibility $\chi_{\mathbf{q}}$ across the Brillouin
zone for different temperatures. At high temperatures, fluctuations
near the two stripe magnetic ordering vectors are equally strong,
$\left\langle \Delta_{X}^{2}\right\rangle =\left\langle \Delta_{Y}^{2}\right\rangle $.
Above the magnetic ordering temperature $T_{N}$ but below the Ising-nematic
ordering temperature $T_{s}$, fluctuations associated with one of
the stripe states become stronger (in the figure, $\left\langle \Delta_{X}^{2}\right\rangle >\left\langle \Delta_{Y}^{2}\right\rangle $)
and the tetragonal symmetry is broken inside the unit cell. Stronger
fluctuations around one ordering vector yield stronger intensity and
narrower peaks. }

\label{fig_nematic} 
\end{figure}

In the itinerant picture which we adopt here, the nature of the Ising-nematic
phase has a clear interpretation in terms of magnetic fluctuations.
The system has two degenerate stripe magnetic ground states with ordering
vectors $(\pi,0)$ and $(0,\pi)$, described by the two order parameters
$\boldsymbol{\Delta}_{X}$ and $\boldsymbol{\Delta}_{Y}$. At high
temperatures $\left\langle \boldsymbol{\Delta}_{X}\right\rangle =\left\langle \boldsymbol{\Delta}_{Y}\right\rangle =0$,
and the fluctuations of each order parameter have equal strength,
i.e., $\left\langle \boldsymbol{\Delta}_{X}^{2}\right\rangle =\left\langle \boldsymbol{\Delta}_{Y}^{2}\right\rangle $.
The Ising-nematic phase emerges when fluctuations associated with
one of the ordering vectors become stronger than the other, $\left\langle \boldsymbol{\Delta}_{X}^{2}\right\rangle \neq\left\langle \boldsymbol{\Delta}_{Y}^{2}\right\rangle $,
while still $\left\langle \boldsymbol{\Delta}_{X}\right\rangle =\left\langle \boldsymbol{\Delta}_{Y}\right\rangle =0$
(see Fig. \ref{fig_nematic}). Because there are two possible choices,
$\left\langle \boldsymbol{\Delta}_{X}^{2}\right\rangle >\left\langle \boldsymbol{\Delta}_{Y}^{2}\right\rangle $
or $\left\langle \boldsymbol{\Delta}_{X}^{2}\right\rangle <\left\langle \boldsymbol{\Delta}_{Y}^{2}\right\rangle $,
the symmetry which breaks down at the Ising-nematic phase transition
is a $Z_{2}$ symmetry. Once magnetic fluctuations around $(\pi,0)$
become stronger (or weaker) than the fluctuations around $(0,\pi)$,
the equivalence between $x$ and $y$ directions inside the unit cell
breaks down. The Ising-nematic transition then triggers the transition
from a tetragonal to an orthorhombic phase, and also imposes orbital
order because the Fermi pockets centered at $(\pi,0)$ and $(0,\pi)$
have different orbital character. Furthermore, this anisotropy in
the spectrum of magnetic fluctuations leads to anisotropic scattering
of electrons, resulting in different in-plane resistivities along
$x$ and $y$ directions \cite{Fernandes11}.

The structure of the paper is the following. In Section \ref{sec:model}
we derive the effective action for the nematic order parameter using
a two-stage Hubbard-Stratonovich transformation. In Sec. \ref{sec:solution}
we extend the effective action to arbitrary $N$, where $N$ is the
number of components of the magnetic order parameter ($N=3$ in the
original $O(3)$-isotropic model). We solve for the nematic order
in the mean-field approximation, justified in the $N\rightarrow\infty$
limit, obtain different regimes of system behavior and discuss each
regime in some detail. In Sec.\ref{sec:RG} we study the effective
action at a finite $N$ using the renormalization group (RG) technique.
We find that the system behavior for $N=3$ is the same as for large
$N$, but there is a change in the RG flow structure at $N\leq2$.
In Sec. \ref{sec:conseq} we discuss the interplay between Ising-nematic,
orbital, and structural order, and show how the nematic order gives
rise to pseudogap behavior despite the fact that the $Z_{2}$ order
parameter has zero momentum and does not by itself reconstructs low-energy
fermionic states. In Sec. \ref{sec:comparison} we compare our theory
with the experiments and with other theoretical works on the nematic
order. We present our conclusions in Sec.\ref{sec:conclusions}.

\section{Microscopic model and the effective action \label{sec:model}}

We start from the minimal four-band model with two hole pockets $\Gamma$
at the center of the Brillouin zone and two electron pockets $X$
and $Y$ at $\mathbf{Q}_{1}=\left(\pi,0\right)$ and $\mathbf{Q}_{2}=\left(0,\pi\right)$,
respectively (see Fig. \ref{fig_nematic}). We follow Ref. \cite{Eremin10}
and consider that only one of the two hole pockets participate in
the SDW state. The extension to the case when both $\Gamma$ pockets
are involved in the SDW reconstruction is straightforward and does
not change the analysis below except for the renormalization of the
couplings.

For simplicity, we consider parabolic dispersions with $\varepsilon_{\Gamma,\mathbf{k}}=\varepsilon_{0}-\frac{k^{2}}{2m}-\mu$,
$\varepsilon_{X,\mathbf{k+Q_{1}}}=-\varepsilon_{0}+\frac{k_{x}^{2}}{2m_{x}}+\frac{k_{y}^{2}}{2m_{y}}-\mu$,
and $\varepsilon_{Y,\mathbf{k+Q_{2}}}=-\varepsilon_{0}+\frac{k_{x}^{2}}{2m_{y}}+\frac{k_{y}^{2}}{2m_{x}}-\mu$,
where $m_{i}$ denotes the band masses, $\varepsilon_{0}$ is the
offset energy, and $\mu$ is the chemical potential. Near the Fermi
energy and for small ellipticity, the dispersions can be approximated
by $\varepsilon_{\Gamma,\mathbf{k}}=-\varepsilon$, $\varepsilon_{X,\mathbf{k+Q_{1}}}=\varepsilon-\delta_{0}+\delta_{2}\cos2\theta$,
$\varepsilon_{Y,\mathbf{k+Q_{2}}}=\varepsilon-\delta_{0}-\delta_{2}\cos2\theta$,
with $\delta_{0}=2\mu$, $\delta_{2}=\varepsilon_{0}m(m_{x}-m_{y})/(2m_{x}m_{y})$,
and $\theta=\tan^{-1}k_{y}/k_{x}$ \cite{Vorontsov10}.

Electrons with spin $\alpha$ of the band $i$ are created by the
operators $c_{i,\mathbf{k}\alpha}^{\dagger}$, yielding the non-interacting
Hamiltonian: \begin{equation}
\mathcal{H}_{0}=\sum\limits _{i,\mathbf{k}}\varepsilon_{i,\mathbf{k}}c_{i,\mathbf{k}\alpha}^{\dagger}c_{i,\mathbf{k}\alpha}\label{H_0}\end{equation}

Here and for the rest of the paper the summation over repeated spin
indices is assumed, and we shift the momenta of the fermions near
the $X$ and $Y$ Fermi pockets by $\mathbf{Q}_{1}$ and $\mathbf{Q}_{2}$,
respectively, i.e. $\varepsilon_{X,\mathbf{k+Q_{1}}}\rightarrow\varepsilon_{X,\mathbf{k}}$,
$\varepsilon_{Y,\mathbf{k+Q_{2}}}\rightarrow\varepsilon_{Y,\mathbf{k}}$.

As discussed in Ref. \cite{Maiti10}, this model has eight fermionic
interactions $u_{n}$ which can be decomposed into the spin density-wave
(SDW), the charge density-wave (CDW) and the pairing channels. Since
our goal is to study stripe magnetism and the accompanied Ising-nematic
order, we keep only the interactions in the spin channel with momenta
near $\mathbf{Q}_{1}$ and $\mathbf{Q}_{2}$, restricting the interacting
Hamiltonian to \begin{equation}
\mathcal{H}_{\mathrm{int}}=-\frac{1}{2}u_{\mathrm{spin}}\sum\limits _{i,\mathbf{q}}\mathbf{s}_{i,\mathbf{q}}\cdot\mathbf{s}_{i,-\mathbf{q}}\label{H_int}\end{equation}
 where $\mathbf{s}_{i,\mathbf{q}}=\sum_{k}c_{\Gamma,\mathbf{k+q}\alpha}^{\dagger}\boldsymbol{\sigma}_{\alpha\beta}c_{i,\mathbf{k}\beta}$
is the electronic spin operator, with Pauli matrices $\boldsymbol{\sigma}_{\alpha\beta}$.
The coupling $u_{\mathrm{spin}}$ is the combination of density-density
and pair-hopping interactions between hole and electron states ($u_{1}$
and $u_{3}$ terms in the notation of Ref. \cite{Chubukov08}):

\begin{eqnarray}
u_{1}c_{\Gamma,\alpha}^{\dagger}c_{\Gamma,\alpha}c_{X,\beta}^{\dagger}c_{X,\beta} & = & -\frac{u_{1}}{2}c_{\Gamma,\alpha}^{\dagger}\boldsymbol{\sigma}_{\alpha\beta}c_{X,\beta}\cdot c_{X,\gamma}^{\dagger}\boldsymbol{\sigma}_{\gamma\delta}c_{\Gamma,\delta}\nonumber \\
 &  & +(\cdots)\nonumber \\
u_{3}c_{\Gamma,\alpha}^{\dagger}c_{X,\alpha}c_{\Gamma,\beta}^{\dagger}c_{X,\beta} & = & -\frac{u_{3}}{2}c_{\Gamma,\alpha}^{\dagger}\boldsymbol{\sigma}_{\alpha\beta}c_{X,\beta}\cdot c_{X,\gamma}^{\dagger}\boldsymbol{\sigma}_{\gamma\delta}c_{\Gamma,\delta}\nonumber \\
 &  & +(\cdots)\label{SDW_channel}\end{eqnarray}
 where the dots stand for the terms with $\delta_{\alpha,\beta}\delta_{\gamma,\delta}$,
which only contribute to the CDW channel. Combining the two contributions
for the SDW channel, we find $u_{\mathrm{spin}}=u_{1}+u_{3}$. Once
$u_{\mathrm{spin}}$ exceeds some critical value (which gets smaller
when $\delta_{0}$ and $\delta_{2}$ decrease), static magnetic susceptibility
diverges at $(0,\pi)$ and $(\pi,0)$, and the system develops long-range
magnetic order. An excitonic-type SDW instability in Fe-pnictides,
resulting from the interaction between hole and electron pockets,
has been considered by several authors \cite{Eremin10,Cvetkovic09,Gorkov08,Timm09,Kuroki08,DungHaiLee08,Platt08,Vorontsov09,Fernandes_Schmalian,Knolle10}.

Our calculations are done in two steps. In the first step, we introduce
the bosonic fields $\boldsymbol{\Delta}_{(X,Y)}\propto\sum_{\mathbf{k}}c_{\Gamma,\mathbf{k}\alpha}^{\dagger}\boldsymbol{\sigma}_{\alpha\beta}c_{(X,Y),\mathbf{k}\beta}$
for the collective magnetic degrees of freedom, integrate out the
fermions, and obtain a Ginzburg-Landau (GL) action for $\boldsymbol{\Delta}_{X}$
and $\boldsymbol{\Delta}_{Y}$. We show, in agreement with earlier
results \cite{Eremin10}, that in a mean-field approximation only
one of the magnetic order parameters - either $\left\langle \boldsymbol{\Delta}_{X}\right\rangle $
or $\left\langle \boldsymbol{\Delta}_{Y}\right\rangle $ - becomes
non-zero in the magnetically ordered state. This leads to stripe-type
SDW order in which spins are ordered ferromagnetically in one direction
and antiferromagnetically in the other, i.e. the ordering momentum
is either $(\pi,0)$ or $(0,\pi)$. In the second step, we include
fluctuations of the $\boldsymbol{\Delta}_{X,Y}$ fields, introduce
the collective Ising-nematic bosonic variable $\phi\propto\Delta_{X}^{2}-\Delta_{Y}^{2}$
together with $\psi\propto\boldsymbol{\Delta}_{X}^{2}+\boldsymbol{\Delta}_{Y}^{2}$,
integrate over $\boldsymbol{\Delta}_{X}$ and $\boldsymbol{\Delta}_{Y}$,
and obtain an effective action in terms of $\phi$ and $\psi$. We
analyze this action and check whether the system develops an instability
towards $\left\langle \phi\right\rangle \neq0$ before $\left\langle \boldsymbol{\Delta}_{X}\right\rangle $
or $\left\langle \boldsymbol{\Delta}_{Y}\right\rangle $ becomes non-zero.

\subsection{The action in terms of $\boldsymbol{\Delta}_{X}$ and $\boldsymbol{\Delta}_{Y}$}

A straightforward way to obtain the action in terms of $\left\langle \boldsymbol{\Delta}_{X}\right\rangle $
and $\left\langle \boldsymbol{\Delta}_{Y}\right\rangle $ is to start
with the fermionic Hamiltonian $\mathcal{H}=\mathcal{H}_{\mathrm{0}}+\mathcal{H}_{\mathrm{int}}$
in Eqs. (\ref{H_0}) and (\ref{H_int}), write the partition function
as the integral over Grassmann variables:

\begin{equation}
Z\propto\int dc_{i,\mathbf{k}}dc_{i,\mathbf{k}}^{\dagger}\mathrm{e}^{-\beta\mathcal{H}}\label{ch_X}\end{equation}
 and then decouple the quartic term in fermionic operators using the
Hubbard-Stratonovich transformation:

\begin{equation}
\mathrm{e}^{\frac{ax^{2}}{2}}=\frac{1}{\sqrt{2\pi a}}\int dy\,\mathrm{e}^{\left(-\frac{y^{2}}{2a}+yx\right)}\label{Hubbard_Stratononich}\end{equation}
 where, in our case, $x=\mathbf{s}_{i,\mathbf{0}}=\sum_{k}c_{\Gamma,\mathbf{k}\alpha}^{\dagger}\boldsymbol{\sigma}_{\alpha\beta}c_{(X,Y),\mathbf{k}\beta}$
and $y=\boldsymbol{\Delta}_{(X,Y)}$. We then integrate Eq. (\ref{ch_X})
over fermionic variables using the fact that after the Hubbard-Stratonovich
transformation the effective action becomes quadratic with respect
to the fermionic operators. The result of the integration is recast
back into the exponent and the partition function is expressed as:

\begin{equation}
Z\propto\int d\boldsymbol{\Delta}_{X}d\boldsymbol{\Delta}_{Y}\mathrm{e}^{-S_{\mathrm{eff}}\left[\boldsymbol{\Delta}_{X},\boldsymbol{\Delta}_{Y}\right]}\label{aux_action_1}\end{equation}

If the relevant $\boldsymbol{\Delta}_{X}$ and $\boldsymbol{\Delta}_{Y}$
are small, which we assume to hold even if the magnetic transition
is first-order (we present the conditions on the parameters below),
one can expand $S_{\mathrm{eff}}\left[\boldsymbol{\Delta}_{X},\boldsymbol{\Delta}_{Y}\right]$
in powers of $\boldsymbol{\Delta}_{X}$ and $\boldsymbol{\Delta}_{Y}$
and obtain the Ginzburg-Landau type of action for the order parameters
$\boldsymbol{\Delta}_{X},\boldsymbol{\Delta}_{Y}$. For uniform $\boldsymbol{\Delta}_{i}$,
the most generic form of $S_{\mathrm{eff}}\left[\boldsymbol{\Delta}_{X},\boldsymbol{\Delta}_{Y}\right]$
is \begin{eqnarray}
S_{\mathrm{eff}}\left[\boldsymbol{\Delta}_{X},\boldsymbol{\Delta}_{Y}\right] & = & r_{0}\left(\boldsymbol{\Delta}_{X}^{2}+\boldsymbol{\Delta}_{Y}^{2}\right)+\frac{u}{2}\left(\boldsymbol{\Delta}_{X}^{2}+\boldsymbol{\Delta}_{Y}^{2}\right)^{2}\nonumber \\
 &  & -\frac{g}{2}\left(\boldsymbol{\Delta}_{X}^{2}-\boldsymbol{\Delta}_{Y}^{2}\right)^{2}+v\left(\boldsymbol{\Delta}_{X}\cdot\boldsymbol{\Delta}_{Y}\right)^{2}\label{action}\end{eqnarray}

Carrying out this procedure, we obtain the coefficients $r_{0}$,
$u$, $g$, and $v$ in terms of the non-interacting fermionic propagators
convoluted with Pauli matrices (details can be found in Appendix A).
The coefficient $v$ vanishes in our model because of the anti-commutation
property of the Pauli matrices: $\sigma^{i}\sigma^{j}+\sigma^{j}\sigma^{i}=0$
for $i\neq j$. To get a non-zero $v$, one needs to include direct
interactions between the two electron pockets \cite{Eremin10}. The
other three prefactors are expressed via fermionic propagators $G_{j,\mathbf{k}}^{-1}=i\omega_{n}-\xi_{j,\mathbf{k}}$
as \begin{eqnarray}
r_{0} & = & \frac{2}{u_{\mathrm{spin}}}+2\int_{k}G_{\Gamma,k}G_{X,k}\nonumber \\
u & = & \frac{1}{2}\int_{k}G_{\Gamma,k}^{2}\left(G_{X,k}+G_{Y,k}\right)^{2}\nonumber \\
g & = & -\frac{1}{2}\int_{k}G_{\Gamma,k}^{2}\left(G_{X,k}-G_{Y,k}\right)^{2}\label{aux_action}\end{eqnarray}
 where $\int_{k}=T\sum_{n}\int\frac{d^{d}k}{\left(2\pi\right)^{d}}$
and $k=\left(\mathbf{k},\omega_{n}\right)$, with momentum $\mathbf{k}$
and Matsubara frequency $\omega_{n}=\left(2n+1\right)\pi T$. Similar
coefficients were found in Ref. \cite{Brydon11}, which focused on
the magnetic instabilities in a two-band model.

Evaluating the momentum integrals and summing over Matsubara frequencies
we obtain that $\int_{k}G_{\Gamma,k}G_{X,k}$ is negative and at perfect
nesting diverges as $N_{F}\log\Lambda/T$, where $N_{F}\approx m/(2\pi)$
is the density of states of the hole pocket and $\Lambda$ is the
upper cutoff for the low-energy theory - i.e. the scale at which corrections
to the parabolic dispersion become of the order one. Away from perfect
nesting, the logarithm saturates at $T=0$ at some finite but still
large value \cite{Tesanovic09}. As a result, $r_{0}$ decreases with
decreasing $T$ and, if the amplitude of the interaction $u_{\mathrm{spin}}$
is not too small, changes sign at some temperature $T=T_{N,0}$. This
is the mean-field transition temperature below which magnetic order
appears. Near $T_{N,0}$ one can expand $r_{0}$ as $r_{0}=a(T-T_{N,0})$,
with $a>0$.

The GL action can be straightforwardly extended to include the momentum
and frequency dependence of the $\boldsymbol{\Delta}$ fields, $\boldsymbol{\Delta}_{i}\to\boldsymbol{\Delta}_{i}(q)$.
This extension does not modify in any relevant way the pre-factors
of the quartic terms, but changes the pre-factor for the quadratic
term at small $q$ to:

\begin{eqnarray}
r_{0}\rightarrow\chi_{i,q}^{-1} & = & \frac{2}{u_{\mathrm{spin}}}+2\int_{k}G_{\Gamma,k}G_{i,k+q}\nonumber \\
 & = & r_{0}+\gamma\left|\nu_{n}\right|+f_{\mathbf{q}}\label{chi_i}\end{eqnarray}

Here $i=X,Y$, $\nu_{n}=2\pi Tn$ is the bosonic Matsubara frequency,
$\gamma$ is the Landau damping coefficient, and $f_{\mathbf{q}}$
is in general an anisotropic function of $\mathbf{q}$, $f_{\mathbf{q}}=q_{x}^{2}(1\pm\eta)+q_{y}^{2}(1\mp\eta)+\eta_{z}q_{z}^{2}$,
with $-1<\eta<1$ and upper (lower) signs referring to $\Delta_{X}$
($\Delta_{Y}$). We verified that all results which we obtain below
do not depend on whether $\eta$ is finite or zero (see Appendix B).
To simplify the presentation, we then set $\eta=0$ and approximate
$\chi_{i,q}^{-1}$ by $r_{0}+\gamma|\nu_{n}|+q_{\parallel}^{2}+\eta_{z}q_{z}^{2}$,
the same for $\boldsymbol{\Delta}_{X}$ and $\boldsymbol{\Delta}_{Y}$.
Below we consider this anisotropic 3D dispersion and also an isotropic
dispersion in dimensions $2\leq d\leq4$, in which case $q_{\parallel}^{2}+\eta_{z}q_{z}^{2}$
is replaced by the $d$-dimensional momentum amplitude $q^{2}$.

For the two fourth-order terms in (\ref{aux_action}) we obtain at
a finite $T$, expanding around perfect nesting to leading order in
the chemical potential $\mu$ and in the mass anisotropy $\delta m=m(m_{x}-m_{y})/(2m_{x}m_{y})$:

\begin{eqnarray}
u & \approx & \frac{7\zeta\left(3\right)N_{F}}{4\pi^{2}T^{2}}\nonumber \\
g & \approx & 0.024u\left(\frac{\varepsilon_{0}\delta m}{T}\right)^{2}\label{parameters_quartic}\end{eqnarray}
 for $\delta m\ll T/\varepsilon_{0}\ll1$. At $T=0$, we obtain \begin{eqnarray}
u & \approx & \frac{N_{F}}{4\mu^{2}}\nonumber \\
g & \approx & u\left(\frac{\varepsilon_{0}\delta m}{2\mu}\right)^{2}\label{parameters_quartic_1}\end{eqnarray}
 for $\delta m\ll\left|\mu\right|/\varepsilon_{0}\ll1$. In general,
$u>0$ and $u>g$, i.e. the action (\ref{action}) increases when
either $\Delta_{X}$ or $\Delta_{Y}$ becomes large. If this condition
was not satisfied, the expansion of $S_{\mathrm{eff}}$ in powers
of $\boldsymbol{\Delta}_{X}$ and $\boldsymbol{\Delta}_{Y}$ would
only make sense if higher order terms were considered. The crucial
result for our consideration is that $g$ vanishes for circular electron
Fermi surfaces ($\delta m=0$), but is positive for any non-zero ellipticity,
independent on the sign of $\epsilon_{0}$ and on whether $m_{x}$
is larger or smaller than $m_{y}$.

The action $S_{\mathrm{eff}}$ is exact and includes all fluctuations
of the two bosonic fields. Before we consider these fluctuations,
let us analyze Eq. (\ref{action}) in the mean-field approximation.
The conventional way to justify mean-field theory is to extend the
original one-flavor fermionic model to $L$ flavors, such that $\boldsymbol{\Delta}_{X}$
and $\boldsymbol{\Delta}_{Y}$ couple to $L$ clones of $\mathbf{s}_{i,\mathbf{q=0}}$,
and take the limit $L\to\infty$. To do this properly, one has to
rescale simultaneously $u_{\mathrm{spin}}\to u_{\mathrm{spin}}/L$.
After this rescaling, the effective action can be written as $S_{\mathrm{eff}}=L\tilde{S}_{\mathrm{eff}}$,
where $\tilde{S}_{\mathrm{eff}}$ is the same as Eq. (\ref{action}),
but with rescaled $u_{\mathrm{spin}}$. Because of the overall factor
$L$ the action $\tilde{S}_{\mathrm{eff}}$ can be approximated by
its value at the minimum, as corrections to the partition function
from fluctuations of $\boldsymbol{\Delta}_{i}$ are small in $\left(\log L\right)/L$.

After solving for the minimum of $S_{\mathrm{eff}}\left[\boldsymbol{\Delta}_{X},\boldsymbol{\Delta}_{Y}\right]$
in Eq. (\ref{action}), we find that, when $g=0$, the ground state
has a huge degeneracy because any configuration $\boldsymbol{\Delta}=\left\langle \boldsymbol{\Delta}_{X}\right\rangle \mathrm{e}^{i\mathbf{Q}_{1}\cdot\mathbf{r}}+\left\langle \boldsymbol{\Delta}_{Y}\right\rangle \mathrm{e}^{i\mathbf{Q}_{1}\cdot\mathbf{r}}$
with $\left\langle \boldsymbol{\Delta}_{X}\right\rangle ^{2}+\left\langle \boldsymbol{\Delta}_{Y}\right\rangle ^{2}=-r_{0}/u$
minimizes $\tilde{S}_{\mathrm{eff}}$. A non-zero $g$ gives rise
to the additional coupling $2g\Delta_{X}^{2}\Delta_{Y}^{2}$, which
breaks this degeneracy. For a positive $g$, this term favors the
states in which only one order parameter has a nonzero value, i.e.
configurations with either $\left\langle \boldsymbol{\Delta}_{X}\right\rangle \neq0$
or $\left\langle \boldsymbol{\Delta}_{X}\right\rangle \neq0$, but
not both.

To relate $\left\langle \boldsymbol{\Delta}_{X}\right\rangle $ or
$\left\langle \boldsymbol{\Delta}_{Y}\right\rangle $ to the magnetic
ordering, we return to the effective action written in terms of the
double functional integral over the fermionic and $\Delta$ fields.
Assuming that the electronic spin $\mathbf{s}_{i,\mathbf{q}=0}=\sum_{\mathbf{k}}c_{\Gamma,\mathbf{k}\alpha}^{\dagger}\boldsymbol{\sigma}_{\alpha\beta}c_{i,\mathbf{k}\beta}$
acquires a non-zero magnitude $\left\langle s_{i,\mathbf{0}}\right\rangle \neq0$,
we again minimize $S_{\mathrm{eff}}$, but this time with $\left\langle s_{i,\mathbf{0}}\right\rangle $
as a parameter, yielding $\left\langle s_{i,\mathbf{0}}\right\rangle =\left\langle \Delta_{i}\right\rangle /u_{\mathrm{spin}}$.
Since a non-zero $s_{X,\mathbf{0}}$ ($s_{Y,0}$) implies magnetic
order with the momentum $\mathbf{Q}_{1}=\left(\pi,0\right)$ ($\mathbf{Q}_{2}=\left(0,\pi\right)$),
the fact that only one of the $\boldsymbol{\Delta}_{i}$ orders means
that the magnetic ordering has a particular momentum. One can easily
verify that in such a state spins order ferromagnetically along one
direction and antiferromagnetically along the other one.

\subsection{The action in terms of the Ising-nematic order parameter}

Since the action (\ref{action}) is invariant with respect to the
interchange between $\boldsymbol{\Delta}_{X}$ and $\boldsymbol{\Delta}_{Y}$,
the onset of either $(\pi,0)$ or $(0,\pi)$ SDW state breaks not
only the conventional $O(3)$ spin-rotational symmetry, but also the
additional $Z_{2}$ (Ising) symmetry associated with choosing between
$\boldsymbol{\Delta}_{X}$ and $\boldsymbol{\Delta}_{Y}$ (Refs. \cite{chandra,Fang08,Xu08}).
The issue we now consider is whether the $Z_{2}$ symmetry breaking
preempts the $O(3)$ symmetry breaking, i.e. it happens before the
Ginzburg-Landau parameter $r_{0}$ changes sign and the magnetic order
sets in.

Such a $Z_{2}$ symmetry breaking without magnetic order would imply
that fluctuations associated with one of the bosonic fields are larger
than the fluctuations associated with the other one, e.g., $\left\langle \Delta_{X}^{2}\right\rangle >\left\langle \Delta_{Y}^{2}\right\rangle $
while $\left\langle \Delta_{X}\right\rangle =\left\langle \Delta_{Y}\right\rangle =0$.
A direct experimental detection of this state could be done by performing
inelastic neutron scattering in detwinned samples and measuring the
spectrum at $(\pi,0)$ and $(0,\pi)$.

That the action (\ref{action}) can potentially lead to a preemptive
Ising-nematic instability is evident from the presence of the term
$g\left(\Delta_{X}^{2}-\Delta_{Y}^{2}\right)^{2}$, which can give
rise to an ordered state with $\left\langle \Delta_{X}^{2}\right\rangle -\left\langle \Delta_{Y}^{2}\right\rangle \neq0$
in a way similar to how the $\boldsymbol{s}_{i,\mathbf{q}}\boldsymbol{s}_{i,-\mathbf{q}}$
term in the Hamiltonian (\ref{H_int}) gives rise to a state with
non-zero $\left\langle \boldsymbol{s}_{i,0}\right\rangle \neq0$.
The preemptive Ising-nematic instability, however, does not appear
in the mean-field limit of $L\rightarrow\infty$ fermionic flavors
simply because magnetic fluctuations are absent at $L=\infty$, and
a non-zero $\left\langle \Delta_{i}^{2}\right\rangle \neq0$ appears
simultaneously to $\left\langle \Delta_{i}\right\rangle \neq0$, once
$r_{0}$ changes sign. However, it may well happen once we return
to the original model with $L=1$ fermionic flavor and include magnetic
fluctuations.

To study a potential preemptive $Z_{2}$ transition, we need to introduce
collective variables of the fields $\boldsymbol{\Delta}_{X}$ and
$\boldsymbol{\Delta}_{Y}$. Let us introduce auxiliary scalar fields
$\phi$ for $\boldsymbol{\Delta}_{X}^{2}-\boldsymbol{\Delta}_{Y}^{2}$
and $\psi$ for $\boldsymbol{\Delta}_{X}^{2}+\boldsymbol{\Delta}_{Y}^{2}$.
The field $\psi$ always has a non-zero expectation value $\left\langle \psi\right\rangle \neq0$,
which describes Gaussian corrections to the magnetic susceptibility
$\chi_{i,q}^{-1}$ in Eq. \ref{chi_i}. Meanwhile, the field $\phi$
may or may not have a non-zero expectation value. If it does, it generates
a non-zero value of $\left\langle \boldsymbol{\Delta}_{X}^{2}-\boldsymbol{\Delta}_{Y}^{2}\right\rangle $
and the system develops an Ising-nematic order.

The effective action in terms of $\phi$ and $\psi$ is obtained by
using again the Hubbard-Stratonovich transformation of Eq. (\ref{ch_X}),
but this time the variable $x$ is either $\Delta_{X}^{2}+\Delta_{Y}^{2}$
or $\Delta_{X}^{2}-\Delta_{Y}^{2}$. After applying this transformation,
we express the partition function in terms of double integrals over
the fields $\left(\boldsymbol{\Delta}_{X},\boldsymbol{\Delta}_{Y}\right)$
and $(\phi,\psi)$:

\begin{equation}
Z\propto\int d\boldsymbol{\Delta}_{X}d\boldsymbol{\Delta}_{Y}d\phi\, d\psi\,\mathrm{e}^{-S_{\mathrm{eff}}\left[\boldsymbol{\Delta}_{i},\phi,\psi\right]}\label{aux_S_Delta_phi}\end{equation}
 where: \begin{eqnarray}
S_{\mathrm{eff}}\left[\boldsymbol{\Delta}_{i},\phi,\psi\right] &  & =\int_{q}\chi_{q}^{-1}\left(\Delta_{X}^{2}+\Delta_{Y}^{2}\right)+\int_{x}\left(\frac{\phi^{2}}{2g}-\frac{\psi^{2}}{2u}\right)+\nonumber \\
 &  & \int_{x}\psi\left(\Delta_{X}^{2}+\Delta_{Y}^{2}\right)+\int_{x}\phi\left(\Delta_{X}^{2}-\Delta_{Y}^{2}\right)\label{S_Delta_phi}\end{eqnarray}

Once $\phi$ becomes non-zero, we have from (\ref{S_Delta_phi}) \begin{equation}
\left\langle \Delta_{X}^{2}\right\rangle -\left\langle \Delta_{Y}^{2}\right\rangle =\frac{\phi}{g}\label{dr_1}\end{equation}

If the magnetic long-range order is not developed, i.e. $\left\langle \Delta_{i}\right\rangle =0$,
then it is straightforward to integrate over the fields $\left(\boldsymbol{\Delta}_{X},\boldsymbol{\Delta}_{Y}\right)$.
Carrying out the integration we obtain the effective action in terms
on $\phi$ and $\psi$ in the form \begin{equation}
S_{\mathrm{eff}}\left[\phi,\psi\right]=\int_{q}\left\{ \frac{\phi^{2}}{2g}-\frac{\psi^{2}}{2u}+\frac{3}{2}\log\left[\left(\chi_{q}^{-1}+\psi\right)^{2}-\phi^{2}\right]\right\} \label{action_phi}\end{equation}

We later modify the derivation of $S_{\mathrm{eff}}\left[\phi,\psi\right]$
to the case where the system has magnetic order. In the next two sections
we will study the effective action (\ref{action_phi}) in its own
mean-field theory and then using a renormalization-group (RG) formalism.

\section{Mean-field theory \label{sec:solution}}

To justify a mean-field treatment of $S_{\mathrm{eff}}$ from (\ref{action_phi})
we do a trick similar to the one before, but with the bosonic rather
than the fermionic variables. Namely, we extend the number of components
of the fields $\boldsymbol{\Delta}_{X,Y}$ from the original value
$N=3$ to arbitrary $N$, assume that the $\phi$ and $\psi$ fields
interact equally with all components of $\boldsymbol{\Delta}_{X,Y}$,
and rescale the coupling constants to $g\to g/N$ and $u\to u/N$.
The effective action (\ref{action_phi}) then has an overall pre-factor
$N$ implying that for large $N$ it can be analyzed by just taking
its value at the extremum.

We emphasize that the extension to $N\rightarrow\infty$ components
is a different approximation than the previous one, in which we made
$L\rightarrow\infty$ copies of the fermionic fields. For $L\rightarrow\infty$,
the terms of the effective action containing the field $\phi$ have
an overall pre-factor of $1/L\rightarrow0$. This means that fluctuations
of $\phi$ are large, preventing a preemptive Ising-nematic order.
On the other hand, by abandoning the large $L$ limit and taking instead
the $N\rightarrow\infty$ limit, we allow strong fluctuations of both
$\Delta_{X}^{2}$ and $\Delta_{Y}^{2}$, but eliminate fluctuations
of the scalar fields $\phi$ and $\psi$ which only account for a
small correction, of order $\left(\log N\right)/N$, to the partition
function. Rewriting the action of Eq. (\ref{action_phi}) for $N$
components of the $\Delta_{i}$ field gives:

\begin{equation}
S_{\mathrm{eff}}\left[\phi,\psi\right]=N\int_{q}\left\{ \frac{\phi^{2}}{2g}-\frac{\psi^{2}}{2u}+\frac{1}{2}\log\left[\left(\chi_{q}^{-1}+\psi\right)^{2}-\phi^{2}\right]\right\} \label{S_phi_N}\end{equation}

The mean-field theory for this action is the saddle-point solution
of (\ref{S_phi_N}), i.e. the minimum of $S_{\mathrm{eff}}\left[\phi,\psi\right]$
with respect to $\phi$, and the maximum with respect to $\psi$.
That $e^{-S_{\mathrm{eff}}\left[\phi,\psi\right]}$ increases when
$\psi$ gets larger is related to the fact that Gaussian corrections
to $r_{0}$ coming from the quartic term $(\Delta_{X}^{2}+\Delta_{Y}^{2})^{2}$
are confined to the upper cutoff of the theory. These corrections
just renormalize $r_{0}$ by a constant, independent on $\phi$, and
play no role in our analysis. What matters to us is how $\psi$ is
affected by $\phi$ and vise versa. These mutually influencing terms
are independent on the upper cutoff and are therefore well described
within the mean-field theory. Differentiating Eq. (\ref{S_phi_N})
with respect to $\phi$ and $\psi$ and taking $\partial S_{\mathrm{eff}}\left[\phi,\psi\right]/\partial\phi=\partial S_{\mathrm{eff}}\left[\phi,\psi\right]/\partial\psi=0$,
we obtain two non-linear coupled equations for $\phi$ and $\psi$:
\begin{eqnarray}
\frac{\psi}{u} & = & \int_{q}\frac{r_{0}+\psi+q^{2}+\gamma|\nu_{n}|}{\left(r_{0}+\psi+q^{2}+\gamma|\nu_{n}|\right)^{2}-\phi^{2}}\nonumber \\
\frac{\phi}{g} & = & \int_{q}\frac{\phi}{\left(r_{0}+\psi+q^{2}+\gamma|\nu_{n}|\right)^{2}-\phi^{2}}\label{self_cons}\end{eqnarray}

It is convenient to re-express the first formula as a self-consistent
equation for the renormalized mass of the bosonic field, $r=r_{0}+\psi$.
In the paramagnetic phase, $r\propto\xi^{-2}$, where $\xi$ is the
magnetic correlation length. It is also useful to remove the high-energy
contribution to $\psi$ (i.e. the contribution coming from the upper
cutoff), by incorporating it into the renormalization of $r_{0}$.
Specifically, we rewrite the first equation as: \begin{eqnarray}
r & = & r_{0}+\psi=r_{0}+u\int_{q}\frac{r+q^{2}+\gamma|\nu_{n}|}{\left(r+q^{2}+\gamma|\nu_{n}|\right)^{2}-\phi^{2}}\label{aux_self_cons}\\
 & = & \bar{r}_{0}+u\int_{q}\left[\frac{r+q^{2}+\gamma|\nu_{n}|}{\left(r+q^{2}+\gamma|\nu_{n}|\right)^{2}-\phi^{2}}-\frac{1}{q^{2}+\gamma|\nu_{n}|}\right]\nonumber \end{eqnarray}
 where $\bar{r}_{0}=r_{0}+u\int_{q}\frac{1}{q^{2}+\gamma|\nu_{n}|}.$
For classical systems, only the $\nu_{n}=0$ term matters. The remaining
momentum integral in the last line in (\ref{aux_self_cons}) is infrared
and ultraviolet convergent for $2<d<4$, such that the upper limit
of integration can be safely extended to infinity. Since $r_{0}$
and $\bar{r}_{0}$ differ only by a constant, $\bar{r}_{0}$ is also
a monotonic increasing function of $T$ and can be expressed as $\bar{r}_{0}=\bar{a}\left(T-\bar{T}_{N,0}\right)$.
For quantum systems such a renormalization of $r_{0}$ is not enough
in $d>2$, and the $d+1$ dimensional integral over momenta and frequency
in (\ref{aux_self_cons}) is still generally confined to the upper
cutoff. In this situation, we use additional renormalizations (see
below) to restrict the consideration to small energies, at which the
effective action describes universal low energy behavior.

For $\phi=0$, magnetic order emerges when $r=0$, i.e. when the static
susceptibility for the fields $\Delta_{X}$ and $\Delta_{Y}$, $\chi_{0}^{-1}=r$,
diverges. The relationship between $r$ and $\bar{r}_{0}$ follows
from (\ref{aux_self_cons}):

\begin{equation}
r=\bar{r}_{0}+u\int_{q}\left[\frac{1}{r+q^{2}+\gamma|\nu_{n}|}-\frac{1}{q^{2}+\gamma|\nu_{n}|}\right]\label{aux_self_cons2}\end{equation}

For classical systems the integrals are infrared divergent for $d\leq2$,
meaning that $r$ never reaches zero - this is nothing but the Mermin-Wagner
theorem. For $d>2$ we immediately find from (\ref{aux_self_cons2})
that $r$ and $\bar{r}_{0}$ vanish simultaneously, i.e. in the absence
of a preemptive Ising instability long-range magnetic order appears
at $r=\bar{r}_{0}=0$.

Suppose now that a non-zero solution for $\phi$ appears already at
$r>0$, i.e. when the system is still in the paramagnetic phase. Once
$\phi$ becomes non-zero, the static magnetic susceptibilities for
the fields $\Delta_{X}$ and $\Delta_{Y}$ become non-equivalent:\begin{eqnarray}
\chi_{X}\left(q=0\right) & = & \frac{1}{r-\phi}\nonumber \\
\chi_{Y}\left(q=0\right) & = & \frac{1}{r+\phi}\label{magnetic_suscept}\end{eqnarray}

Now the magnetic transition occurs when $r=\left|\phi\right|$, i.e.
at a temperature larger than without Ising-nematic order. In other
words, a preemptive Ising-nematic order increases the magnetic transition
temperature.

In what follows, we analyze the phase diagrams resulting from the
set of non-linear equations (\ref{self_cons}) in three different
regimes: the classical regime, where thermal fluctuations dominate,
$T\gamma\gg\bar{r}_{0}$ (Section III-A); the quantum regime, $T=0$
(Section III-B); and the regime of intermediate temperature $T\gamma\sim\bar{r}_{0}$,
where thermal and quantum fluctuations are equally important (Section
III-C). In all cases, the key parameter that controls the characters
of the magnetic and nematic transitions is the dimensionless ratio:

\begin{equation}
\alpha\equiv\frac{u}{g}\label{alpha}\end{equation}
 which measures the strength of the nematic coupling $g$ in units
of the magnetic coupling constant $u$. As it is evident from Eqs.
(\ref{parameters_quartic}) and (\ref{parameters_quartic_1}), $\alpha$
depends on the parameters describing the band structure dispersions.
It is therefore affected by changes in the chemical potential and
in the ellipticity - see Section VI for a systematic analysis of $\alpha$
as function of doping, pressure, and lattice stiffness.

In the classical regime (Section III-A) and in the intermediate regime
(Section III-C), the other independent variable in the phase diagrams
(besides $\alpha$) is the temperature. In the former case, the temperature
dependence appears only via the difference from the mean-field Neel
transition temperature, $T-T_{N,0}$. In the latter, we consider the
explicit dependency on $T$. On the other hand, in the $T=0$ quantum
regime (Section III-B), the independent variable is the distance to
the mean-field critical point at $r_{0}=0$.

Our primary interest is to obtain the phase diagram in the most general
case of a quasi-two-dimensional system with anisotropic magnetic interactions.
To better understand the results in this case, we first consider a
few limiting cases:

\begin{itemize}
\item We first analyze the classical $d=2$ case in Section III-A-1. In
this situation, there is no finite temperature magnetic transition,
but there is always an intermediary paramagnetic phase with nematic
order. At small $\alpha$ (large nematic coupling $g$), the nematic
transition is first-order, whereas at larger $\alpha$ (smaller $g$),
it is second-order. The $d=2$ case is investigated also in the quantum
limit (Section III-B-1) and in the intermediate-temperature regime
(Section III-C-1). The behavior in the intermediate regime is the
same as in the classical regime. At $T=0$, we find instead that the
magnetic and the nematic transitions are simultaneous and first-order.\textbf{ }
\item Next we analyze the classical $d=3$ case with isotropic magnetic
interactions in Section III-A-2. We find that, regardless of the value
of the dimensionless coupling $\alpha$, the magnetic and nematic
transitions are always simultaneous and first-order. We obtain a similar
result in the quantum case (Section III-B-2) and in the intermediary
temperature regime (Section III-C-2).\textbf{ }
\item To model the more general case of anisotropic quasi-2D systems, we
first consider in Section III-A-3 the classical system with isotropic
dispersion in arbitrary dimensionality $2<d<3$. We find that, for
large values of $\alpha$ (small $g$), the system behavior is the
same as in the $d=2$ case, with split second-order magnetic and nematic
transitions. On the other hand, for small $\alpha$ (large $g$),
the behavior is similar to the $d=3$ case, with simultaneous first-order
transitions. The two regimes are separated by two tri-critical points
and an intermediary regime with a first-order nematic transition split
from a lower-temperature second-order magnetic transition. The quantum
and intermediary-temperature regimes are analyzed in Sections III-B-3
and III-C-3, respectively.\textbf{ }
\item In Section III-A-4, we obtain the classical phase diagram of a quasi-two-dimensional
system with anisotropic out-of-plane magnetic dispersion, characterized
by the parameter $\eta_{z}$. For small $\eta_{z}$ (weak anisotropy),
the phase diagram is the same as in the previous case of intermediary
dimension $2<d<3$ and isotropic interactions. For larger $\eta_{z}$
(moderate anisotropy), we still obtain a regime of split second-order
nematic and magnetic transitions at large $\alpha$ and a regime of
simultaneous first-order transitions at small $\alpha$. However,
the two regimes are now separated by an intermediary regime with a
higher-temperature second-order nematic transition split from a first-order
magnetic transition. In Sections III-B-4 and III-C-4, we present the
results for the system behavior in the quantum and intermediary-temperature
regimes, respectively.\textbf{ }
\item Finally, in Section III-A-5, we analyze in more details the intermediary
regime of split second-order nematic and first-order magnetic transitions,
which appear in the classical phase diagram of the quasi-two-dimensional
system with moderate out-of-plane anisotropy. In particular, we show
that the first-order magnetic transition is simultaneous to a meta-nematic
transition, where the nematic order parameter jumps between two finite
values. 
\end{itemize}

\subsection{ Classical phase diagram as a function of $u/g$}

When the temperature is high enough ($T\gamma\gg\bar{r}_{0}$), the
dominant contribution to the sum over Matsubara frequencies, $\int_{q}=T\sum_{\nu_{n}}\int d^{d}q/(2\pi)^{d}$
in Eqs. (\ref{self_cons}), comes from the term with zero Matsubara
frequency. This approximation substantially simplifies Eqs. (\ref{self_cons})
as the remaining momentum integrals can be evaluated exactly. We begin
with the $d=2$ case.

\subsubsection{The case $d=2$}

For $d=2$, the integration over momentum in Eqs. (\ref{self_cons})
yields:

\begin{eqnarray}
r & = & \bar{r}_{0}-\frac{\bar{u}}{4}\log\left(r^{2}-\phi^{2}\right)\nonumber \\
r & = & \phi\coth\left(\frac{2\phi}{\bar{g}}\right)\label{sa_2}\end{eqnarray}
 where we defined the renormalized parameters $\bar{r}_{0}=r_{0}+\bar{u}\log\Lambda$,
$\bar{u}=uT/(2\pi)$, and $\bar{g}=gT/(2\pi)$, with $\Lambda$ denoting
the upper momentum cutoff. The second equation implies that a solution
with $\phi\neq0$ is only possible when $r>0$. Eliminating $r$ from
these equations, we obtain after further rescaling:

\begin{equation}
\phi^{*}\coth\phi^{*}+\alpha\log\left(\frac{\phi^{*}}{\sinh\phi^{*}}\right)=\bar{\bar{r}}_{0}\label{aux_sa_2}\end{equation}
 where $\phi^{*}=2\phi/\bar{g}$, $\bar{\bar{r}}_{0}=2\bar{r}_{0}/\bar{g}-\alpha\log\left(\bar{g}/2\right)$,
and $\alpha=u/g$ (see Eq. \ref{alpha}).

\begin{figure}
\begin{centering}
\includegraphics[width=0.7\columnwidth]{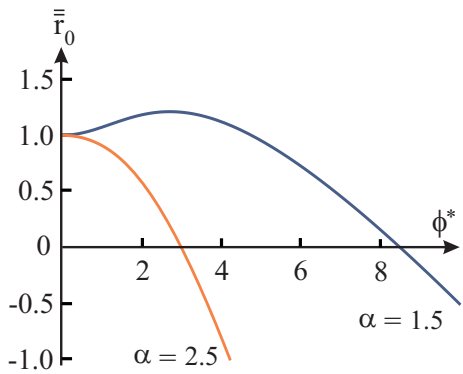} 
\par\end{centering}

\caption{(Color online) $\bar{\bar{r}}_{0}$ as the function of the Ising-nematic
order parameter $\phi^{*}=2\phi/\bar{g}$ for two representative values
$1<\alpha<2$ (blue/dark line) and $\alpha>2$ (orange/light line).
The value of $\phi^{*}$ at which the solution first appears upon
decreasing $\bar{\bar{r}}_{0}$ sets the type of the nematic transition.
For $\alpha>2$, the solution first emerges at $\phi^{*}=0$, and
the transition is second-order. For $1<\alpha<2$, $\bar{\bar{r}}_{0}$
is non-monotonic function of $\phi^{*}$ and the solution first emerges
at a non-zero $\phi^{*}$. The nematic transition then becomes first-order. }

\label{fig_2D_r0} 
\end{figure}

\begin{figure}[htp]
\begin{centering}
\includegraphics[width=0.8\columnwidth]{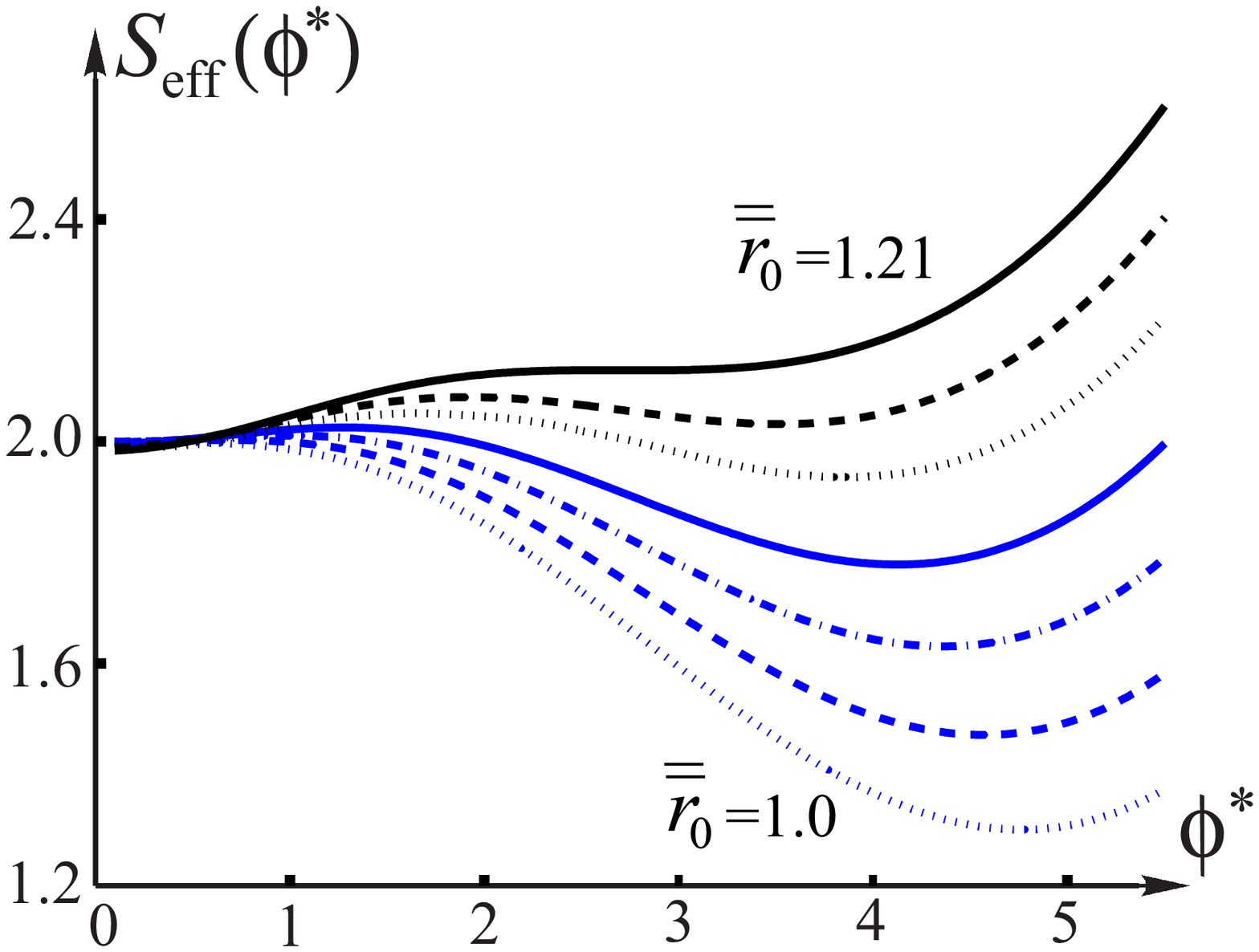} 
\par\end{centering}

\caption{(Color online) The effective action $\tilde{S}_{\mathrm{eff}}\left[\alpha,\,\phi^{*},\bar{\bar{r}}_{0}\right]$
from Eq. \ref{S_d2} as a function of $\phi^{*}$ for $\alpha=1.5$
and several $\bar{\bar{r}}_{0}$. The evolution of $\tilde{S}_{\mathrm{eff}}$
is typical of a first-order phase transition. At large $\bar{\bar{r}}_{0}$,
$\tilde{S}_{\mathrm{eff}}\left[1.5,\,\phi^{*},\bar{\bar{r}}_{0}\right]$
monotonically increases with increasing $\phi^{*}$. At $\bar{\bar{r}}_{0}^{max}=1.21$,
the effective action develops an inflection point (black/dark solid
curve) which at smaller $\bar{\bar{r}}_{0}$ splits into a maximum
and a minimum. At $\bar{\bar{r}}_{0}^{cr}=1.15$, the values of $\tilde{S}_{\mathrm{eff}}$
at $\phi^{*}=0$ and at the local minimum at finite $\phi^{*}$ become
equal (black dotted curve), and the system undergoes a first-order
Ising-nematic transition. The local minimum in $S_{\mathrm{eff}}$
at $\phi^{*}=0$ survives down to $\bar{\bar{r}}_{0}^{min}=1$ (blue/light
gray dotted curve).}

\label{fig:Sphi} 
\end{figure}

Recall that the original model is constrained to $u>g$, i.e. $\alpha>1$.
The variable $\bar{\bar{r}}_{0}$ decreases with decreasing $T$,
since it only differs by a constant and by an overall factor from
$r_{0}$. The leading instability of the system upon decreasing $T$
is into a state with $\phi^{*}$ corresponding to the maximum of the
left-hand side of Eq. (\ref{aux_sa_2}). A simple analysis shows that
the maximum is at $\phi^{*}=0$ for $\alpha\geq2$ and at a finite
$\phi^{*}$ for $1<\alpha<2$ (see Fig. \ref{fig_2D_r0}). The implication
is that, for $\alpha\geq2$, the system undergoes a second-order Ising-nematic
transition at $\bar{\bar{r}}_{0}^{cr}=1$ ($r=\bar{g}/2$), while
for $1<\alpha<2$ the Ising-nematic transition becomes first-order
and the solution for $\phi^{*}$ first appears at a larger $\bar{\bar{r}}_{0}=\bar{\bar{r}}_{0}^{max}$
(see Figs. \ref{fig_2D_r0} and \ref{fig:Sphi}). The value $\phi_{cr}^{*}$
at which the left-hand side of Eq. (\ref{aux_sa_2}) has a maximum
gradually increases as $\alpha$ decreases, approaching infinity as
$\alpha\to1$.

The actual $\bar{\bar{r}}_{0}^{cr}$ at which the first-order Ising-nematic
transition occurs is somewhat smaller than $\bar{\bar{r}}_{0}^{max}$
because at $\bar{\bar{r}}_{0}^{max}$ the effective action $S_{\mathrm{eff}}$
only develops a local minimum at a non-zero $\phi_{cr}^{*}$. The
actual transition occurs when the value of the effective action at
the local minimum becomes equal to $S_{\mathrm{eff}}(\phi^{*}=0)$.
To obtain $\bar{\bar{r}}_{0}^{cr}$ we then need to evaluate the effective
action at both minima $\phi^{*}\neq0$ and $\phi^{*}=0$, and find
$\bar{\bar{r}}_{0}^{cr}$ at which the two terms become equal. For
better clarity, we compute $S_{\mathrm{eff}}$ not only at the extrema
(when the self-consistent equations (\ref{sa_2}) are valid), but
for arbitrary $\phi^{*}$ at a given $\bar{\bar{r}}_{0}$. To do this,
we solve the equation $\partial S_{\mathrm{eff}}/\partial\psi=0$
to obtain $r\left(\alpha,\,\phi^{*},\bar{\bar{r}}_{0}\right)$, substitute
it back into the effective action and obtain $S_{\mathrm{eff}}\left[\alpha,\,\phi^{*},\bar{\bar{r}}_{0}\right]$.
Carrying out the calculations, we obtain, neglecting a constant term,
\begin{equation}
S_{\mathrm{eff}}\left[\alpha,\,\phi^{*},\bar{\bar{r}}_{0}\right]=\frac{\bar{g}}{8}\,\tilde{S}_{\mathrm{eff}}\left[\alpha,\,\phi^{*},\bar{\bar{r}}_{0}\right],\end{equation}
 where \begin{eqnarray}
 &  & \tilde{S}_{\mathrm{eff}}\left[\alpha,\,\phi^{*},\bar{\bar{r}}_{0}\right]=\left(\phi^{*}\right)^{2}+r^{*}\left(2-\log\left[\left(r^{*}\right)^{2}-\left(\phi^{*}\right)^{2}\right]\right)\nonumber \\
 &  & -\frac{\alpha}{4}\log^{2}\left[\left(r^{*}\right)^{2}-\left(\phi^{*}\right)^{2}\right]-\phi^{*}\log\left(\frac{r^{*}+\phi^{*}}{r^{*}-\phi^{*}}\right)\label{S_d2}\end{eqnarray}
 and $r^{*}=r^{*}\left(\phi^{*},\alpha,\bar{\bar{r}}_{0}\right)$
is the solution of the equation \begin{equation}
r^{*}+\frac{\alpha}{2}\log\left[\left(r^{*}\right)^{2}-\left(\phi^{*}\right)^{2}\right]=\bar{\bar{r}}_{0}\label{S_d22}\end{equation}

In Fig. \ref{fig:Sphi} we plot $\tilde{S}_{\mathrm{eff}}\left[\alpha,\,\phi^{*},\bar{\bar{r}}_{0}\right]$
for $\alpha=1.5$ as a function of $\phi^{*}$ for several values
of $\bar{\bar{r}}_{0}$. We see that for $\bar{\bar{r}}_{0}>1.21$,
$S_{\mathrm{eff}}$ has a minimum only at $\phi^{*}=0$, and $S_{\mathrm{eff}}\left[1.5,\,\phi^{*},\bar{\bar{r}}_{0}\right]$
monotonically increases with increasing $\phi^{*}$. However, once
$\bar{\bar{r}}_{0}$ becomes smaller than $\bar{\bar{r}}_{0}^{max}$,
which for this value of $\alpha$ is $\bar{\bar{r}}_{0}^{max}=1.21$,
the function $S_{\mathrm{eff}}\left[1.5,\,\phi^{*},\bar{\bar{r}}_{0}\right]$
develops an inflection point at $\phi_{cr}^{*}\approx2.72$. At smaller
$\bar{\bar{r}}_{0}$, this inflection point gradually splits into
a maximum at $\phi<\phi_{cr}^{*}$ and a minimum at $\phi>\phi_{cr}^{*}$.
At $\bar{\bar{r}}_{0}^{cr}=1.15$ the values of $\tilde{S}_{\mathrm{eff}}$
at $\phi^{*}=0$ and at the local minimum $\phi^{*}\neq0$ become
equal, and the system undergoes a first-order Ising-nematic transition.
The local minimum of $S_{\mathrm{eff}}$ at $\phi^{*}=0$ survives
down to $\bar{\bar{r}}_{0}^{min}=1$. Below this temperature, the
effective action has only one minimum at a finite $\phi^{*}$.

The jump in $\phi$ at the first-order transition affects the susceptibilities
associated with the two magnetic order parameters $\Delta_{X}$ and
$\Delta_{Y}$, which become non-equivalent once $\phi$ becomes finite,
see Eq. (\ref{magnetic_suscept}). This implies that the static susceptibility
and the magnetic correlation length change discontinuously at the
first-order Ising transition, even though there is no magnetic instability
(the value to which $\phi$ jumps is always smaller than $r$, see
the second equation in (\ref{sa_2})). Actually, as we already mentioned,
magnetic order never emerges for $d=2$ at a finite temperature, so
$Z_{2}$ is the only symmetry that gets broken.

\begin{figure}
\begin{centering}
\includegraphics[width=1\columnwidth]{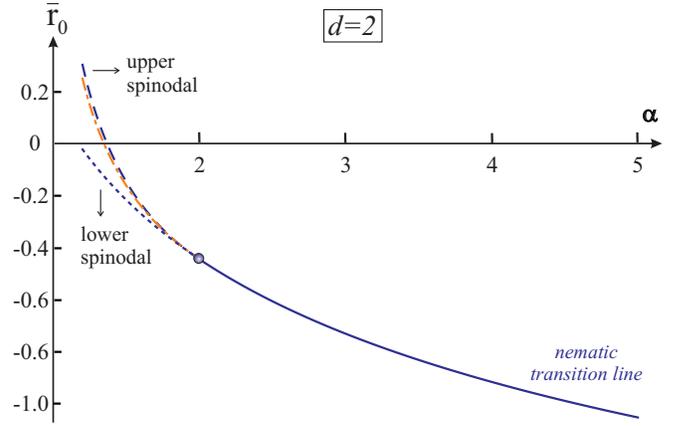} 
\par\end{centering}

\caption{(Color online) Calculated phase diagram in the $\left(\alpha,\bar{r}_{0}\right)$
plane for the $d=2$ case and high enough temperatures, where classical
fluctuations dominate. In the region $\alpha<2$, the nematic transition
is first-order. The dashed blue/dark line represents the upper spinodal
(where the ordered state becomes metastable), while the dotted blue/dark
line denotes the lower spinodal (where the disordered state ceases
to be metastable), and the dashed-dotted orange/light line is the
actual transition line where the global minimum of $S_{\mathrm{eff}}$
shifts from $\phi=0$ to $\phi\neq0$. We recall there is no magnetic
order at a finite temperature in $d=2$. In this figure, we fixed
$\bar{u}=1$ and changed $\bar{g}$. }

\label{fig_2D_phase_diagram} 
\end{figure}

To obtain the phase diagram in the $\left(\alpha,\bar{\bar{r}}_{0}\right)$
plane, we need to analyze the behavior of the system once the Ising
order sets in. We found that the Ising-nematic order parameter continuously
increases with decreasing $\bar{\bar{r}}_{0}$ for all $\alpha>1$,
implying that there is no other first-order transition line in the
phase diagram besides the one at which the Ising-nematic order develops.
The phase diagram is presented in Fig. \ref{fig_2D_phase_diagram}.
The upper spinodal in this figure corresponds to $\bar{r}_{0}^{max}$
where a local minimum of $S_{\mathrm{eff}}$ appears at $\phi^{*}\neq0$,
and the lower spinodal refers to $\bar{r}_{0}^{min}$ where $\phi^{*}=0$
ceases to be a local minimum of $S_{\mathrm{eff}}$. A first-order
transition happens in between the upper and lower spinodal lines.

\subsubsection{The case $d=3$}

A very different phase diagram emerges in three dimensions. The momentum
integration in Eqs. (\ref{self_cons}) now yields

\begin{eqnarray}
r & = & \bar{r}_{0}-\frac{\bar{u}}{4}\left(\sqrt{r+\phi}+\sqrt{r-\phi}\right)\nonumber \\
\phi & = & \frac{\bar{g}}{4}\left(\sqrt{r+\phi}-\sqrt{r-\phi}\right)\label{sa_d3}\end{eqnarray}
 where $\bar{r}_{0}=r_{0}-2\Lambda\bar{u}/\pi$, and, as before, $\bar{u}=uT/\left(2\pi\right)$,
and $\bar{g}=gT/\left(2\pi\right)$. It is convenient to re-express
these equations in terms of $r$ and $z\equiv\phi/r$ ($0\leq z\leq1$).
Eliminating $r$ from (\ref{sa_d3}) we then obtain:

\begin{equation}
\bar{r}_{0}=\frac{\bar{g}^{2}}{8}\left(\alpha+\frac{1}{1+\sqrt{1-z^{2}}}\right)\label{aux_sa_d3}\end{equation}

The right-hand side of (\ref{aux_sa_d3}) is an increasing function
of $z$ for all values of $\alpha$. Therefore, the first instability
of the system is at $\bar{r}_{0}=\bar{g}^{2}\left(1+\alpha\right)/8$
towards the state with the largest possible value of $z$, namely,
$z=1$. As a result, the order parameter $\phi$ jumps at the Ising-nematic
transition to $\phi=r$. For such $\phi$, the static susceptibility
associated with one of the magnetic order parameters diverges, see
Eqs. (\ref{magnetic_suscept}), implying that the first-order Ising-nematic
transition instantaneously brings the system to the verge of magnetic
order. The divergence of the susceptibility reflects the fact that
the Goldstone modes of the magnetically ordered state are gapless.
In the large-$N$ limit, the impact of longitudinal fluctuations,
where the susceptibility remains finite, is negligible.

In this situation, we have to extend our analysis and investigate
the possibility that $\Delta_{X}$ jumps to a non-zero value at the
Ising transition. To do this, we go back to the effective action (\ref{aux_S_Delta_phi})
written in terms of double integrals over both $\Delta_{X,Y}$ and
their collective variables $\phi$ and $\psi$, and allow one of components
of $\Delta_{X}$ to have a non-zero mean value $M=\left\langle \Delta_{X}^{i}\right\rangle $.
Expressing $\Delta_{X,\mathbf{q}}^{j}=M\delta_{\mathbf{q},0}\delta^{i,j}+\tilde{\Delta}_{X,\mathbf{q}}^{j}$
and integrating over $\tilde{\boldsymbol{\Delta}}_{X}$ and $\boldsymbol{\Delta}_{Y}$
we obtain: \begin{eqnarray}
 &  & S_{\mathrm{eff}}\left[\phi,\psi,M\right]=S_{\mathrm{eff}}\left[\phi,\psi\right]+\nonumber \\
 &  & M^{2}\left(r-\phi\right)\left[1+\left(r-\phi\right)^{2}\int_{q}\frac{1}{q^{2}\left(r-\phi+q^{2}\right)}\right]\label{action_Xm}\end{eqnarray}

In the spirit of the $N\rightarrow\infty$ approximation, we rescale
$M^{2}\rightarrow M^{2}N$ and differentiate $S_{\mathrm{eff}}\left[\phi,\psi,M\right]$
over all three variables. This yields the set of three self-consistent
equations for $\phi$, $r$, and $M$. The equation for $M$ is:

\begin{equation}
M\left(r-\phi\right)\left(1+\left(r-\phi\right)^{2}\int_{q}\frac{1}{q^{2}\left(r-\phi+q^{2}\right)}\right)=0.\label{sa_m}\end{equation}

The solution is either $M=0$ or $r=\phi$. We take $M\neq0$ and
$r=\phi$. The other two equations for $r=\phi$ are\begin{eqnarray}
\phi & = & \bar{r}_{0}-\frac{\bar{u}}{4}\sqrt{2\phi}+\bar{u}\bar{M}^{2},\nonumber \\
\phi & = & \frac{\bar{g}}{4}\sqrt{2\phi}+\bar{g}\bar{M}^{2},\label{self_cons_m}\end{eqnarray}
 where $\bar{M}=M\sqrt{2\pi/\bar{T}}$. Solving the first equation
for $\phi$ we obtain:

\begin{equation}
\sqrt{\phi}=\frac{\bar{g}}{4\sqrt{2}}\left(1+\sqrt{1+\frac{32\bar{M}^{2}}{\bar{g}}}\right).\label{aux_self_cons_m}\end{equation}

Substituting this solution into the second equation yields $\bar{r}_{0}$
as a function of $M$:

\begin{eqnarray}
\bar{r}_{0} & = & \frac{\bar{g}^{2}}{16}\left[-16\frac{\bar{M}^{2}}{\bar{g}}\left(\alpha-1\right)+\right.\label{r0_m}\\
 &  & \left.\left(1+\alpha\right)\left(1+\sqrt{1+\frac{32\bar{M}^{2}}{\bar{g}}}\right)\right]\nonumber \end{eqnarray}

Following the same strategy as before, we look for the value of $\bar{M}$
correspondent to the largest $\bar{r}_{0}$, i.e. we determine the
value of $\bar{M}$ that emerges at the largest temperature. A straightforward
calculations shows that it is non-zero and equals to \begin{equation}
\bar{M}=\frac{\sqrt{2\bar{g}\alpha}}{4(\alpha-1)}.\label{aux_r0_m}\end{equation}

This implies that the first-order Ising-nematic transition at which
$\phi$ reaches its maximum value ($=r$) triggers a first-order magnetic
transition into a state with a finite $M=\left\langle \Delta_{X}^{i}\right\rangle $.
We emphasize, however, that in this and similar cases which we consider
below, the magnitude of the nematic order parameter $\phi$ is larger
than $g\bar{M}^{2}$, i.e. the magnetic transition is secondary to
the nematic transition. Indeed, substituting $\bar{M}$ from (\ref{aux_r0_m})
into (\ref{aux_self_cons_m}) we obtain \begin{equation}
\phi=\alpha\left(\bar{g}\bar{M}^{2}\right)>\bar{g}\bar{M}^{2}.\label{dr_2}\end{equation}

We emphasize that, without nematic instability, a {}``pure'' magnetic
transition would take place at a smaller temperature, when the condition
$r=0$ is satisfied, instead of $r=\phi$. Therefore, even though
both transitions are simultaneous and first-order, the nematic one
is the primary transition, and the magnetic transition is induced
by the feedback from the nematic order.

\begin{figure}
\begin{centering}
\includegraphics[width=1\columnwidth]{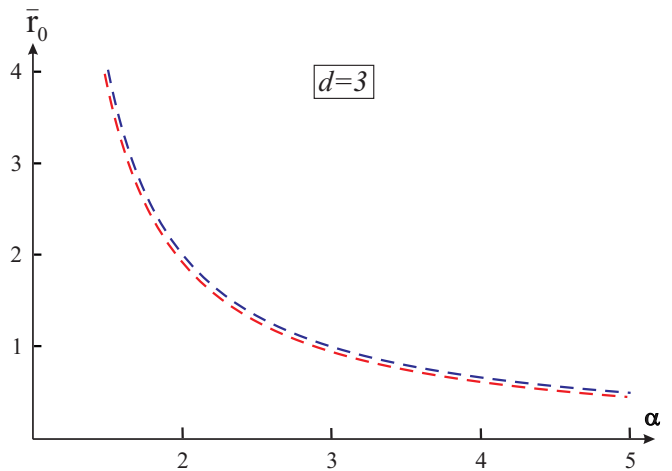} 
\par\end{centering}

\caption{(Color online) Classical phase diagram in the $\left(\alpha,\bar{r}_{0}\right)$
plane for the $d=3$ case. For any value of $\alpha$, the nematic
and magnetic transition are simultaneous and first-order. }

\label{fig_3D_phase_diagram} 
\end{figure}

Note also that in the mean-field theory the susceptibility associated
with the field $\boldsymbol{\Delta}_{X}$ remains massless ($\chi_{X}^{-1}=r-\phi=0$)
despite that magnetic order develops and enforces a gap in the spectrum
of longitudinal fluctuations. This is a consequence of taking the
$N\rightarrow\infty$ limit, in which one longitudinal mode is negligible
compared to $N-1$ gapless transverse modes.

Like in the $d=2$ case, the value of $\bar{r}_{0}$, at which the
effective action develops an inflection point and the solution with
$\phi=r$ and $M\neq0$ first appears, is the upper spinodal $\bar{r}_{0}^{max}$.
The actual first-order transition occurs at a smaller $\bar{r}_{0}^{cr}$,
at which the values of $S_{\mathrm{eff}}$ at $\phi=M=0$ and at the
local minimum at $\phi=r$, $M\neq0$ become equal (for $r=\phi$
the local minimum is with respect to variations of $M$).

The phase diagram for $d=3$ is presented in Fig. \ref{fig_3D_phase_diagram}.
As in the $d=2$ case, there are no additional transition lines, i.e.
after jumping to finite values at the first-order transition, both
$\phi$ and $M$ continuously increase with decreasing $\bar{r}_{0}$.
The similarities between the $d=2$ and $d=3$ cases, however, stop
here because the phases below the critical line are very different.
In $d=3$, Ising-nematic order immediately triggers a magnetic order,
such that there is no regime where the $Z_{2}$ symmetry is broken
but the $O(3)$ symmetry is unbroken. In $d=2$, there is no magnetic
order below the Ising-nematic transition, which can be either first-order
or second-order, depending on $\alpha$.

We now consider what happens in anisotropic three dimensional systems.
There are several ways to model the anisotropy: one can either consider
the dimension $d$ to be an arbitrary number between $2$ and $3$
or one can keep $d=3$ but consider an anisotropic magnetic dispersion
with different stiffness along $q_{z}$ and along $q_{x},\, q_{y}$.
It turns out that the system behavior is universal at small and large
$\alpha$, but at intermediate $\alpha$ it depends on the choice
of the model. This will lead us to the two phase diagrams shown in
Fig. \ref{fig_exp_phase_diagram}. We first consider arbitrary $2<d<3$
and then an anisotropic $d=3$ dispersion.

\subsubsection{The case $2<d<3$.}

We assume first that $M=0$ and later extend the formalism to include
a non-zero magnetization. For $2<d<3$, the integration over momentum
in Eqs. (\ref{self_cons}) yields \begin{eqnarray}
r & = & \bar{r}_{0}-\frac{\bar{u}}{4}\left[\left(r+\phi\right)^{\frac{d-2}{2}}+\left(r-\phi\right)^{\frac{d-2}{2}}\right]\nonumber \\
\phi & = & \frac{\bar{g}}{4}\left[\left(r+\phi\right)^{\frac{d-2}{2}}-\left(r-\phi\right)^{\frac{d-2}{2}}\right],\label{self_consistent_eq}\end{eqnarray}
 where:

\begin{eqnarray}
\frac{\bar{u}}{u}=\frac{\bar{g}}{g} & = & \frac{2TS_{d}}{\left(2\pi\right)^{d}}\int_{0}^{\infty}dx\,\frac{x^{d-3}}{1+x^{2}}\label{aux_self_consistent_eq}\\
 & = & \frac{T}{2^{d-1}\pi^{\frac{d-2}{2}}\Gamma\left(\frac{d}{2}\right)\sin\left(\frac{\left(d-2\right)\pi}{2}\right)}\nonumber \end{eqnarray}
 and $S_{d}=2\pi^{d/2}/\Gamma(d/2)$ is the area of a $d$-dimensional
sphere with unit radius. Introducing, as before, $z\equiv\phi/r$
($0\leq z\leq1$), we solve the second equation for $r\left(z\right)$,
substitute the solution into the first equation, and obtain:

\begin{equation}
\bar{r}_{0}=\left(\frac{\bar{g}}{4}\right)^{\frac{2}{4-d}}Q\left(z,\alpha\right),\label{r0_alpha}\end{equation}
 with:

\begin{eqnarray}
 &  & Q\left(z,\alpha\right)=\left[\frac{\left(1+z\right)^{\frac{d-2}{2}}-\left(1-z\right)^{\frac{d-2}{2}}}{z}\right]^{\frac{d-2}{4-d}}\times\nonumber \\
 &  & \left[\left(1+z\right)^{\frac{d-2}{2}}\left(\alpha+\frac{1}{z}\right)+\left(1-z\right)^{\frac{d-2}{2}}\left(\alpha-\frac{1}{z}\right)\right].\label{Q_alpha_z}\end{eqnarray}

\begin{figure}
\begin{centering}
\includegraphics[width=0.7\columnwidth]{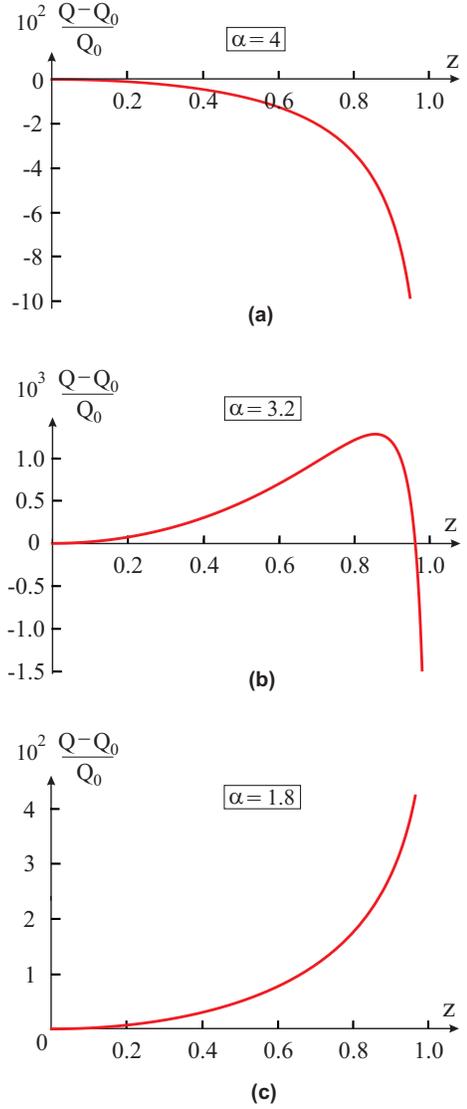} 
\par\end{centering}

\caption{(Color online) $Q\left(z,\alpha\right)\propto\bar{r}_{0}$, as defined
in Eqs. (\ref{r0_alpha}) and (\ref{Q_alpha_z}), plotted as function
of $z\equiv\phi/r$ for three representative values of $\alpha$ at
$d=2.5$: (a) $\alpha>\alpha_{c2}=3.5$, (b) $2=\alpha_{c1}<\alpha<\alpha_{c2}$,
and (c) $\alpha<\alpha_{c1}$. The maximum of $Q$ shifts from $z=0$
at large $\alpha$ to intermediate $z$ at intermediate $\alpha$,
and to $z=1$ at small $\alpha$. }

\label{fig_Q_z} 
\end{figure}

We determine the leading instability of the system by again looking
at the location of the maximum of $Q\left(z,\alpha\right)$ at a given
$\alpha$. In Fig. \ref{fig_Q_z}, we plot $Q(z,\alpha)$ for $d=2.5$
and three different values of $\alpha$. We find three different regimes
for the behavior of the system: For $1<\alpha<\alpha_{c1}=2$, the
first instability is at $z=1$; for $\alpha_{c1}<\alpha<\alpha_{c2}=3.5$,
the first instability is at an intermediate value $0<z<1$; for $\alpha>\alpha_{c2}$,
the first instability is at $z=0$. Expanding $Q\left(z,\alpha\right)$
near $z=0$ and $z=1$ and analyzing the sign of the slope, we find
the expressions for $\alpha_{c1}$ and $\alpha_{c2}$ in an arbitrary
dimension:

\begin{equation}
\alpha_{c1}=\frac{1}{3-d}\,,\;\alpha_{c2}=\frac{6-d}{6-2d}.\label{alpha_c}\end{equation}

\begin{figure}
\begin{centering}
\includegraphics[width=1\columnwidth]{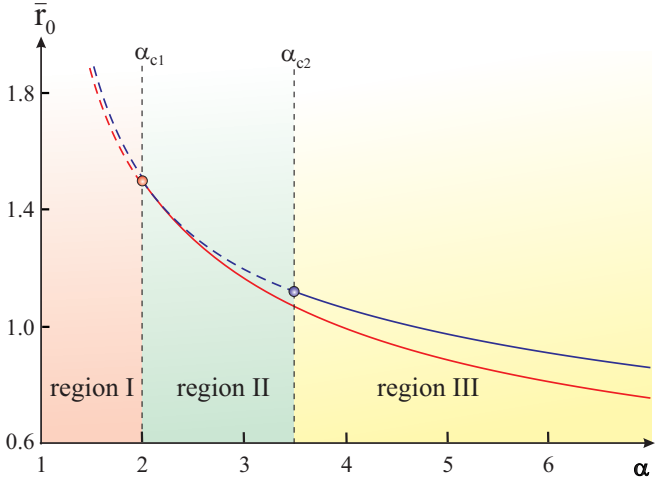} 
\par\end{centering}

\caption{(Color online) The calculated phase diagram in the $\left(\alpha,\bar{r}_{0}\right)$
plane for the representative case $d=2.5$. In region I, there is
a simultaneous nematic and magnetic first-order transition. In region
II, the two transitions split, but the nematic transition remains
first-order, while the magnetic transition becomes second-order. In
region III, the nematic transition also becomes second-order. The
three regions are separated by two tri-critical points. The temperature
behavior of the nematic and magnetic order parameter in each region
is shown in Fig. \ref{fig_order_parameters} below. }

\label{fig_phase_diagram_regions} 
\end{figure}

For $d=2$ we have $\alpha_{c1}=1$ and $\alpha_{c2}=2$, i.e. the
regime $1<\alpha<\alpha_{c1}$ disappears, in agreement with Fig.
\ref{fig_2D_phase_diagram}. On the other hand, for $d=3$, both $\alpha_{1}$
and $\alpha_{2}$ tend to infinity, and the region $1<\alpha<\alpha_{c1}$
extends to all values of $\alpha$, in agreement with Fig. \ref{fig_3D_phase_diagram}.
For $d$ between $2$ and $3$ all three regions are present, as shown
in the phase diagram of Fig. \ref{fig_phase_diagram_regions}.

In the first region $1<\alpha<\alpha_{c1}$ (region I in Fig. \ref{fig_phase_diagram_regions}),
the behavior of the system is the same as we found in $d=3$: $\phi$
jumps at the Ising-nematic transition to the largest possible value
$\phi=r$, triggering a simultaneous magnetic transition. To determine
whether the latter is also first order, we extend the analysis of
the effective action in the same way as we did for $d=3$: we introduce
magnetic long-range order via the order parameter $M=\left\langle \Delta_{X}^{i}\right\rangle $
and obtain the set of three coupled equations for $M$, $\phi$, and
$\psi$. The equation for $M$ again gives either $M=0$ or $\phi=r$.
We choose $r=\phi$ and re-express the other two equations as \begin{eqnarray}
\phi & = & \bar{r}_{0}-\frac{\bar{u}}{4}\left(2\phi\right)^{\frac{d-2}{2}}+\bar{u}\bar{M}^{2}\nonumber \\
\phi & = & \frac{\bar{g}}{4}\left(2\phi\right)^{\frac{d-2}{2}}+\bar{g}\bar{M}^{2}\label{aux_r0_m_sc}\end{eqnarray}
 with $\bar{M}/M=\sqrt{u/\bar{u}}$ (see Eq. \ref{aux_self_consistent_eq}).
For $\alpha\leq\alpha_{c1}$, we expand in $\bar{M}$ and obtain an
explicit equation relating $\bar{r}_{0}$ to the magnetization $\bar{M}$:

\begin{equation}
\bar{r}_{0}=\left(\frac{\bar{g}}{2}\right)^{\frac{2}{4-d}}\left(\frac{1+\alpha}{2}\right)+\frac{\bar{g}}{2(4-d)}\bar{Q}\left(\bar{M},\alpha\right),\label{r0_m_d_23}\end{equation}
 with:

\begin{eqnarray}
\bar{Q}\left(\bar{M},\alpha\right) & \approx & 4\left(3-d\right)\left(\alpha_{c1}-\alpha\right)\bar{M}^{2}\nonumber \\
 &  & -8\left(\frac{\bar{g}}{2}\right)^{\frac{2-d}{4-d}}\left(1+\alpha\right)\left(\frac{d-2}{4-d}\right)\bar{M}^{4}.\label{aux_r0_m_d_23}\end{eqnarray}

For $\alpha\lesssim\alpha_{c_{1}}$ the maximum of $\bar{Q}\left(\bar{M},\alpha\right)$
is at a finite magnetization:

\begin{equation}
\bar{M}=\frac{1}{2}\left[\frac{\left(4-d\right)\left(3-d\right)\left(2\bar{g}\right)^{\frac{d-2}{4-d}}}{\left(1+\alpha\right)\left(d-2\right)}\right]^{1/2}\sqrt{\alpha_{c1}-\alpha}.\label{m_d_23}\end{equation}

\begin{figure}
\begin{centering}
\includegraphics[width=1\columnwidth]{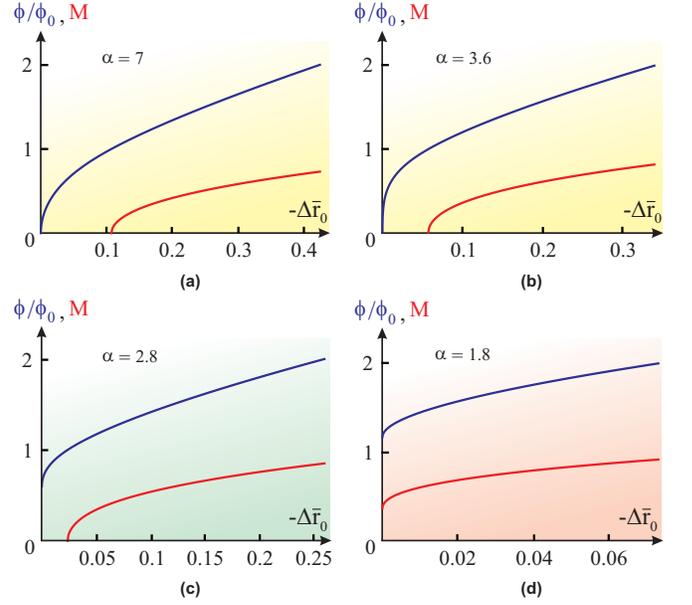} 
\par\end{centering}

\caption{(Color online) Nematic order parameter $\phi$ (blue/dark gray curves)
and magnetic order parameter $M$ (red/light gray curves) as function
of $-\Delta\bar{r}_{0}=r_{0,cr}-r_{0}\propto(T_{s}-T)$ for four different
values of $\alpha$ at $d=2.5$. In (a) and (b) we use $\alpha>\alpha_{c2}=3.5$
(region III of the phase diagram), in (c) we use $\alpha_{c2}>\alpha>\alpha_{c1}=2$
(region II of the phase diagram), and in (d) we use $\alpha<\alpha_{c1}$
(region I of the phase diagram). $\phi_{0}$ is the value of $\phi$
corresponding to $\phi=r$. }

\label{fig_order_parameters} 
\end{figure}

Therefore, in region I, the first-order Ising transition triggers
the first-order magnetic transition. We again emphasize, however,
that the Ising-nematic order parameter is \textit{larger} than the
square of the magnetic order parameter, i.e. the magnetic transition
is the secondary transition, triggered by the preemptive Ising transition.
This is most clearly seen from Eq. (\ref{m_d_23}): the jump in the
magnetization approaches zero as $\alpha$ approaches $\alpha_{c1}$,
while the jump in the Ising-nematic order parameter remains finite
and reaches $\phi=(1/2)(\bar{g}/2)^{2/(4-d)}$. Thus, $\alpha_{c1}$
corresponds to a \emph{magnetic tri-critical point,} while a preemptive
first-order Ising transition exists on both sides of it. In Fig. \ref{fig_order_parameters}
d, we present numerical results for $\phi$ and $M$ as functions
of $\bar{r}_{0}$ in region I.

In the second region $\alpha_{c1}<\alpha<\alpha_{c2}$ (region II
in Fig. \ref{fig_phase_diagram_regions}), the Ising-nematic transition
is still first-order, but the magnitude of the jump of $\phi$ is
smaller than the value required to trigger a magnetic transition,
i.e. $\phi<r$. As a result, in this region the magnetic and Ising
transitions are split, with the former occurring at a smaller $\bar{r}_{0}$
(i.e. at a smaller temperature) than the latter. The magnetic transition
then becomes second-order, as the maximum of $\bar{Q}\left(M,\alpha\right)$
remains at $M=0$ for $\alpha\gtrsim\alpha_{c1}$, see Eq. (\ref{aux_r0_m_d_23}).
In Fig. \ref{fig_order_parameters} c, we present the numerical solution
for $\phi$ and $M$ in this region, as function of temperature.

To determine what happens as the system approaches $\alpha_{c2}$,
we expand $Q\left(z,\alpha\right)$ in Eq. (\ref{Q_alpha_z}) around
$z=0$ and obtain: \begin{eqnarray}
 &  & \frac{Q(z,\alpha)}{\left(d-2\right)^{\frac{2}{4-d}}}\approx\left(\frac{2\alpha}{d-2}+1\right)+\frac{\left(3-d\right)}{6}\times\nonumber \\
 &  & \left[\left(\alpha_{c2}-\alpha\right)z^{2}+\frac{\left(15+2d\right)\left(6-d\right)}{120}\left(\kappa\alpha_{c2}-\alpha\right)z^{4}\right]\label{Q_expansion_1}\end{eqnarray}
 with a constant $\kappa\equiv\frac{90-7d-d^{2}}{\left(6-d\right)\left(15+2d\right)}<1$.
Clearly, the jump of the Ising-nematic order parameter across the
transition approaches zero as $\alpha$ increases towards $\alpha_{c2}$,
$z_{max}\sim\sqrt{\alpha_{c2}-\alpha}$. Therefore, $\alpha_{c2}$
is a \emph{nematic tri-critical point,} beyond which the Ising-nematic
transition becomes second-order (region III in Fig. \ref{fig_phase_diagram_regions}).
In this region, a non-zero $\phi$ gradually develops below the transition
line, and the magnetic transition splits even further from the Ising-nematic
transition. We present the numerical solution for $\phi$ and $M$
in region III in Fig. \ref{fig_order_parameters} b,a.

\begin{figure}
\begin{centering}
\includegraphics[width=0.8\columnwidth]{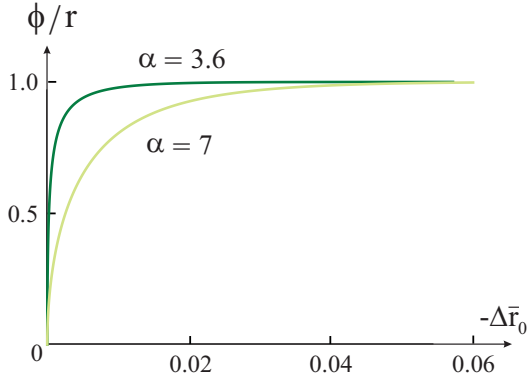} 
\par\end{centering}

\caption{(Color online) Nematic order parameter $\phi$, in units of the renormalized
magnetic mass term $r$, as function of the reduced temperature $-\Delta\bar{r}_{0}=r_{0,cr}-r_{0}\propto(T_{s}-T)$
at $d=2.5$ for two different values of $\alpha$ : $\alpha=7$, which
is far from the nematic tri-critical point $\alpha_{c2}=3.5$ (light-green/
light gray curve), and $\alpha=3.6$, which is very close to the nematic
tri-critical point (dark-green/dark gray curve). In both cases, $\phi$
evolves continuously from zero, but the slope is very large for $\alpha$
only slightly above $\alpha_{c2}$. }

\label{fig_nematic_increase} 
\end{figure}

Note that the slope with which $\phi$ increases below the instability
remains very high over some range of $\alpha$ in region III, as illustrated
in Fig. \ref{fig_order_parameters}(b) and, more transparently, in
Fig. \ref{fig_nematic_increase}. From the practical point of view,
the Ising transition can then still be viewed as almost first-order.
This can also be seen from Eq. \ref{Q_expansion_1} because in coefficient
of the $z^{4}$ term, $\kappa$ remains very close to one for all
$2<d<3$ ($0.048<1-\kappa<0.054$). As a result, if $\alpha$ is not
too far from $\alpha_{c2}$, $\phi$ increases very rapidly. For instance,
the value $\bar{r}_{0}/\bar{r}_{0}^{max}$ for which $\phi$ increases
to $\phi=r/2$ is $\left|\bar{r}_{0}/\bar{r}_{0}^{max}-1\right|\lesssim10^{-3}\left(\alpha-\alpha_{c2}\right)$,
almost independent on the dimensionality.

We emphasize once again that for $\alpha<\alpha_{c2}$, when the nematic
transition is first order, the value of $\bar{r}_{0}$ at which the
solution of Eq. (\ref{r0_alpha}) first appears is the upper spinodal
of the system, $\bar{r}_{0}^{max}$. The lower spinodal $\bar{r}_{0}^{min}$
is the value of $\bar{r}_{0}$ below which $\phi=0$ is no longer
the local minimum of $S_{\mathrm{eff}}$. The actual phase transition
in the presence of fluctuations is between $\bar{r}_{0}^{max}$ and
$\bar{r}_{0}^{min}$, and is close to $\bar{r}_{0}^{min}$ if fluctuations
are weak. If we use $\bar{r}_{0}^{min}$ instead of $\bar{r}_{0}^{max}$
for the value of $\bar{r}_{0}^{cr}$ at which the first-order transition
occurs, we find that $\alpha_{c2}$ remains intact, but $\alpha_{c1}$
moves to a larger value $\alpha_{c1}^{\prime}$ given by:

\begin{equation}
\alpha_{c1}^{\prime}=\frac{\left(\frac{2}{d-2}\right)^{\frac{d-2}{4-d}}-\left(d-2\right)}{2-\left(\frac{2}{d-2}\right)^{\frac{d-2}{4-d}}}\label{alpha_c3}\end{equation}

One can easily check that for $2<d<3$, $\alpha_{c1}<\alpha_{c1}^{\prime}<\alpha_{c2}$.
Thus, the magnetic tri-critical point does not merge with the nematic
tri-critical point, even if the Ising-nematic transition occurs at
the lowest possible $\bar{r}_{0}^{min}$.

\begin{figure}
\begin{centering}
\includegraphics[width=0.8\columnwidth]{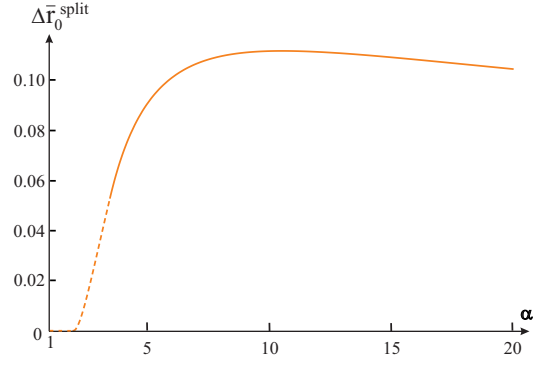} 
\par\end{centering}

\caption{(Color online) Splitting between the nematic and magnetic transitions,
$\Delta\bar{r}_{0}^{\mathrm{split}}\equiv\bar{r}_{0}^{\mathrm{nem}}-\bar{r}_{0}^{\mathrm{mag}}$,
as function of $\alpha\equiv u/g$, for the case $d=2.5$. The dashed
line refers to regions I and II of the phase diagram of Fig. \ref{fig_phase_diagram_regions},
whereas the solid line refers to region III. }

\label{fig_splitting} 
\end{figure}

Notice that in the phase diagram of Fig. \ref{fig_phase_diagram_regions}
the splitting between the magnetic and nematic transitions initially
increases with increasing $\alpha$, once the system crosses the nematic
tri-critical point. However, for small values of the nematic coupling
$g$ (large values of $\alpha$), this splitting is expected to decrease,
since it should tend to zero for infinitesimal $g$. This non-monotonic
behavior is a consequence of the fact that, at smaller $\alpha$,
the strong first-order nematic transition brings in a simultaneous
magnetic transition, despite the fact that the magnetic correlation
length is rather small immediately above the nematic transition. This
correlation length, meanwhile, increases monotonically as $\alpha$
increases. We show the non-monotonic behavior of the splitting between
the magnetic and nematic transitions as function of $\alpha$ in Fig.
\ref{fig_splitting} for the case $d=2.5$. Note that the splitting
begins to slowly decrease only at $\alpha\geq10$.

\subsubsection{Anisotropic $d=3$ case}

We now show that the phase diagram obtained for the case $2<d<3$
is qualitatively the same as the phase diagram of the $d=3$ model
with strongly anisotropic, quasi-2D magnetic dispersion. However,
for less anisotropic systems, we find a different phase diagram, with
new behavior at intermediate $\alpha$.

To model the anisotropy, we consider the system composed of stacked
two-dimensional layers. In each layer $n$, we define the magnetic
order parameters $\boldsymbol{\Delta}_{i}^{n}$, with $i=X,Y$. The
coupling between different layers adds the term $-\frac{\eta_{z}}{2}\boldsymbol{\Delta}_{i}^{n}\cdot\boldsymbol{\Delta}_{i}^{n+1}$
to the action with $\eta_{z}<1$. As a result, the magnetic susceptibility
in Eq. (\ref{chi_i}) becomes:\begin{equation}
\chi_{i,q}^{-1}=r_{0}+q_{\parallel}^{2}+\eta_{z}\sin^{2}\frac{q_{z}}{2}\label{an_suscept}\end{equation}
 with $0<q_{z}<2\pi$ and $\mathbf{q}_{\parallel}=\left(q_{x},q_{y}\right)$.
An alternative possibility is to consider anisotropic but still quadratic
dispersion \begin{equation}
\chi_{i,q}^{-1}=r_{0}+q_{\parallel}^{2}+\beta_{}^{2}q_{z}^{2}\label{an_suscept_1}\end{equation}
 with $0<\beta_{}<1$, and set the same cutoff $\Lambda$ for all
three components of ${\bf q}$. We found that the phase diagram is
\textit{the same} regardless of whether we use Eq. (\ref{an_suscept})
or (\ref{an_suscept_1}) for the bosonic susceptibility. For definiteness,
here we present the results for $\chi_{q}$ given by Eq. (\ref{an_suscept})
and consider the case of Eq. (\ref{an_suscept_1}) in Appendix \ref{append_quadr}.

We plug Eq. (\ref{an_suscept}) into the self-consistent equations
(\ref{self_cons}) and evaluate the three-dimensional integrals. Absorbing
the cutoff $\Lambda$ into $r_{0}$, $\bar{r}_{0}=r_{0}+2\bar{u}\log\left(2\Lambda\right)$,
we obtain in the paramagnetic phase:

\begin{eqnarray}
r & = & \bar{r}_{0}-\bar{u}\log\left(\sqrt{r+\phi}+\sqrt{\eta_{z}+r+\phi}\right)\nonumber \\
 &  & -\bar{u}\log\left(\sqrt{r-\phi}+\sqrt{\eta_{z}+r-\phi}\right)\nonumber \\
\phi & = & \bar{g}\,\log\left(\frac{\sqrt{r+\phi}+\sqrt{\eta_{z}+r+\phi}}{\sqrt{r-\phi}+\sqrt{\eta_{z}+r-\phi}}\right)\label{an_self_cons}\end{eqnarray}
 where, as before, $\bar{u}=uT/\left(2\pi\right)$ and $\bar{g}=gT/\left(2\pi\right)$.
>From the second equation, we obtain $r$ as function of $\phi$:

\begin{equation}
r\left(\phi\right)=-\frac{\eta_{z}}{2}+\cosh\left(\frac{\phi}{g}\right)\sqrt{\frac{\eta_{z}^{2}}{4}+\frac{\phi^{2}}{\sinh^{2}\left(\frac{\phi}{g}\right)}}\label{an_r_phi_1}\end{equation}

Substituting (\ref{an_r_phi_1}) into the first equation in (\ref{an_self_cons})
we obtain $\bar{r}_{0}\left(\phi\right)$, whose maximum determines
the first instability of the system. At a non-zero $\bar{M}$ we find,
as before, $r=\phi$ and \begin{eqnarray}
 &  & \bar{r}_{0}\left(\phi\right)=\nonumber \\
 &  & \phi+\bar{u}\left[\log\left(\sqrt{2\phi\eta_{z}}+\sqrt{\eta_{z}^{2}+2\phi\eta_{z}}\right)-\bar{M}^{2}\right]\label{an_r0_phi_2}\end{eqnarray}
 where the magnetization is given by:

\begin{equation}
\bar{M}^{2}=\frac{\phi}{g}-\log\left(\frac{\sqrt{2\phi}+\sqrt{\eta_{z}+2\phi}}{\sqrt{\eta_{z}}}\right)\label{an_m_r}\end{equation}

Notice the characteristic logarithmic dependence on the anisotropy,
$\log\eta_{z}$. Using these expressions, we obtain the phase diagram
of the anisotropic system by analyzing the first instability upon
decreasing $\bar{r}_{0}$.

\begin{figure}
\begin{centering}
\includegraphics[width=1\columnwidth]{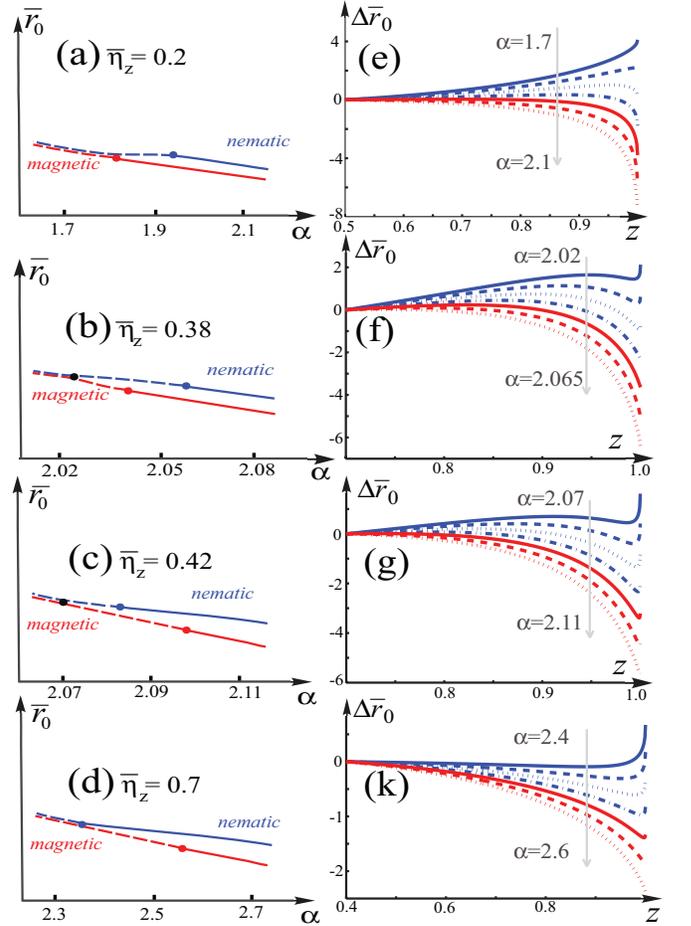} 
\par\end{centering}

\caption{(Color online) Left panels: the four different phase diagrams for
the anisotropic 3D model with the magnetic susceptibility given by
Eq. (\ref{an_suscept}). The behavior found in isotropic systems for
a fractional $2<d<3$ only holds at small $\bar{\eta}_{z}\equiv\eta_{z}/\bar{g}$.
At larger $\bar{\eta}_{z}$, the magnetic and nematic tri-critical
points $\alpha_{c1}$ and $\alpha_{c2}$ interchange and a new behavior
emerges at intermediate $\alpha$. The behavior presented in panels
(a) and (d) takes place over sizable ranges of small and moderate
$\bar{\eta}_{z}$, respectively. The behavior presented in panels
(b) and (c) takes place only over a narrow range of $\bar{\eta}_{z}$,
around the $\bar{\eta}_{z}$ value for which $\alpha_{c1}$ and $\alpha_{c2}$
cross. Right panels: the behavior of $\bar{r}_{0}(z)$ for several
different $\alpha$ for each phase diagram ($z=\phi/r$). The key
difference between the regimes (a) and (d) is that in (a) $\bar{r}_{0}(z)$
has only one maximum at every $\alpha$, while in (d) $\bar{r}_{0}(z)$
has two distinct maxima and a minimum in between. \label{fig_ilya}}

\end{figure}

We show our results for various $\eta_{z}$ in Fig.\ref{fig_ilya},
where we present four possible phase diagrams together with their
respective $\bar{r}_{0}(z,\alpha)$ profiles for different values
of $\alpha$ (recall that $z=\phi/r$). The phase diagrams shown in
panels (a) and (d) exist over sizable ranges of small and moderate
$\eta_{z}$, respectively. On the other hand, the phase diagrams in
panels (b) and (c) exist only over rather narrow ranges of $\eta_{z}$
and just show how the system actually evolves from the regime (a)
into the regime (d).

\begin{figure}
\begin{centering}
\includegraphics[width=0.85\columnwidth]{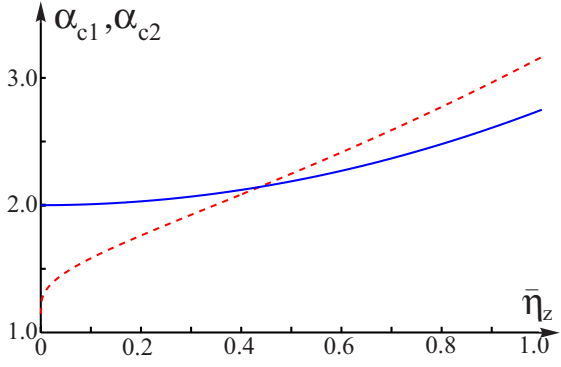} 
\par\end{centering}

\caption{(Color online) Magnetic and nematic tri-critical points $\alpha_{c1}$
(red/light gray dotted line) and $\alpha_{c2}$ (blue/dark gray solid
line) as functions of $\bar{\eta}_{z}$ for the 3D anisotropic model
with the magnetic susceptibility given by Eq. (\ref{an_suscept}).
Notice that $\alpha_{c1}$ and $\alpha_{c2}$ cross at $\bar{\eta}_{z}\sim0.43$.
\label{ilya_eta}}

\end{figure}

The behavior at small $\eta_{z}$ shown in Fig.\ref{fig_ilya}(a)
is the same as for a fractional dimension $2<d<3$, Fig. \ref{fig_phase_diagram_regions}.
Namely, there is a magnetic tri-critical point at $\alpha_{c1}$ and
a nematic tri-critical point at $\alpha_{c2}>\alpha_{c1}$, with the
regime of split first-order nematic and second-order magnetic transition
in between. The behavior at larger $\eta_{z}$ is, however, different.
We see from Fig.\ref{fig_ilya}(d) that, in the regime of intermediate
$\alpha$, there is a second-order nematic followed by a first-order
magnetic transition.

This change in the system behavior upon increasing $\eta_{z}$ can
be better understood by considering how the ratio of the two tri-critical
points $\alpha_{c1}/\alpha_{c2}$ evolves with $\eta_{z}$. In 2D,
when $\eta_{z}=0$, we have $\alpha_{c2}/\alpha_{c1}=2$. For the
isotropic system in $2<d<3$, we have from (\ref{alpha_c}) that $\alpha_{c2}/\alpha_{c1}=3-d/2$,
implying that $\alpha_{c2}$ remains larger than $\alpha_{c1}$ for
all $2<d<3$, what gives rise to the phase diagram of Fig. \ref{fig_phase_diagram_regions}.
However, there is no requirement that $\alpha_{c2}$ must remain larger
than $\alpha_{c1}$ as the system approaches the isotropic 3D regime
- in fact, the only requirement is that in this limiting case both
$\alpha_{c1}$ and $\alpha_{c2}$ diverge. In Fig. \ref{ilya_eta}
we plot $\alpha_{c1}$ and $\alpha_{c2}$ as a function of $\eta_{z}$
for the 3D system with anisotropic dispersion. We see that $\alpha_{c1}$
and $\alpha_{c2}$ cross at a certain $\eta_{z}$, beyond which the
nematic tri-critical point occurs at a smaller $\alpha$ than the
magnetic tri-critical point. This leads to the phase diagram of Fig.
\ref{fig_ilya}(d).

We can show analytically that $\alpha_{c1}$ and $\alpha_{c2}$ cross
upon increasing $\eta_{z}$. To obtain $\alpha_{c2}$, we substitute
$r(\phi)$ from (\ref{an_r_phi_1}) in (\ref{an_self_cons}) and expand
the right-hand side in powers of $\phi$. Since the quadratic $\phi^{2}$
term vanishes at $\alpha_{c2}$, we obtain: \begin{equation}
\alpha_{c2}=2+\frac{3\bar{\eta}_{z}^{2}}{4}\label{alpha_c2}\end{equation}
 where $\bar{\eta}_{z}\equiv\eta_{z}/\bar{g}$. To obtain $\alpha_{c1}$,
we express $\phi$ in terms of $\bar{m}$ using Eq. (\ref{an_m_r}),
substitute the result in Eq. (\ref{an_r0_phi_2}) and expand the right-hand
side in powers of $\bar{M}$. Since the quadratic $\bar{M}^{2}$ term
vanishes at $\alpha_{c1}$, we find: \begin{equation}
\alpha_{c1}=\left(1-\frac{2}{\sqrt{2\bar{\phi}_{c1}}\sqrt{\bar{\eta}_{z}+2\bar{\phi}_{c1}}}\right)^{-1}\label{alpha_c1}\end{equation}
 where $\bar{\phi}_{c1}$ is the solution of \begin{equation}
\bar{\phi}_{c1}=\log\left(\frac{\sqrt{2\bar{\phi}_{c1}}+\sqrt{\bar{\eta}_{z}+2\bar{\phi}_{c1}}}{\sqrt{\bar{\eta}_{z}}}\right)\label{phi_c1}\end{equation}

A simple analysis then shows that $\alpha_{c2}$ becomes larger than
$\alpha_{c1}$ at $\bar{\eta}_{z}^{c}=0.43$. One has to be careful
to properly determine $\alpha_{c2}$ for $\bar{\eta}_{z}>\bar{\eta}_{z}^{c}$,
since the actual nematic tri-critical point takes place when $z=0$
is the global maximum of $\bar{r}_{0}(z)$, and not only a local maximum.
This subtlety does not affect the result that $\alpha_{c1}$ and $\alpha_{c2}$
cross, nor the regime $\bar{\eta}_{z}<\bar{\eta}_{z}^{c}$. In the
phase diagram shown in Fig. \ref{fig_anisotropic_weak}, $\alpha_{c2}$
is the actual nematic tri-critical point.

The phase diagram shown in Fig. \ref{fig_ilya}(d) contains a new
type of a phase transition not seen in quasi-two dimensional systems,
namely, a meta-nematic transition. We explore it in more detail in
the next subsection.

\subsubsection{ Meta-nematic transition}

\begin{figure}
\begin{centering}
\includegraphics[width=1\columnwidth]{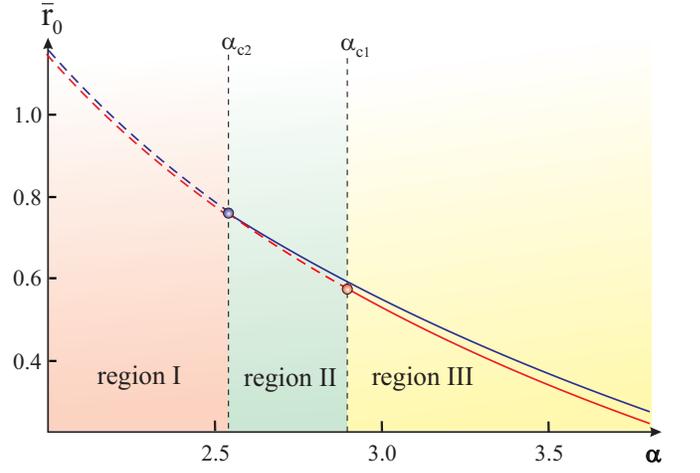} 
\par\end{centering}

\caption{(Color online) Calculated phase diagram in the $\left(\alpha,\bar{r}_{0}\right)$
plane for moderately anisotropic $d=3$ system with $\eta_{z}=0.3\bar{u}$.
Three different regions of system behavior are marked in the same
way as in Fig. \ref{fig_phase_diagram_regions}. Notice that the positions
of the tri-critical points are reversed with respect to the case of
strongly anisotropic systems. \label{fig_anisotropic_weak}}

\end{figure}

\begin{figure}
\begin{centering}
\includegraphics[width=1\columnwidth]{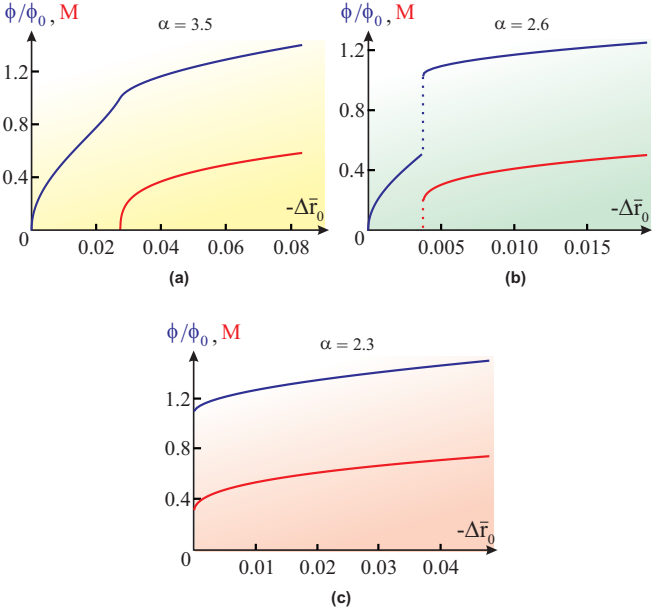} 
\par\end{centering}

\caption{(Color online) Nematic order parameter $\phi$ (blue/dark gray curves)
and magnetic order parameter $m$ (red/light gray curves) as function
of $-\Delta\bar{r}_{0}=r_{0,cr}-r_{0}\propto(T_{s}-T)$ for three
different values of $\alpha$ in the case of a moderately anisotropic
system with $\eta_{z}=0.3\bar{u}$ (see Fig. \ref{fig_anisotropic_weak}).
Panel (a) is for $\alpha>\alpha_{c1}\approx2.91$ (region III of the
phase diagram), panel (b) for $\alpha_{c2}\approx2.54<\alpha<\alpha_{c1}$
(region II of the phase diagram), and panel (c) is for $\alpha<\alpha_{c2}$
(region I of the phase diagram). In panel (b), the dotted line marks
the meta-nematic transition, coincident with the magnetic transition.
$\phi_{0}$ is the value of $\phi$ corresponding to $\phi=r$.\label{fig_metanematic}}

\end{figure}

In Fig. \ref{fig_anisotropic_weak} we explicitly show the three different
regimes that appear in Fig. \ref{fig_ilya}(d), as we did in Fig.
\ref{fig_phase_diagram_regions}. The behavior of the nematic and
the magnetic order parameters as functions of the distance to the
transitions in all three regimes is shown in Fig. \ref{fig_metanematic}.
For these particular figures, for convenience, we considered $\eta_{z}=0.3\bar{u}$
instead of $\eta_{z}=0.7\bar{g}$, which does not change the properties
of the phase diagram.

A careful analysis of the profile of $\bar{r}_{0}(z,\alpha)$ as a
function of $z$ for the phase diagram in Fig. \ref{fig_ilya} (d)
shows that, at small $\alpha$, the maximum of $\bar{r}_{0}(z,\alpha)$
is at $z=1$ (i.e. at $\phi=r$). This implies that upon decreasing
$\bar{r}_{0}$ the system undergoes a first-order nematic transition,
which triggers a simultaneous first-order magnetic transition. This
system behavior takes place in region I of Fig. \ref{fig_anisotropic_weak},
and is displayed in Fig. \ref{fig_metanematic} (c).

However, as $\alpha$ increases, $\bar{r}_{0}(z,\alpha)$ develops
a local maximum at $z=0$ as well. At the nematic tri-critical point
$\alpha_{c2}$, $\bar{r}_{0}(0,\alpha)$ becomes equal to $\bar{r}_{0}(1,\alpha)$,
and $\bar{r}_{0}(z,\alpha)$ has a minimum for $0<z<1$. Once $\alpha$
becomes larger than $\alpha_{c2}$, the absolute maximum of $\bar{r}_{0}(z,\alpha)$
shifts to $z=0$, i.e. the nematic transition becomes second-order.
As $\bar{r}_{0}$ decreases, a solution with small and finite $z\neq0$
develops. But because the profile of $\bar{r}_{0}(z,\alpha)$ has
two maxima, another solution with $z=1$ necessarily appears once
$\bar{r}_{0}$ gets smaller than some critical value. We checked that
the value of $S_{\mathrm{eff}}$ at $z=1$ eventually becomes larger
than the value corresponding to the solution at small $z$. As a result,
at some $\bar{r}_{0}$, the nematic order parameter undergoes a \emph{meta-nematic
transition}, where it jumps from some small value $\phi\neq0$ to
$\phi=r$ ($z=1$). In accordance to what we found earlier, we explicitly
confirmed that such a transition triggers a first-order magnetic transition.
Therefore, for $\alpha>\alpha_{c2}$, the second-order nematic and
meta-nematic transitions split. The second-order transition occurs
first, and the meta-nematic transition occurs at a smaller $\bar{r}_{0}$,
triggering a first-order magnetic transition. This system behavior
takes place in region II of the phase diagram of Fig. \ref{fig_anisotropic_weak},
as shown in Fig. \ref{fig_metanematic} (b).

As $\alpha$ increases further, the magnetic transition remains first-order
up to $\alpha=\alpha_{c1}>\alpha_{c2}$. At this point, the position
of the maximum of $\bar{r}_{0}(M,\, z=1,\alpha)$ shifts to $M=0$,
i.e the magnetic transition becomes second order. We found both analytically
and numerically that at exactly the same $\alpha=\alpha_{c1}$, the
local minimum of $\bar{r}_{0}(z,\alpha)$ moves to $z=1$, meaning
that $\bar{r}_{0}(z,\alpha)$ becomes a monotonically decreasing function
of $z$ for all $0\leq z\leq1$. As a result, for $\alpha>\alpha_{c1}$,
the nematic order parameter $\phi$ monotonically increases with decreasing
$\bar{r}_{0}$ until it reaches the value $\phi=r$, where the system
undergoes a second-order magnetic transition. This system behavior
corresponds to region III in Fig. \ref{fig_anisotropic_weak}, and
is the same as region III in Fig. \ref{fig_phase_diagram_regions}.
Notice from Fig. \ref{fig_metanematic} (a), however, that for $\alpha\gtrapprox\alpha_{c1}$,
the proximity to the meta-nematic transition line produces a kink
in the temperature dependence of the nematic order parameter.

We acknowledge that, for values of $\eta_{z}$ that are not very small
- such as the one considered here - the magnetic susceptibility of
the anisotropic 3D system may have extra terms not contemplated by
our formalism, such as corrections to the continuous in-plane dispersion.
However, as we show in details in Appendix \ref{append_quadr}, the
same behavior is obtained for an anisotropic quadratic dispersion
with equal momentum cutoff along all three momentum directions. This
gives extra confidence that the phase diagram of Fig. \ref{fig_anisotropic_weak}
may be realized in at least some moderately anisotropic systems.

We point out that a phase diagram with the intermediate regime of
a second-order nematic transition and first-order magnetic transition
was earlier obtained in a semi-phenomenological model for the interaction
between the structural and magnetic degrees of freedom \cite{Cano10}.
Kim \textit{et al} considered a microscopic version of that model
\cite{Kim11}, showing that anharmonic elastic terms may bring the
system into the regime of split second-order nematic and first-order
magnetic transitions (we discuss the coupling between the lattice
and the nematic degrees of freedom in Section \ref{sec:conseq} below).
Our results show that such a behavior can be obtained in a purely
magnetic model, even if the coupling to structural degrees of freedom
is negligibly small.

\subsection{Quantum phase diagram as a function of $u/g$}

To complement our analysis of thermal fluctuations, we now consider
the opposite limit of $T=0$, when the Matsubara frequency become
a continuous variable and $\int_{q}=\int d^{d}qd\nu_{n}/(2\pi)^{(d+1)}$.
Now $r_{0}$ is a function of some control parameter, e.g. doping,
pressure, or applied field. We show that no new phases appear in the
$T=0$ limit, compared to the three phases that we found previously
in the classical limit. As before, we consider first $d=2$, then
$d=3$, and then arbitrary $d$ between $2$ and $3$ and anisotropic
3D systems.

\subsubsection{$d=2$}

Integrating the self-consistent equations for $\phi$ and $r$ in
(\ref{self_cons}) over both $\nu_{n}$ and $\mathbf{q}$, we obtain
for $d=2$ \begin{eqnarray}
1 & = & \tilde{g}\left(\log\frac{\Lambda^{2}}{\sqrt{r^{2}-\phi^{2}}}+1-\frac{r}{\phi}\tanh^{-1}\frac{\phi}{r}\right)\nonumber \\
r & = & \bar{r}_{0}-\tilde{u}\left(r\log\frac{\Lambda^{2}}{\sqrt{r^{2}-\phi^{2}}}+r-\phi\tanh^{-1}\frac{\phi}{r}\right)\label{sa_before_17}\end{eqnarray}
 where $\tilde{g}=g/\left(4\pi^{2}\gamma\right)$, $\tilde{u}=u/\left(4\pi^{2}\gamma\right)$
are dimensionless couplings, and $\Lambda$ is the upper limit of
the integral over momentum. The parameter $\bar{r}_{0}$ again decreases
as the system approaches the magnetic transition. Note that the dependence
on $\Lambda$ is still present even after we absorbed the $r-$ and
$\phi$-independent contribution of the right-hand side of Eq. (\ref{aux_self_cons})
into $\bar{r}_{0}$. This dependence is eliminated only after we rescale
all variables by $\Lambda^{2}$. The implication is that the relevant
$\bar{r}_{0}$, $r$, and $\phi$ are all of the order of $\Lambda^{2}$
if the dimensionless $\tilde{g}$ and $\tilde{u}$ are of order one.
Thus, unlike the classical case, where the dependence on $\Lambda$
was fully absorbed into $\bar{r}_{0}$, here the low-energy behavior
becomes non-universal. To proceed - and to make later comparison with
the RG results - here we consider the case of $\bar{r}_{0}$ much
smaller than $\Lambda$. This is achieved by taking the dimensionless
couplings $\tilde{g}$ and $\tilde{u}$ to be small, which also implies
that $r$ and $\phi$ are small compared to $\Lambda^{2}$.

In the absence of the preemptive Ising-nematic instability, the magnetic
instability would again occur at $\bar{r}_{0}=0$. Searching for the
solution with $\phi\neq0$, we introduce as before $z=\phi/r$, solve
the first equation for $r(z)$ and substitute the result into the
second equation. This gives the relation between $\bar{r}_{0}$ and
$z$. For small $\tilde{g}$ this relation takes the form \begin{equation}
\bar{r}_{0}=\Lambda^{2}\mathrm{e}^{-1/\tilde{g}}\left(1+\alpha\right)f_{2}(z)\label{sa_17}\end{equation}
 where \begin{equation}
f_{2}(z)=\frac{\mathrm{e}}{\sqrt{1-z^{2}}}\times\left(\frac{1-z}{1+z}\right)^{\frac{1}{2z}}.\label{sa_18}\end{equation}

\begin{figure}
\begin{centering}
\includegraphics[width=0.7\columnwidth]{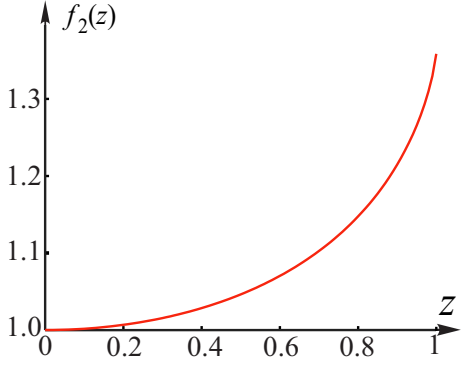} 
\par\end{centering}

\caption{ (Color online) The function $f_{2}(z)$ from Eq.(\ref{sa_18}). }

\label{fig:fz2} 
\end{figure}

Note that $\bar{r}_{0}\ll\Lambda^{2}$ for $\tilde{g}\ll1$, as we
anticipated. The dependence on $z$ comes through $f_{2}(z)$. This
function, which we plot in Fig. \ref{fig:fz2}, monotonically increases
with increasing $z$, implying that the first instability occurs at
$z=1$. Because $f(z)$ does not depend on $\alpha$, this result
obviously holds for any $\alpha$.

When $\tilde{g}$ and $\tilde{u}$ are not small, the relation between
$r_{0}$ and $z$ is more complex, but the result is the same - the
first instability upon decreasing $\bar{r}_{0}$ is into the state
with the largest $\phi=r$. As in earlier cases, an instability with
$\phi=r$ implies that the Ising and magnetic transitions occur simultaneously
and that both are first-order transitions. We analyzed the whole phase
diagram and again found that there is only one transition line at
which both $\phi$ and $M$ jump to finite values. At smaller $\bar{r}_{0}$,
$\phi$ and $M$ monotonically increase.

\subsubsection{$d=3$}

For $d=3$, the dependence on the upper cutoff is more severe - it
is power-law rather than logarithmic. The set of two equations on
$r$ and $\phi$ becomes, after we integrate over frequencies \begin{eqnarray}
\phi & = & \frac{\tilde{g}}{2\pi}\int_{0}^{\Lambda^{2}}dy\sqrt{y}\tanh^{-1}\frac{\phi}{y+r}\nonumber \\
r & = & \bar{r}_{0}+\frac{\tilde{u}}{4\pi}\int_{0}^{\Lambda^{2}}dy\sqrt{y}\log\frac{y^{2}}{(y+r)^{2}-\phi^{2}}\label{sa_19}\end{eqnarray}

The dependence on $\Lambda$ can be eliminated by rescaling $(r,\phi,\bar{r}_{0})\to\Lambda^{2}(r,\phi,\bar{r}_{0})$
and $(\tilde{g},\tilde{u})\to(\tilde{g},\tilde{u})/(2\pi\Lambda)$.
Introducing, as before $z=\phi/r$, we rewrite Eqs. (\ref{sa_19})
in rescaled variables as \begin{eqnarray}
z & = & \tilde{g}\sqrt{r}\int_{0}^{\frac{1}{r}}du\sqrt{u}\tanh^{-1}\frac{z}{u+1}\nonumber \\
r & = & \bar{r}_{0}+\frac{\tilde{u}}{2}r\sqrt{r}\int_{0}^{\frac{1}{r}}du\sqrt{u}\log\frac{u^{2}}{(u+1)^{2}-z^{2}}\label{sa_20}\end{eqnarray}

One can easily make sure that the the first equation in (\ref{sa_20})
has a solution only when the rescaled $\tilde{g}$ is above the threshold
$\tilde{g}_{cr}=1/2$. Once $\tilde{g}$ is above $\tilde{g}_{cr}$,
the relevant rescaled $\bar{r}_{0}$ is of order one, i.e. the actual
$\bar{r}_{0}$ is of order $\Lambda^{2}$. While the model remains
perfectly well defined, universal predictions with respect to the
low energy behavior cannot be made. Still, like in $d=2$, one can
make the relevant $\bar{r}_{0}$ to be much smaller than $\Lambda^{2}$.
For this, one has to place $\tilde{g}$ close to the critical value,
$\tilde{g}=1/2+\epsilon$, and consider $\epsilon\ll1/\Lambda$. Expanding
in $\epsilon$ in (\ref{sa_20}) and relating $r$ to $z$ we obtain,
in the original variables, \begin{equation}
\bar{r}_{0}=\frac{2}{\pi}\Lambda^{3}\left(\tilde{g}-\tilde{g}_{cr}\right)\left(1+\alpha\right)f_{3}(z)\label{sa_21}\end{equation}
 where \begin{equation}
f_{3}(z)=\left[\pi-\frac{1}{z}\int_{0}^{\infty}du\sqrt{u}\left(-\frac{z}{u+1}+\tanh^{-1}\frac{z}{u+1}\right)\right]^{-1}\label{sa_22}\end{equation}

\begin{figure}
\begin{centering}
\includegraphics[width=0.7\columnwidth]{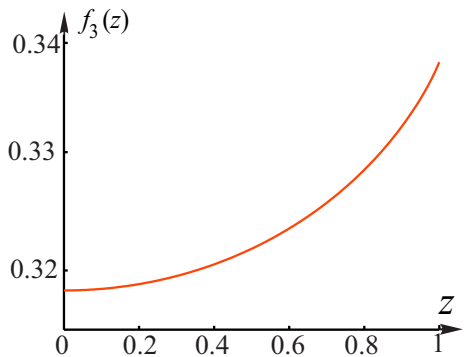} 
\par\end{centering}

\caption{(Color online) The function $f_{3}(z)$ from Eq.(\ref{sa_22}). }

\label{fig:fz3} 
\end{figure}

This function, which we plot in Fig.\ref{fig:fz3}, increases with
$z$ such that the first instability occurs into the state with $z=1$,
i.e. $\phi=r$. This implies that the Ising-nematic and magnetic orders
appear simultaneously, via a first-order transition. We see therefore
that at $T=0$ there is no difference between $d=3$ and $d=2$ -
a first-order simultaneous Ising/magnetic transition occurs in both
cases.

\subsubsection{$2<d<3$}

We also analyzed the case of arbitrary $d$ between $2$ and $3$
and found the same result as for $d=2$ and $d=3$: there is a first-order
Ising-nematic transition which triggers a simultaneous first-order
magnetic transition. One can easily show that this result holds for
all $d>1$ in the quantum limit, and for all $d>3$ in the classical
limit. The analogy between quantum systems in $d>1$ and classical
systems in $d>3$ is not surprising because the dynamical exponent
is $z_{dyn}=2$, meaning that the behavior of a quantum system in
$d$ dimensions is the same as that of a classical system with $d+2$
dimensions.

\subsubsection{Anisotropic $d=3$ case}

The same result - a simultaneous first-order nematic and magnetic
transition, holds also for anisotropic $d=3$ systems with the susceptibility
$\chi_{q}$ given by Eq. (\ref{an_suscept}). In this respect, at
$T=0$ there is no difference between the system behavior in $2<d<3$
and for anisotropic 3D dispersion, no matter what the degree of anisotropy
is. We stress again that this conclusion is not universally true and
that it could be possible to construct models with same order parameter
symmetry that display second order transitions (see, for instance,
Ref. \cite{Millis10}).

\subsection{Phase diagram at arbitrary $T$}

We now combine the quantum and classical analysis and consider the
phase diagram at a finite $T\gamma\sim\bar{r}_{0}$ when both quantum
and classical fluctuations are equally relevant.

\subsubsection{$d=2$}

For $d=2$, we have at high temperatures a second order Ising transition
at $\alpha>2$ and a first-order Ising transition into $\phi<r$ at
$1<\alpha<2$. At $T=0$, we have instead a first order transition
into $\phi=r$ and $\bar{m}\neq0$ for all $\alpha$. A simple analysis
shows that the behavior at any finite $T$ remains the same as at
high temperatures simply because at any $T>0$ classical fluctuations
do not allow a non-zero magnetic order. The value of $\alpha$ at
which the first-order Ising transition becomes second order changes
with $T$, but the phase diagram at any $T$ still consists of a single
line along which the system undergoes either first-order or second-order
Ising transition.

\subsubsection{$d=3$}

At $d=3$, the behavior at large $T$ and at $T=0$ are identical
- in both cases the first instability is into a state with $\phi=r$
(a simultaneous first-order Ising/magnetic transition). A simple analysis
shows that this behavior holds for any $T$, no matter how small or
large.

\subsubsection{$2<d<3$}

This case is the most interesting one. At high temperatures, all three
types of transitions are realized, depending on $\alpha$, while at
$T=0$ the system only undergoes a phase transition into a state with
$\phi=r$. As a result, at a given $d$ between $2$ and $3$, the
character of the transition changes as a function of $\alpha$ at
a fixed $T$, and as a function of $T$ at a fixed $\alpha$.

We verified that the phase diagrams do not change if we impose upper
cutoff on the frequency summation rather than on the integration over
momentum. The former is more convenient for numerical calculations,
and below we use frequency rather than momentum cutoff. Physically,
the frequency cutoff $\Lambda_{\nu}$ becomes more important than
the momentum cutoff $\Lambda=\Lambda_{q}$ if the frequency dependence
of the bosonic $\chi(\mathbf{q},\nu_{n})$ becomes stronger than $\gamma|\nu_{n}|$
at energies smaller than $\Lambda_{q}$.

\begin{figure}[!t]
 \includegraphics[width=1\linewidth]{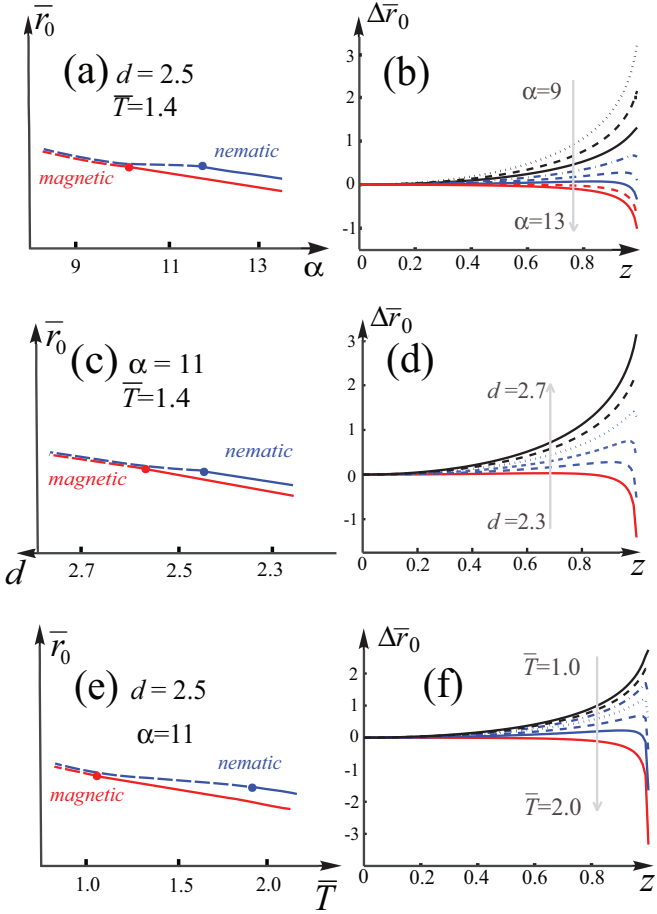}

\caption{(Color online) The phase diagram as a function of the ratio of the
couplings $\alpha$ at fixed reduced temperature $\bar{T}$ and dimensionality
$d$ (panels (a)-(b)), as a function of $d$ at fixed $\alpha$ and
$\bar{T}$ (panels (c)-(d)), and as a function of $\bar{T}$ at fixed
$\alpha$ and $d$ (panels (e)-(f)). The parameters $\alpha$ and
$\bar{T}$ are defined in the text. We set $n_{max}=100$. }

\label{fig:figfiniteT} 
\end{figure}

For a generic $2<d<3$ and an arbitrary $T$, the equations for $r$
and $\phi$ become, after integrating over momentum

\begin{eqnarray}
r & = & \bar{r}_{0}-\frac{\bar{u}}{4}\sum_{\nu_{n}}\left[-2\left(\gamma|\nu_{n}|\right)^{\frac{d-2}{2}}+\left(r+\phi+\gamma|\nu_{n}|\right)^{\frac{d-2}{2}}\right.\nonumber \\
 &  & \left.+\left(r-\phi+\gamma|\nu_{n}|\right)^{\frac{d-2}{2}}\right]\nonumber \\
\phi & = & \frac{\bar{g}}{4}\sum_{\nu_{n}}\left[\left(r+\phi+\gamma|\nu_{n}|\right)^{\frac{d-2}{2}}\right.\nonumber \\
 &  & \left.-\left(r-\phi+\gamma|\nu_{n}|\right)^{\frac{d-2}{2}}\right]\label{sa_quantum}\end{eqnarray}
 where $\bar{u}$ and $\bar{g}$ are defined in (\ref{aux_self_consistent_eq}).
The frequency summation extends up to $n=n_{max}=\Lambda_{\nu}/(2\pi T)$.
Introducing as before $z=\phi/r$ and rescaling in addition the temperature
$\bar{T}=\gamma T/\bar{r}_{0}$, we numerically extract $r$ as a
function of $z$ from the second equation, substitute into the first
equation, and obtain $\bar{r}_{0}$ as a function of $z$. In Fig.
\ref{fig:figfiniteT} we plot $\bar{r}_{0}(z)$ and the phase diagrams
upon varying $\alpha$, $d$ and $\bar{T}$. We see the same trend
as in the classical phase diagram, namely, as $\alpha$ gets larger
at some fixed $\bar{T}$ and $d$, the Ising and magnetic transitions
split, and at even larger $\alpha$ the Ising-nematic transition eventually
becomes second order (panels (a)-(b)). The same trend holds upon the
reduction of dimensionality (i.e., make the system more two-dimensional)
at a fixed $\alpha$ and $\bar{T}$ (panels (c)-(d)) and also upon
decreasing $\bar{T}$ at a fixed $\alpha$ and $d$ (panels (e)-(f)).

Overall, we see that the phase diagram as a function of $\alpha$
does not change qualitatively between high temperatures, when the
classical approximation is valid, and $\bar{T}=O(1)$, when both classical
and quantum fluctuations are relevant. The only real difference is
the change in the critical values $\alpha_{c1}$ at which the Ising-nematic
and magnetic transitions split, and $\alpha_{c2}$ at which the Ising-nematic
transition becomes second-order. As $\bar{T}$ gets smaller, both
critical $\alpha$ become larger. Note that the rescaling of $T$
to $\bar{T}=\gamma T/\bar{r}_{0}$ is important for obtaining the
correct temperature dependence of the transitions due to hidden $T$
dependence in $z=\phi/r$ via $r=r(T)$.

\subsubsection{Anisotropic $d=3$ case}

We analyzed the two cases of anisotropic magnetic dispersion, Eqs.
(\ref{an_suscept}) and (\ref{an_suscept_1}), at a finite $T$. In
each case we found the same behavior as at large $T$, with different
phase diagrams depending on the degree of anisotropy, as discussed
in Fig. \ref{fig_ilya}. For strongly anisotropic, quasi-2D spin susceptibility
$\chi_{q}$, the results are similar to the case $2<d<3$, i.e the
magnetic tri-critical point is located on the left of the nematic
tri-critical point, and in the intermediate region the nematic transition
is first-order, while the magnetic transition is second-order (recall
Fig. \ref{fig:figfiniteT}). For weaker anisotropy, however, the two
tri-critical points interchange, as we observed in the purely classical
analysis, see Fig. \ref{fig_ilya}. Once this happens, in the intermediate
region between the two tri-critical points the nematic transition
is second-order, while the magnetic transition is first-order. The
only difference with respect to the purely classical case is that
both $\alpha_{c1}$ and $\alpha_{c2}$ shift to larger values at a
smaller $T$.

\section{RG analysis at a finite $N$ \label{sec:RG}}

The mean-field analysis is quite straightforward, but it is rigorously
justified only in the artificial limit of large $N$, where, we remind,
$N$ is the number of components of the $\boldsymbol{\Delta}$ fields.
The actual number of spin components is $N=3$, and it is by no means
guaranteed that the behavior at $N=3$ is the same as at large $N$.
To verify this, we need to return back to the effective action $S_{\mathrm{eff}}\left[\boldsymbol{\Delta}_{X},\boldsymbol{\Delta}_{Y}\right]$,
Eq. (\ref{action}), and use a complementary approach which is not
restricted to large $N$. One such approach, commonly used to study
phase transitions, is the renormalization group (RG) technique. In
RG, one progressively integrates out contributions from high energies
down to $E$ and analyzes how the parameters of the effective model
vary with $L\equiv\log W/E$, where $W$ is the bandwidth (the highest
energy scale in the problem). In our case, the parameters are $u$
and $g$, and in the RG approach one studies the flow of the running
couplings $u_{L}$ and $g_{L}$ and of any other coupling generated
by the RG flow. Alternatively, one can vary the distance to a transition,
i.e. vary $\bar{r}_{0}$, in which case $L=\log mW/\bar{r}_{0}$.

Quite generally, the RG flow may lead to three types of behavior depending
on the structure of the RG equations and on the bare values $u$ and
$g$. One possibility is that the couplings $u_{L}$ and $g_{L}$
flow to zero, which implies that there is no preemptive Ising-nematic
transition. Another possibility is that $g_{L}$ and $u_{L}$ flow
to infinity (more accurately, to strong coupling) in such a way that
the stability condition for the effective model $u_{L}>g_{L}$ is
not broken. In this situation, the system undergoes a second-order
Ising-nematic transition at the scale $L_{cr}=\log mW/\bar{r}_{0}^{cr}$
at which $g_{L}$ diverges. The third possibility is that the stability
condition $u_{L}>g_{L}$ gets broken at some $L^{*}<L_{cr}$. In this
case, the effective action becomes unstable with respect to a discontinuous
variation of $\boldsymbol{\Delta}_{X,Y}^{2}$, and the system undergoes
a first-order transition.

The RG approach is still a weak-coupling approach in the sense that
the bare couplings $u$ and $g$ are assumed to be much smaller than
the bandwidth. The advantage of the RG technique is that it can be
applied to any $N$, and, from this perspective, it goes beyond mean-field
approximation. However, the RG approach has its own limitations -
it can be rigorously justified only in the marginal dimension $d_{eff}=d+z_{dyn}=4$,
when the renormalizations are logarithmic. In our case $z_{dyn}=2$,
meaning that the marginal behavior takes place at $T=0$ and $d=2$.
Still, we obtained the $N=\infty$ phase diagram at $T=0$ and $d=2$
in the previous section, and it is instructive to compare that phase
diagram with the RG phase diagram at arbitrary $N$ to verify whether
or not the behavior at $N=3$ is the same as at $N=\infty$, at least
in this particular case. We remind that our $N=\infty$ quantum analysis
in $d=2$ shows that the system undergoes a first-order transition
into the magnetic state with $\phi=r$ for any $\alpha>1$. We now
analyze how the phase diagram looks like for arbitrary $N$.

\begin{figure}
\begin{centering}
\includegraphics[width=1\columnwidth]{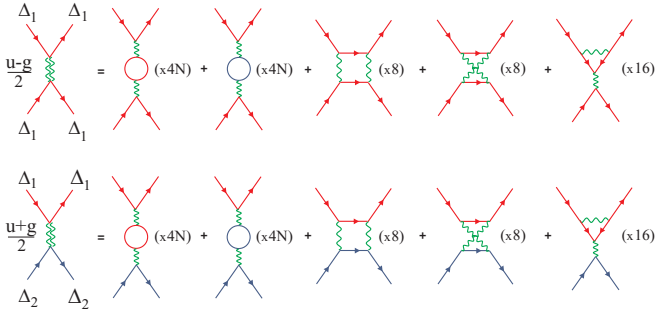} 
\par\end{centering}

\caption{(Color online) The one-loop diagrams responsible for the renormalization
of the two bosonic vertices $u_{11}=\left(u-g\right)/2$ and $u_{12}=\left(u+g\right)/2$,
with their respective combinatorial factors. The wavy lines correspond
to the interactions, whereas the continuous lines correspond to $\Delta_{1}$
and $\Delta_{2}$ (red/light gray and blue/dark gray lines, respectively).
Notice that the two first diagrams in each line contain a closed bosonic
loop and, therefore, an overall pre-factor of $N$. }

\label{fig_RG_diagrams} 
\end{figure}

We derived the one-loop RG equations for the flow of the running couplings
$u_{L}$ ad $g_{L}$ by the momentum-shell method (see, for instance,
Ref. \cite{Ma_book}) and also derived them by analyzing the parquet
diagrams for the renormalization of the four-boson vertices, which
are presented in Fig. \ref{fig_RG_diagrams} together with their respective
combinatorial pre-factors. Notice the special role of the diagrams
which contain a closed bosonic loop. Summation over the internal bosonic
indices yields a factor $N$, which does not appear in the diagrams
without closed loops. The mean-field results are reproduced if we
consider only these diagrams.

For further convenience, we rescale the coupling constants to $u,g\rightarrow2N\left(u,g\right)$.
Combining logarithmic contributions from all parquet diagrams we can
cast the renormalizations of $u$ and $g$ into the form of differential
RG equations\begin{eqnarray}
\dot{u}_{L} & = & -2\left(1+\frac{4}{N}\right)u_{L}^{2}-\frac{4}{N}g_{L}^{2}+\frac{4}{N}u_{L}g_{L}\nonumber \\
\dot{g}_{L} & = & 2\left(1+\frac{2}{N}\right)g_{L}^{2}-\frac{12}{N}u_{L}g_{L}\label{RG}\end{eqnarray}
 where $\dot{X}_{L}=dX_{L}/dL$.

For the particular cases $N=3$ and $N=6$, these equations reduce
to those obtained in Refs.\cite{Qi09} and \cite{Millis10} for the
similar model. We checked that no other couplings allowed by symmetry
are generated by the RG flow if only $u$ and $g$ have non-zero initial
values. In particular, if the third coupling $v$ in Eq. (\ref{action})
is initially zero, it is not generated by RG.

We recall that both $u_{L}$ and $g_{L}$ are initially positive (bare
values $u$ and $g$ are given by Eqs. \ref{aux_action}), and that
the bosonic action (\ref{action}) is stable as long as $u_{L}/g_{L}>1$.
If this condition breaks down at some $L$, one of the coefficients
of the quartic terms becomes negative and the system undergoes a first-order
transition into a state with a non-zero $\phi$.

For $N\rightarrow\infty$ the equations for $u_{L}$ and $g_{L}$
decouple and we can easily solve them and obtain \begin{eqnarray}
u_{L} & = & \frac{u}{1+2uL}\nonumber \\
g_{L} & = & \frac{g}{1-2gL}\label{solution_infty}\end{eqnarray}

We see that $u_{L}$ flows to zero, while $g_{L}$ increases under
the RG flow, diverging at $L_{cr}=\frac{1}{2g_{0}}$. If $g_{L}$
would be the only parameter in the problem, this divergence would
indicate a preemptive second-order Ising-nematic instability, since
the susceptibility of the Ising-nematic order parameter diverges at
$L_{cr}$. However, in our case there are two couplings, and the action
is stable only as long as $u_{L}>g_{L}$. This condition breaks down
at a smaller $L^{*}=\frac{1}{4g_{0}}-\frac{1}{4u_{0}}$, before $g_{L}$
diverges. The outcome is that for $N=\infty$ and $d=2$ the system
undergoes a first-order Ising-nematic transition at $T=0$. This is
in agreement with the mean-field analysis.

At a finite $N$, the two equations are coupled and both $u_{L}$
and $g_{L}$ can diverge. To understand what happens in this case,
it is convenient to define the ratio $p_{L}=u_{L}/g_{L}$ and re-express
the flow equations in terms of $g_{L}$ and $p_{L}$: \begin{eqnarray}
\dot{p}_{L} & = & 2g_{L}\left[\left(-1+\frac{2}{N}\right)p_{L}^{2}-p_{L}-\frac{2}{N}\right]\label{RG_alt_1}\\
\dot{g}_{L} & = & 2g_{L}^{2}\left(1+\frac{2}{N}-\frac{6}{N}p_{L}\right)\label{RG_alt}\end{eqnarray}

It is straightforward to verify that this set of RG equations has
several fixed trajectories along which $p_{L}$ is a constant and
$g_{L}$ evolves. The fixed trajectories are obtained by setting $\dot{p}_{L}=0$
in Eq. (\ref{RG_alt_1}). Solving the quadratic equation $\left(-1+2/N\right)p_{L}^{2}-p_{L}-2/N=0$
we find two fixed values $p_{1}=-1$ and $p_{2}=-2/\left(N-2\right)$.
The coupling $g_{L}$ diverges along the fixed trajectory with $p_{L}=p_{1}$
as $\dot{g}_{L}=2g_{L}^{2}\left(1+8/N\right)$. Along the second fixed
trajectory $p_{2}$, $g_{L}$ evolves according to $\dot{g}_{L}=2g_{L}^{2}\left(N^{2}+8\right)/\left(N^{2}-2N\right)$,
i.e. it diverges for $N>2$ and tends to zero for $N<2$.

To understand which trajectory is stable and which is not, we consider
small deviations from a fixed trajectory, $p_{L}=p_{i}+\delta p_{iL}$,
and expand the flow equations to lowest order in $\delta p_{iL}$.
We obtain

\begin{eqnarray}
\delta\dot{p}_{1L} & = & 2g_{L}\delta p_{1L}\left(\frac{N-4}{N}\right)\nonumber \\
\delta\dot{p}_{2L} & = & -2g_{L}\delta p_{2L}\left(\frac{N-4}{N}\right)\label{sa_24}\end{eqnarray}

We see that $p_{1}$ is a stable fixed trajectory for $N<4$ and unstable
for $N>4$, while $p_{2}$ is a stable fixed trajectory for $N>4$
and unstable for $N<4$.

There is also the third fixed trajectory $g_{L}=0$ ( $p_{3}=\infty$).
Expanding near $g_{L}=0$ we obtain from Eq. (\ref{RG}) \begin{eqnarray}
\dot{u}_{L} & = & -2\left(1+\frac{4}{N}\right)u_{L}^{2}\nonumber \\
\dot{g}_{L} & = & -\frac{12}{N}u_{L}g_{L}\label{RG_1}\end{eqnarray}
 whose solution is \begin{eqnarray}
u_{L} & = & \frac{u}{1+2\left(1+\frac{4}{N}\right)uL}\nonumber \\
g_{L} & = & g\left(\frac{u_{L}}{u}\right)^{\frac{6}{N+4}}\label{sa_25}\end{eqnarray}

The fixed trajectory $g_{L}=0$ is stable as long as $g_{L}$ remains
small compared to $u_{L}$. Evaluating $p_{L}=u_{L}/g_{L}$ from the
solutions of (\ref{sa_25}) we find \begin{equation}
p_{L}=\frac{u}{g}\left(\frac{u_{L}}{u}\right)^{\frac{N-2}{N+4}}=\frac{u}{g}\left[1+2\left(1+\frac{4}{N}\right)uL\right]^{\frac{2-N}{N+4}}\label{sa_26}\end{equation}

We see that, for $N>2$, $p_{L}$ decreases under the RG flow such
that eventually $g_{L}$ exceeds $u_{L}$. This implies that the fixed
trajectory $g_{L}=0$ is unstable. For $N<2$, however, $p_{L}$ increases
under the RG, and the trajectory $g_{L}=0$ is stable.

Combining this analysis with the numerical solution of Eqs. (\ref{RG_alt_1})
and (\ref{RG_alt}) at intermediate energies, we obtain three different
regimes of system behavior depending on the value of $N$.

\begin{figure}
\begin{centering}
\includegraphics[width=1\columnwidth]{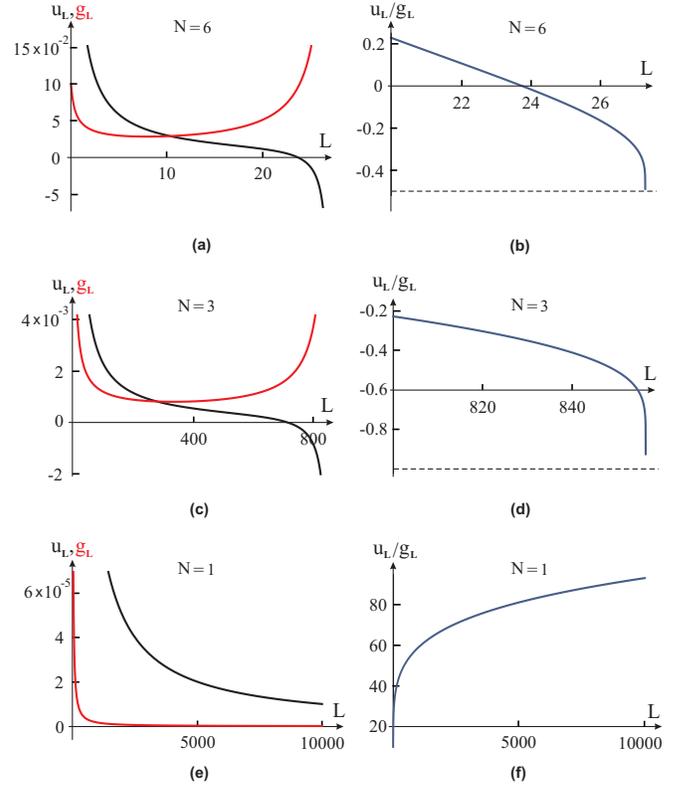} 
\par\end{centering}

\caption{(Color online) RG flow of the running coupling constants $u$ (black
line) and $g$ (red/light gray line), as well as their ratio $u/g$
(blue/dark gray line), as function of $L$. In (a)-(b), we present
the results for $N=6$; in (c)-(d), we display the results for $N=3$;
in (e)-(f) the results shown are for $N=1$. In all cases, the initial
values were $u=1$, $g=0.1$. The dashed lines in (b) and (d) refer
to the stable fixed trajectories $p_{2}=-2/\left(N-2\right)$ and
$p_{1}=-1$, respectively. }

\label{fig_RG_flow} 
\end{figure}

For $N>4$, the RG trajectory is as shown in Fig. \ref{fig_RG_flow}
(a)-(b). For arbitrary $\alpha=u/g$, the system approaches the stable
fixed trajectory $p_{L}=p_{2}=-2/\left(N-2\right)$. Since $p_{2}<0$,
and the bare value of $p$ is positive and larger than one, the running
coupling $p_{L}$ necessarily becomes one at some $L=L^{*}$ along
the RG flow. At this $L^{*}$ the action becomes unstable and the
system undergoes a first-order transition into a state with a non-zero
Ising-nematic order parameter. The only difference between finite
$N>4$ and $N=\infty$ is that, for a finite $N$, $u_{L}$ eventually
flows to infinity while for $N\rightarrow\infty$ it flows to zero.
This difference, however, does not play any role in our consideration
as the RG flow makes sense only as long as $u_{L}/g_{L}$ remains
larger than one.

For $2<N<4$, the fixed trajectory $p_{2}$ becomes unstable and cannot
be reached if the RG flow starts with $u>g>0$. The stable trajectory
to which the system flows is now $p_{1}=-1$, as shown in Fig. \ref{fig_RG_flow}
(c)-(d). Near this fixed trajectory we find from Eqs. (\ref{sa_24})
and (\ref{RG_alt}) that $g_{L}$ increases and diverges at some $L_{cr}$,
while $p_{L}+1=\delta p_{1L}\propto(1/g_{L})^{(4-N)/(N+8)}\to0$.

\begin{figure}
\begin{centering}
\includegraphics[width=0.9\columnwidth]{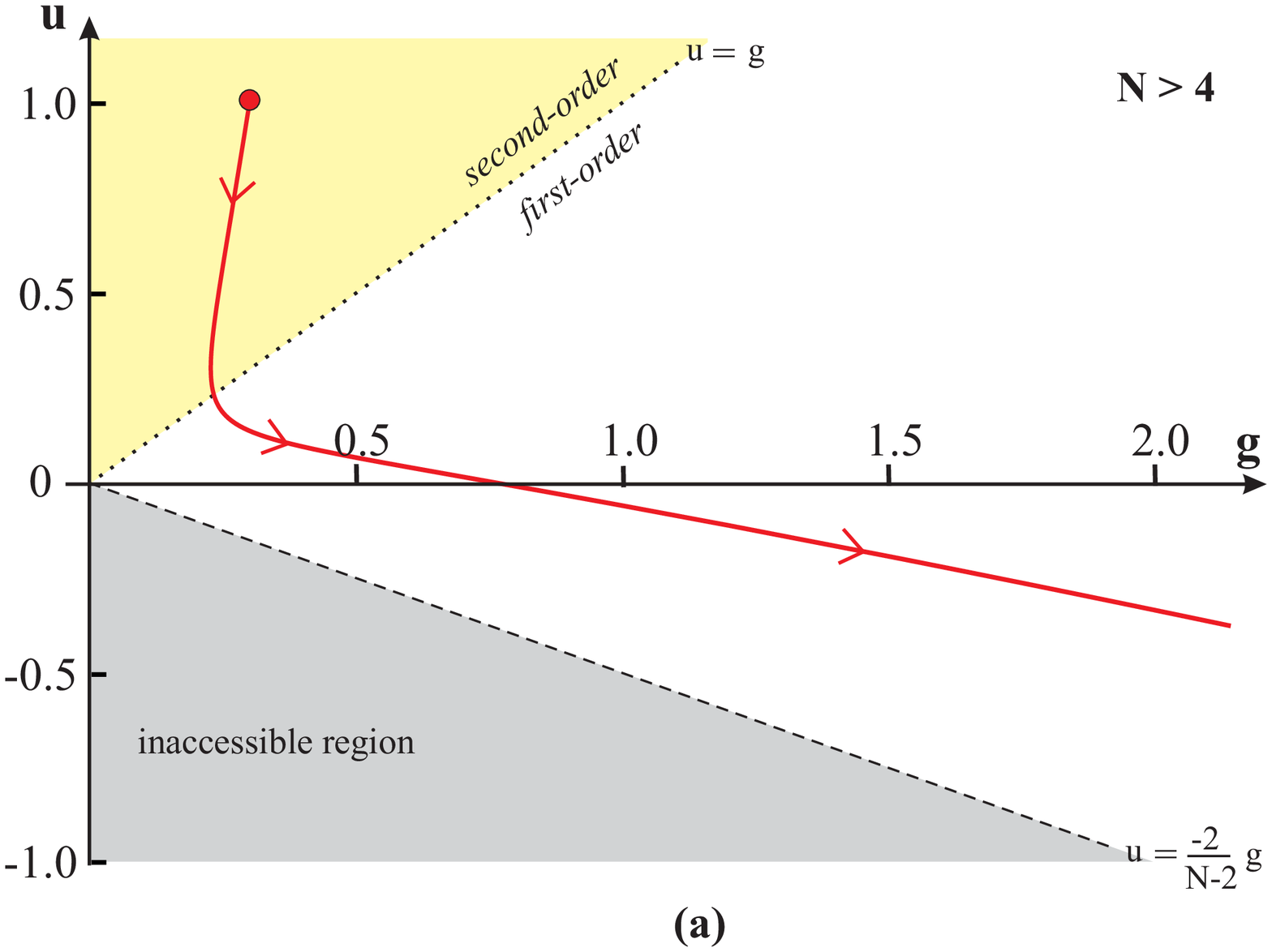} 
\par\end{centering}

\begin{centering}
\includegraphics[width=0.9\columnwidth]{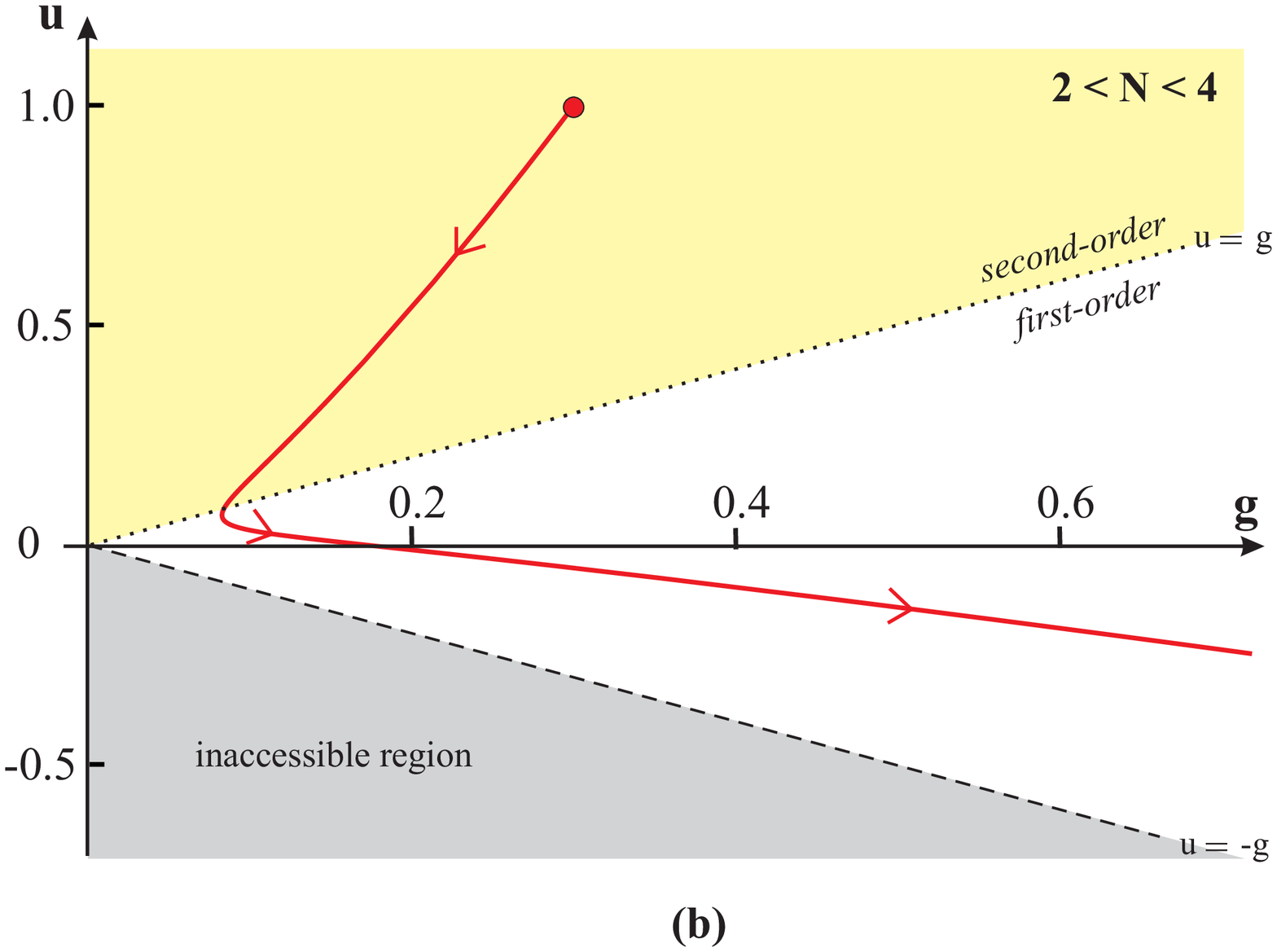} 
\par\end{centering}

\begin{centering}
\includegraphics[width=0.9\columnwidth]{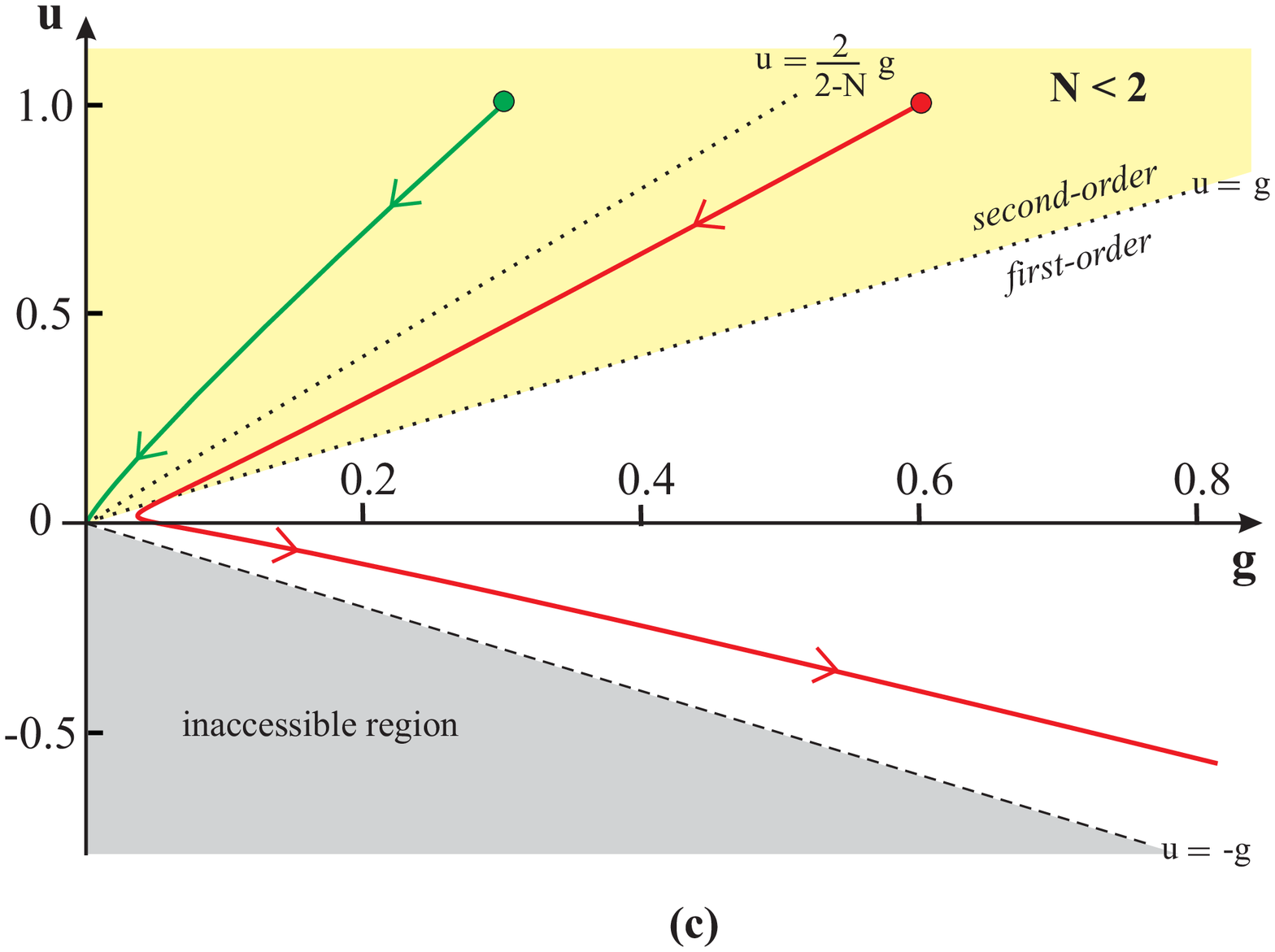} 
\par\end{centering}

\caption{(Color online) Structure of the RG flow in the $\left(g,u\right)$
plane for the representative values (a) $N=6$, (b) $N=3$, and (c)
$N=1$. The fixed trajectories $u=-p_{i}g$ are shown as dashed lines
in the lower-half planes ($p_{1}=-1$ for (b) and (c) and $p_{2}=-2/\left(N-2\right)$
in (a)), and are approached only at very large $L$. In the upper-half
planes, the dotted line $u=g$ separates the second-order and first-order
regimes. In (c), the dotted line $u=2g/\left(2-N\right)$ separates
initial conditions that fall in the basis of attraction of the $g=0$
fixed point ($p_{3}=\infty$, green/light solid line) from those that
flow to the $p_{1}=-1$ fixed trajectory (red/dark solid line). }

\label{fig_RG_structure} 
\end{figure}

While the fixed trajectory is different for $2<N<4$ and for $N>4$,
the behavior relevant to our purposes remains the same for all $N>2$,
namely in the process of RG flow towards a fixed trajectory, the ratio
$u_{L}/g_{L}$ reaches one at some $L=L^{*}$. At this point, the
action becomes unbounded and the system undergoes a first-order phase
transition into a state with a finite Ising-nematic order parameter.
The dependence on $N$ is then only quantitative - as $N$ decreases,
the scale $L^{*}$ gets progressively larger. To see this, we note
that for any finite $N$, the couplings $u_{L}$ and $g_{L}$ first
decrease under the RG flow and only at larger $L$ reverse the trend
and approach the fixed trajectory at $g_{L}\to\infty$ and $u_{L}\to-\infty$.
To get an estimate of how $L^{*}$ evolves with $N$, we use the approximate
Eq.(\ref{sa_26}) and identify $L^{*}$ with the RG scale at which
$p_{L}$ becomes one. We obtain \begin{equation}
L^{*}\propto\left(\frac{u}{g}\right)^{\frac{N+4}{N-2}}\label{sa_27}\end{equation}

For a given $u/g\gg1$, the value $L^{*}$ rapidly increases with
decreasing $N$; for instance, for $u/g=10$, the ratio of $L^{*}$
for $N=3$ and for $N=5$ is $10^{4}$. This means that at smaller
$N$ the system behaves over a wide range of energies as if magnetic
fluctuations were absent, and only very near $T_{N,0}$ it recognizes
that it actually undergoes a first-order nematic transition.

For $N<2$, a new behavior becomes possible, as shown in Fig. \ref{fig_RG_flow}
(e)-(f). Now the fixed trajectory $p_{2}=2/\left(2-N\right)$ crosses
the region $u>g>0$ from where the RG flow begins. Once the bare $u$
and $g$ are such that $u/g>2/\left(2-N\right)$, the RG flow is sandwiched
between the fixed trajectories $p_{2}$ and $p_{3}$ which are respectively
unstable and stable for $N<2$. The RG flow then moves both $g_{L}$
and $u_{L}$ towards $g_{L},u_{L}=0$, keeping $g_{L}<u_{L}$, i.e.
without crossing the first-order instability line. In this situation,
no preemptive nematic instability develops, and the system only undergoes
a mean-field magnetic transition at $T_{N,0}$. If, however, the initial
$u/g<2/\left(2-N\right)$, the system behavior is the same as before,
with the couplings evolving towards the fixed trajectory $p_{1}=-1$,
and $p_{L}$ becoming equal to one at some scale $L^{*}$, at which
the system undergoes a first-order nematic transition. In the formal
limit $N=0$, the whole region $u>g$ falls into the basis of attraction
of the $g_{L}=u_{L}=0$ fixed point, i.e. there is no preemptive nematic
instability for any $u/g>1$. The general structure of the RG flow
equations in the $\left(g,u\right)$ plane is shown in Fig. \ref{fig_RG_structure}.

For $d=2$ and $T=0$ this scenario of no preemptive nematic instability
holds only for $N<2$ and does not affect our actual case of $N=3$,
for which the system behavior under the RG flow is qualitatively the
same as in the mean-field, $N=\infty$ analysis. What happens for
$d>2$ and/or a finite $T$ is unclear because, in the absence of
the logarithmic terms, the approximations leading to the RG equations
are not justified. It is possible in principle that a preemptive nematic
transition does not occur for some large enough $\alpha=u/g$. If
this is the case, then there must be a reentrant behavior for large
$\alpha$ in Fig. \ref{fig_phase_diagram_regions}, i.e. the magnetic
transition temperature $T_{N}$ must reverse trend and come closer
to the nematic instability. However, the more likely scenario is that
the phase diagram which we obtained in the mean-field approximation
survives for the actual $N=3$ component bosonic field for all $\alpha$.

\section{consequences of the Ising-nematic order \label{sec:conseq}}

\subsection{Orbital Order}

\begin{figure}
\begin{centering}
\includegraphics[width=1\columnwidth]{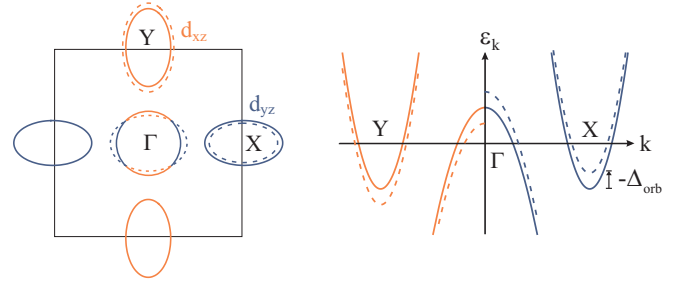} 
\par\end{centering}

\caption{(Color online) Schematic representation of the effect of orbital ordering
on the Fermi surface (left panel) and on the band dispersions (right
panel). The light/orange curves have $d_{xz}$ orbital character,
while the dark/blue curves have $d_{yz}$ orbital character. The solid
lines correspond to the Fermi surface and band dispersions without
orbital order ($T>T_{s}$), whereas the dashed lines refer to temperatures
below the onset of orbital order ($T<T_{s}$). In this figure we considered
$\Delta_{\mathrm{orb}}<0$. }

\label{fig_orbital_order} 
\end{figure}

ARPES measurements on detwinned samples have found that the onset
of resistivity anisotropy is accompanied by the onset of orbital order
in the paramagnetic phase, with different occupations for the $d_{xz}$
and $d_{yz}$ orbitals \cite{Shen11}. One possibility, explored by
several authors in different contexts, is that this orbital ordering
is an intrinsic instability of the system \cite{kruger09,RRPSingh09,Phillips10,w_ku10,bascones,maria,chen10,Lin11,Nevidomskyy11}.
In line with the theme of this work, we explore another possibility,
namely that the orbital order is induced by the Ising-nematic order.
This scenario is generally consistent with the small value of the
measured orbital polarization. To investigate this scenario quantitatively,
we consider a simplified two-orbital model in which the entire $X$
pocket has $d_{yz}$ character, while the entire $Y$ pocket has $d_{xz}$
character \cite{Zhang09}, and assume that there is a splitting $\Delta_{\mathrm{orb}}$
between the on-site energies of the $d_{xz}$ and $d_{yz}$ orbitals.
In the presence of such splitting, the fermionic dispersion becomes
anisotropic, see Fig. \ref{fig_orbital_order}, and the Hamiltonian
acquires the additional terms

\begin{equation}
\mathcal{H}_{\mathrm{orb}}=-\sum_{\mathbf{k}}\Delta_{\mathrm{orb}}\left(c_{X,\mathbf{k}\alpha}^{\dagger}c_{X,\mathbf{k}\alpha}-c_{Y,\mathbf{k}\alpha}^{\dagger}c_{Y,\mathbf{k}\alpha}\right)+\frac{a_{0}}{2}\,\Delta_{\mathrm{orb}}^{2}\label{H_orbital}\end{equation}

Including these two terms into the Hubbard-Stratonovich procedure
and expanding the effective action in powers of $\Delta_{\mathrm{orb}}$
we obtain

\begin{eqnarray}
S\left[\boldsymbol{\Delta}_{X},\boldsymbol{\Delta}_{Y},\Delta_{\mathrm{orb}}\right] & = & S_{\mathrm{eff}}\left[\boldsymbol{\Delta}_{X},\boldsymbol{\Delta}_{Y}\right]+\frac{a}{2}\Delta_{\mathrm{orb}}^{2}\nonumber \\
 &  & -w\left(\Delta_{X}^{2}-\Delta_{Y}^{2}\right)\Delta_{\mathrm{orb}}\label{S_orb}\end{eqnarray}
 with $S_{\mathrm{eff}}\left[\boldsymbol{\Delta}_{X},\boldsymbol{\Delta}_{Y}\right]$
given by Eq. (\ref{action}) and:

\begin{eqnarray}
a & = & a_{0}+4\int_{k}G_{X,k}^{2}\nonumber \\
w & = & 2\int_{k}G_{\Gamma,k}G_{X,k}^{2}\label{aux_S_orb}\end{eqnarray}

Evaluating the integrals and expanding around perfect nesting we find
$a>0$ and $w=-c\mu$ with $c>0$. The $w$ term describes the linear
coupling between the orbital and Ising-nematic order parameters, i.e.
the development of one order triggers the development of the other.
Differentiating Eq. (\ref{S_orb}) with respect to $\Delta_{\mathrm{orb}}$
we obtain \begin{equation}
\left\langle \Delta_{\mathrm{orb}}\right\rangle =\frac{w}{a}\left\langle \Delta_{X}^{2}-\Delta_{Y}^{2}\right\rangle \label{sept_1}\end{equation}

Since $w\propto-\mu$ and $\mu$ scales with doping, the induced orbital
order is expected to be small at small doping. This is in accordance
with the experimental data. The sign of the orbital splitting also
agrees with the data\cite{Shen11}: polarized ARPES measurements on
detwinned electron-doped $\mathrm{Ba\left(Fe_{1-\mathit{x}}Co\mathit{_{x}}\right)_{2}As_{2}}$
samples have shown that when $\left\langle \Delta_{X}^{2}\right\rangle >\left\langle \Delta_{Y}^{2}\right\rangle $,
$\left\langle \Delta_{\mathrm{orb}}\right\rangle $ is negative. This
is consistent with (\ref{sept_1}) since for electron-doped materials
$\mu>0$ and $w<0$. To the best of our knowledge, similar measurements
have not been carried out for hole-doped samples. Interestingly, Eq.
(\ref{sept_1}) shows that the sign of $\left\langle \Delta_{\mathrm{orb}}\right\rangle $
should change for hole-doped materials, which are described in our
model by $\mu<0$. \textbf{ }Note that long-range magnetic order can
also induce orbital polarization \cite{bascones}.

\subsection{Structural order}

The same reasoning also applies to the interplay between the Ising-nematic
and the orthorhombic orders. The structural order is detected experimentally
as the difference between the lattice constants $a$ and $b$ along
the $x$ and $y$ directions of the Fe-plane, respectively. In the
ideal situation, structural order appears only below a particular
structural transition temperature $T_{s}$. In reality, however, some
orthorhombic distortion can be present at any $T$ due to internal
strains. In the case of detwinned samples, a small in-plane strain
is applied along one of the orientations \cite{Fisher11,Tanatar10,Blomberg11,Kuo11}.
Then, strictly speaking, $\varepsilon_{s}=a-b$ is never zero, i.e.
there is no well-defined $T_{s}$ for finite strain. Still, experimentally
one can identify the crossover temperature below which $\varepsilon_{s}$
sharply increases.

The relationship between Ising-nematic and structural orders can be
obtained in the same way as in the previous subsection. Introducing
the orthorhombic order parameter in a way similar to Eq. (\ref{H_orbital})
yields\begin{equation}
\mathcal{H}_{\mathrm{str}}=-\lambda\sum_{\mathbf{k}}\varepsilon_{s}\left(c_{X,\mathbf{k}\alpha}^{\dagger}c_{X,\mathbf{k}\alpha}-c_{Y,\mathbf{k}\alpha}^{\dagger}c_{Y,\mathbf{k}\alpha}\right)+\frac{C_{s}}{2}\epsilon_{s}^{2}\label{H_str}\end{equation}
 where $\lambda$ is a coupling constant and $C_{s}$ is the shear
modulus. Including these two terms into the Hubbard-Stratonovich procedure
we obtain, in the mean-field approximation, the same effective action
as Eq. (\ref{S_orb}), but with $\varepsilon_{s}$ instead of $\Delta_{\mathrm{orb}}$
and with the renormalized shear modulus $C_{s}$ instead of $a$.
Accordingly, the orthorhombic and Ising-nematic order parameters are
linearly related: \begin{equation}
\left\langle \varepsilon_{s}\right\rangle =\frac{\lambda w}{C_{s}}\left\langle \Delta_{X}^{2}-\Delta_{Y}^{2}\right\rangle \label{sept_2}\end{equation}
 i.e. one order immediately triggers the other. Notice, however, that
in distinction to the case of orbital order (Eq. \ref{sept_1}), the
proportionality constant between $\left\langle \varepsilon_{s}\right\rangle $
and the nematic order parameter contains not only the parameter $w$,
but also the magneto-elastic coupling $\lambda$, which can itself
depend on the chemical potential and on additional details of the
band structure.

The linear relation between the two orders has been discussed in a
number of earlier papers \cite{Fang08,FernandesPRL10,Indranil11},
and is not surprising because both orders break the same $Z_{2}$
symmetry between the $x$ and $y$ directions (the orbital order does
the same). It also implies that, in detwinned samples, the applied
strain plays the role of a conjugate field to the Ising-nematic order
parameter\cite{Fernandes11}. We recall that the proportionality coefficient
$w$ scales with $\mu$ and is generally small. As a result, if the
orthorhombic order is induced by the Ising-nematic order at $T=T_{s}$,
the orthorhombic order parameter is initially small and may become
visible only at some distance below the Ising-nematic transition.
This may explain why recent magnetic torque experiments observed nematic
order up to higher temperatures than the structural order \cite{Matsuda11}.
We emphasize, however, that there is only one well-defined transition
temperature below which the tetragonal symmetry is broken.

\begin{figure}
\begin{centering}
\includegraphics[width=0.95\columnwidth]{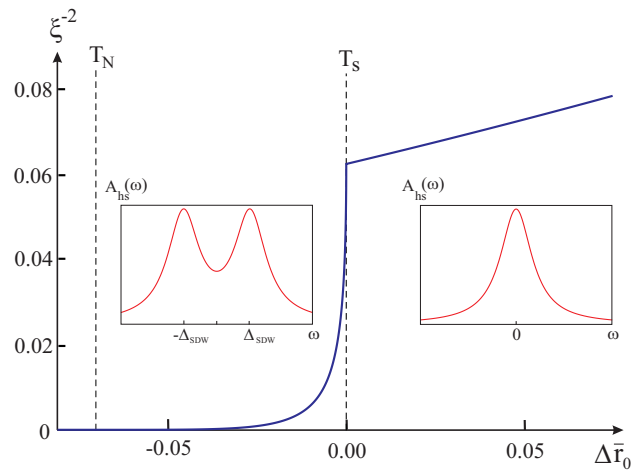} 
\par\end{centering}

\caption{(Color online) Inverse squared magnetic correlation length $\xi^{-2}\propto r$
(in arbitrary units), as function of the reduced temperature $\Delta\bar{r}_{0}\propto T-T_{s}$,
for the case $\alpha=4$ in the phase diagram of Fig. \textbf{\ref{fig_phase_diagram_regions}}.
The dashed lines denote the split second-order nematic ($T_{s}$)
and magnetic ($T_{N}$) transitions. Note the sharp increase and the
discontinuity in the derivative of $\xi$ at the nematic transition.
The insets show schematically, at different temperatures, the frequency-dependence
of the fermionic spectral function $A_{\mathrm{hs}}\left(\omega\right)$
at the hot spots, i.e. the Bragg points of the band dispersion $\varepsilon_{\Gamma,\mathbf{k}}=\varepsilon_{(X,Y),\mathbf{k}+\mathbf{Q}_{i}}$.
Notice the development of peaks at $\omega=\pm\Delta_{\mathrm{SDW}}$
below the nematic transition, where $\Delta_{\mathrm{SDW}}$ denotes
the value of gap that is opened due to SDW order at zero temperature.}

\label{fig_pseudogap} 
\end{figure}

Note that the coupling to structural degrees of freedom also renormalizes
the nematic coupling constant as \cite{FernandesPRL10}

\begin{equation}
\tilde{g}=g+\frac{\lambda^{2}w^{2}}{C_{s}}\label{renormalized_g}\end{equation}

Therefore, even if initially $g=0$, the coupling to the lattice generates
a non-zero nematic coupling and, consequently, nematic order \cite{Indranil11}.

\subsection{Pseudogap behavior \label{sub_pseudogap}}

Recent ARPES data on $\mathrm{NaFeAs}$, whose strongly anisotropic
Fermi surface is very similar to that of the 1111 compounds \cite{Kusakabe_NaFeAs},
show that the reconstruction of the fermionic spectrum begins at temperatures
around $T_{s}\approx54$ K, where the nematic order sets in, rather
than at $T_{N}\approx39$ K, where the stripe magnetic order develops
\cite{He_NaFeAs}. Such a reconstruction increases below $T_{N}$,
which is a good indication that the effect likely has a magnetic origin.

Thermal magnetic fluctuations do give rise to a pseudogap behavior
(often termed as magnetic precursors) by transferring spectral weight
from small frequencies to a frequency comparable to the spectral gap
developed in the magnetically ordered state \cite{magn_pseudogap}.
The only precondition for magnetic precursors is that the magnetic
correlation length $\xi$ must be large enough.

How can the pseudogap develop at the nematic transition? At first
glance, this seems a mere coincidence because the nematic order is
a collective instability in the particle-hole channel with momentum
$q=0$, and a $q=0$ boson cannot reconstruct the fermionic spectrum.
However, one of our key results is that at the nematic transition
the magnetic correlation length increases either discontinuously or
very sharply. Even away from the nematic tri-critical point, where
the Ising-nematic transition is second-order, the magnetic correlation
length has a discontinuous derivative at $T_{s}$, since $\xi^{-2}\rightarrow\xi^{-2}-\phi$,
making it increase faster. This behavior is shown in Fig. \ref{fig_pseudogap},
where we plot the inverse magnetic correlation length $r\propto\xi^{-2}$
as a function of temperature for $\alpha=4$ for the phase diagram
of Fig. \ref{fig_phase_diagram_regions}\textbf{.} Interestingly,
in $\mathrm{NaFeAs}$, as well as in other iron pnictides \cite{nakai_LaFeAsO},
a significant enhancement of magnetic fluctuations was observed \cite{kitagawa_NaFeAs,Ma_NaFeAs}
just below $T_{s}$ by nuclear magnetic resonance (NMR). Once the
correlation length jumps (or sharply increases) to a larger value,
the strength of thermal magnetic fluctuations rapidly increases. As
a result, the fermionic spectral function $A_{\mathbf{k}}\left(\omega\right)$
develops a magnetic pseudogap via the transfer of spectral weight
from zero to finite frequencies at the hot spots $\varepsilon_{\Gamma,\mathbf{k}}=\varepsilon_{(X,Y),\mathbf{k}+\mathbf{Q}_{i}}$
(see Fig. \ref{fig_pseudogap}), what leads to the reconstruction
of the fermionic spectrum, although zero-frequency states appear only
below $T_{N}$\textbf{. }We argue therefore that the jump (or sharp
increase) in $\xi$ at the nematic transition is the {}``glue''
that links the nematic instability and the development of the pseudogap
in the fermionic spectral function.

\section{comparison with experiments and earlier theories \label{sec:comparison}}

\subsection{Experimental phase diagrams}

We now compare our results of Section \ref{sec:solution} for anisotropic
systems to the experimental phase diagrams of the 122 and 1111 compounds.
Our theoretical phase diagram is plotted in Fig. \ref{fig_exp_phase_diagram}
(a) and (b). Our goal is to relate the changes of the system behavior
as function of the parameter $\alpha=u/g$ to the measured changes
imposed by doping, pressure, and alkaline-earth substitution. For
this we need to (i) properly place the parent compounds onto our phase
diagram, (ii) decide which of the two similar but not identical phase
diagrams in Fig. \ref{fig_exp_phase_diagram} is more appropriate,
and (iii) understand how $\alpha$ varies with the experimental parameters.
We first consider 122 materials and then briefly discuss 1111 materials.

We begin with the parent 122 materials. There are three well studied
types of 122 systems, namely, $\mathrm{CaFe_{2}As_{2}}$, $\mathrm{SrFe_{2}As_{2}}$,
and $\mathrm{BaFe_{2}As_{2}}$. The first two materials undergo a
strong simultaneous magneto-structural first-order transition, as
evidenced from several thermodynamic measurements \cite{Colombier09,Torika08,Ni_Ca_firstorder}.
This clearly places both materials well into the region of small $\alpha$,
to the left of the first tri-critical point - either $\alpha_{c1}$
in Fig. \ref{fig_exp_phase_diagram}(a) or $\alpha_{c2}$ in Fig.
\ref{fig_exp_phase_diagram}(b). For $\mathrm{BaFe_{2}As_{2}}$, X-ray
diffraction \cite{Rotundu,Kim11} as well as high-accuracy magnetization
measurements \cite{Birgeneau11} find a very small splitting between
the structural and magnetic transitions ($T_{s}\sim141$ K, $T_{N}\sim140$
K). This places $\mathrm{BaFe_{2}As_{2}}$ in the region of larger
$\alpha$ (i.e. smaller $g$), which can be either slightly to the
right of the magnetic tri-critical point $\alpha_{c1}$ in Fig. \ref{fig_exp_phase_diagram}(a),
or slightly to the right of the nematic tri-critical point $\alpha_{c2}$
in Fig. \ref{fig_exp_phase_diagram}(b). A smaller $\alpha$ for the
parent compounds $\mathrm{CaFe_{2}As_{2}}$, $\mathrm{SrFe_{2}As_{2}}$
compared to the parent compound $\mathrm{BaFe_{2}As_{2}}$ can be
explained by the difference in the values of the shear modulus $C_{s}$.
The analysis of their phase diagrams under pressure \cite{Uhoya_pressure_Ba,Uhoya_pressure_Sr,Goldman_pressure_Ca},
as well as of their mechanical properties\cite{Canfield_pc}, shows
that $\mathrm{CaFe_{2}As_{2}}$ and $\mathrm{SrFe_{2}As_{2}}$ are
softer than $\mathrm{BaFe_{2}As_{2}}$. According to our Eq. (\ref{renormalized_g}),
softer systems with smaller shear modulus $C_{s}$ (and possibly larger
magneto-elastic coupling $\lambda$) have larger $g$, and hence smaller
$\alpha$.

Which of the two phase diagrams in Fig. \ref{fig_exp_phase_diagram}
is more appropriate for $\mathrm{BaFe_{2}As_{2}}$ is a more subtle
issue. The x-ray data \cite{Rotundu,Kim11} show that the orthorhombic
order parameter evolves continuously immediately below $T_{s}$ and
then jumps at the same temperature where the magnetic transition takes
place, as evidenced by the magnetization data \cite{Birgeneau11}.
Neutron diffraction data \cite{sato} do not detect critical magnetic
fluctuations above $T_{N}$, consistent with the idea that the magnetic
transition is first-order and simultaneous to the meta-nematic transition.
This favors the phase diagram of Fig. \ref{fig_exp_phase_diagram}(b).
However, the same neutron data do not detect a clear jump of the magnetic
order parameter at the magnetic transition, as expected at a first-order
transition. So, it is possible in principle that the continuous evolution
of the nematic/orthorhombic order parameter immediately below $T_{s}$
is a secondary effect - due to internal strain, for instance. Then,
the true nematic transition would be first-order, while the magnetic
transition occurring at $T_{N}\leq T_{s}$ would be second order.
This would be consistent with the phase diagram of Fig. \ref{fig_exp_phase_diagram}(a).
In any case, however, the key observation is that the parent compound
$\mathrm{BaFe_{2}As_{2}}$ is in the region where the structural and
magnetic transitions are quite close to each other and at least one
of them is first order.

To investigate the effect of doping and pressure, we need to understand
how $\alpha=u/g$ changes with the chemical potential $\mu$ and the
mass anisotropy of the electron pockets $\delta m=m\left(m_{x}-m_{y}\right)/2m_{x}m_{y}$.
We use Eqs. (\ref{aux_action}) and evaluate $u$ and $g$ for small
chemical potential $\mu/\varepsilon_{0}$ and small $\delta m$. For
the ratio $u/g$ we obtain \begin{equation}
\alpha=\frac{u}{g}\approx42\left(\frac{T}{\varepsilon_{0}\delta m}\right)^{2}\left[1+0.9\left(\frac{\mu}{T}\right)^{2}+0.01\left(\frac{\varepsilon_{0}\delta m}{T}\right)^{2}\right]\label{u_g_ratio}\end{equation}

Electron doping adds carriers to the electron pockets and increases
the magnitude of the chemical potential $\mu$. This, according to
(\ref{u_g_ratio}), \textit{increases} $\alpha$, i.e., under electron
doping the system should move to the regime of split second-order
transitions (see Fig. \ref{fig_exp_phase_diagram}). This is the main
prediction of our theory.

This prediction generally agrees with the experimental phase diagrams
of electron-doped 122 materials, where $\mathrm{Fe}$ is substituted
by $\mathrm{Co,\, Ni,\, Cu,\, Pd,}$ or $\mathrm{Rh}$ \cite{Canfield_transition_metals}.
In particular, for $\mathrm{Ba\left(Fe_{1-\mathit{x}}Co\mathit{_{x}}\right)_{2}As_{2}}$,
magnetic and x-ray measurements demonstrated \cite{Kim11,Birgeneau11}
that the magnetic and structural transitions rapidly split and both
become second-order above $x\approx0.022$. This is particularly evidenced
by the fact that the peak in the derivative of the magnetic susceptibility
$\chi$ (indicative of a first order magnetic transition) is strongly
suppressed beyond this doping value \cite{Birgeneau11}. The splitting
of both transitions upon doping has also been observed in $\mathrm{Ca\left(Fe_{1-\mathit{x}}Co\mathit{_{x}}\right)_{2}As_{2}}$
at $x\approx0.039$ by neutron diffraction data \cite{Prokes11}.
In these systems, the structural transition remains first-order at
least for some doping range after the splitting, what is consistent
with the phase diagram in Fig. \ref{fig_exp_phase_diagram}(a) if
we identify $x\approx0.039$ with $\alpha_{c1}$. The behavior of
the isovalently doped compound $\mathrm{Ba\left(Fe_{1-\mathit{x}}Ru\mathit{_{x}}\right)_{2}As_{2}}$
is also consistent with our theory. In this material, the chemical
potential does not change with doping\cite{Dhaka11}, hence the structural
and magnetic transitions should remain very close for all $x$. Thermodynamic
measurements did indeed find that the two transitions do not split
upon increasing $x$ \cite{Thaler10}.

The interplay between structural and magnetic transitions in the hole-doped
$\mathrm{\left(Ba_{1-\mathit{x}}K_{\mathit{x}}\right)Fe_{2}As_{2}}$
compounds is not so well established, with conflicting reports of
either simultaneous first-order transitions \cite{Avci11}, or split
first-order structural transition and second-order magnetic transition
\cite{Urbano10}. In the context of our model, hole doping adds an
extra complication since at least some effects of doping are absorbed
into the changes of an additional hole pocket at $\left(\pi,\pi\right)$
\cite{Kemper10}. This suggests that the chemical potential in our
effective 4-band model changes at a slower rate than in the case of
electron doping, since extra holes do not necessarily go to the central
pockets. If this is the case, then a first-order structural transition
extends to larger dopings, in agreement with the data.

Consider now the evolution of $\alpha$ with pressure. Band structure
calculations show that pressure reduces the nesting features of the
Fermi surface \cite{Kimber09,Valenti09}, what in our model implies
that $\delta m$ increases. According to Eq. (\ref{u_g_ratio}), $\alpha\propto1/(\delta m)^{2}$
then \textit{decreases}, bringing the system deeper into the regime
of simultaneous first-order structural and magnetic transitions (see
Fig. \ref{fig_exp_phase_diagram}). This agrees with the experimental
results that structural and magnetic transitions do not split under
pressure in $Ae\mathrm{Fe_{2}As_{2}}$ ($Ae=\mathrm{Ba,\, Co,\, Sr}$)
\cite{Colombier09,Goldman_pressure_Ca}.

We now briefly consider the 1111 materials, whose Fermi surfaces are
also similar to the previously discussed $\mathrm{NaFeAs}$ compound.
In these systems, the nematic and magnetic transitions are split and
second-order already in the parent compounds. In our theory, these
compounds should then be placed to the right of $\alpha_{c2}$ in
Fig. \ref{fig_exp_phase_diagram}(a) or to the right of $\alpha_{c1}$
in Fig. \ref{fig_exp_phase_diagram}(b). The reason why $\alpha$
is larger in the 1111 materials is the significantly larger degree
of out-of-plane anisotropy in the 1111 materials compared to the 122
materials, what brings the former closer to the $d=2$ limit. In our
modeling, these systems are then described by a smaller effective
$d$. According to our theory, as $d$ gets smaller, both tri-critical
points $\alpha_{c1,2}$ shift towards smaller values (see Fig. \ref{fig:figfiniteT}
and Eq. (\ref{alpha_c})), extending the regime where the magnetic
and structural transitions are split and second-order. A similar argument
was given in Ref. \cite{Qi09}.

Finally, for the iron chalcogenides $\mathrm{FeTe_{1-x}Se_{x}}$,
our model is applicable in the regime of intermediary Se doping, near
the superconducting dome of the $\left(x,T\right)$ phase diagram.
In this region, ARPES measurements \cite{ZXShen_private} reveal that
the electronic structure is similar to the one considered in our model
(see Fig. \ref{fig_nematic}), and neutron scattering shows \cite{Broholm10}
that magnetic fluctuations are peaked at $\mathbf{Q}_{1}=\left(\pi,0\right)$
and $\mathbf{Q}_{2}=\left(0,\pi\right)$. On the other hand, our model
is not suitable for the undoped $\mathrm{FeTe}$ sample, where the
same ARPES data show the absence of electron pockets centered at $\mathbf{Q}_{1}$
and $\mathbf{Q}_{2}$ \cite{ZXShen_private}.

\subsection{Linear relation between the magnetic and nematic order parameters}

\begin{figure}
\begin{centering}
\includegraphics[width=0.8\columnwidth]{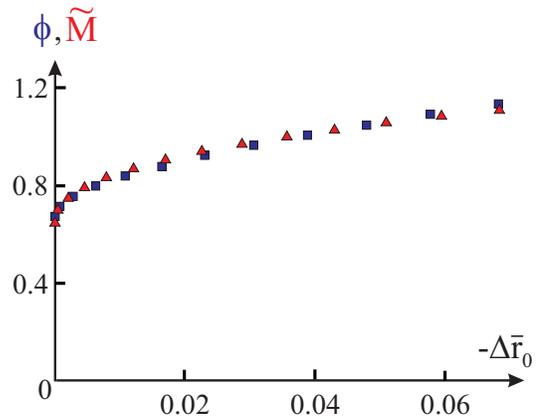} 
\par\end{centering}

\caption{(Color online) Nematic order parameter $\phi$ (square/blue symbols)
and re-scaled magnetic order parameter $\tilde{M}$ (triangles/red
symbols) for $d=2.5$ and $\alpha=1.8$ as function of $-\Delta\bar{r}_{0}=r_{0,cr}-r_{0}\propto(T_{s}-T)$
(the same as in Fig. \ref{fig_order_parameters}d). Here, $\tilde{M}=0.87M+0.33$. }

\label{fig_linear_m_phi_temp} 
\end{figure}

\begin{figure}
\begin{centering}
\includegraphics[width=0.8\columnwidth]{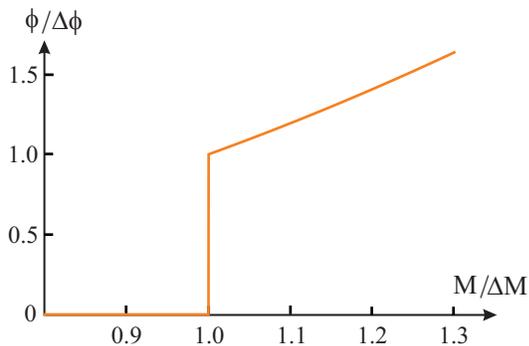} 
\par\end{centering}

\caption{(Color online) Nematic order parameter $\phi$ as function of the
magnetic order parameter $M$ for $d=2.5$ and $\alpha=1.05$. Note
the near linear behavior between them. Here $\Delta\phi$ and $\Delta M$
refer to the jumps at each transition. }

\label{fig_linear_m_phi} 
\end{figure}

Several experimental groups showed that in some 122 materials, most
notably $\mathrm{SrFe_{2}As_{2}}$, the magnetic and nematic order
parameters have very similar temperature dependencies below the simultaneous
first-order magneto-nematic transition \cite{Birgeneau10,Avci11,Geibel08}.
In our analysis, the relationship between $M$ and $\phi$ is given
by Eq. (\ref{aux_r0_m_sc}) and can be readily seen in Fig. \ref{fig_order_parameters}d
for a simultaneous first-order transition. Re-plotting in Fig. \ref{fig_linear_m_phi_temp}
the nematic order parameter $\phi$ and the properly re-scaled magnetic
order parameter $M$ as functions of temperature, we see that both
follow the same trend, indicating that the relationship between the
two order parameters is nearly linear, in agreement with the experimental
data. Even better linear relation is obtained for $\alpha$ closer
to one, as we show in Fig. \ref{fig_linear_m_phi}, where $\phi/\Delta\phi$
is plotted explicitly as function of $M/\Delta M$ ($\Delta\phi$
and $\Delta M$ denote the values of the jumps across the first-order
magneto-nematic transition).

At first glance, this near-linear relation seems non-trivial, because
for split second-order magnetic and nematic transitions, a straightforward
expansion leads to $\phi-\phi\left(m=0\right)\propto M^{2}$. It can,
however, be easily understood analytically by expanding Eq. (\ref{aux_r0_m_sc})
around the jumps $\Delta\phi$ and $\Delta M$. We obtain\begin{equation}
\frac{\phi-\Delta\phi}{\Delta\phi}=\sum_{j=1}^{\infty}a_{j}\left(\frac{M-\Delta M}{\Delta M}\right)^{j}\label{relation_phi_m}\end{equation}

For small $\phi-\Delta\phi$ and $M-\Delta M$, the relationship is
indeed linear, since $a_{1}$ is non-zero for all $2<d\leq3$ (see
Appendix D). In fact, in $\mathrm{SrFe_{2}As_{2}}$, where the linear
relation was experimentally observed\cite{Geibel08}, the measured
temperature dependent orthorhombic and magnetic order parameters are
rather small compared to the magnitude of the jumps. Interestingly,
when $\alpha\rightarrow1$ and the first-order transition gets stronger,
the coefficients of Eq. (\ref{relation_phi_m}) satisfy $a_{1}/a_{2}=2$
and $a_{j+2}=0$ for any $2<d\leq3$, i.e., the dependence has only
linear and quadratic terms. In the same limit, $\Delta\phi$ and $\Delta M$
become large, since both scale as $1/(\alpha-1)$. Thus, a plot of
$\phi$ versus $M$ in absolute units would show a strictly linear
dependence for $\alpha\rightarrow1$.

\subsection{$J_{1}-J_{2}$ and phenomenological models}

The possibility of an Ising-nematic order induced by magnetic fluctuations
was first proposed for the iron pnictides within the localized-moment
scenario \cite{Si08,Fang08,Xu08}, built upon earlier results by Chandra,
Coleman, and Larkin \cite{chandra}. In this localized-moment approach,
one considers spins on a square lattice interacting via a nearest-neighbor
antiferromagnetic exchange $J_{1}$ and a next-nearest neighbor antiferromagnetic
exchange $J_{2}$. It was argued \cite{Si08} that the hybridization
between the Fe and As orbitals in the iron pnictides gives rise to
a rather large $J_{2}$, which can exceed $J_{1}/2$. Once this happens,
the magnetic ground state develops the stripe order with $O(3)\times Z_{2}$
order parameter manifold.

As shown in Ref. \cite{chandra}, one can interpret the stripe order
as composed of two inter-penetrating Neel sublattices, with staggered
magnetization $\mathbf{M}_{1}$ and $\mathbf{M}_{2}$. The configuration
with $\mathbf{M}_{1}$ parallel (anti-parallel) to $\mathbf{M}_{2}$
corresponds to the $\left(\pi,0\right)$ ($\left(0,\pi\right)$) state.
At the mean-field level, the two sublattices are uncoupled. Thermal
and quantum fluctuations, however, induce a coupling between the two
sublattices that favors the collinear configurations. Once this coupling
is included, the effective action takes the form \begin{eqnarray}
S_{J_{1}-J_{2}}\left[\mathbf{M}_{i}\right] & = & J_{2}\int_{q}q^{2}\left(\mathbf{M}_{1,q}\cdot\mathbf{M}_{1,-q}+\mathbf{M}_{2,q}\cdot\mathbf{M}_{2,-q}\right)\nonumber \\
 &  & +J_{1}\int_{q}q_{x}q_{y}\left(\mathbf{M}_{1,q}\cdot\mathbf{M}_{2,-q}\right)\nonumber \\
 &  & -\zeta\frac{J_{1}^{2}}{J_{2}}\int_{x}\left(\mathbf{M}_{1}\cdot\mathbf{M}_{2}\right)^{2}\label{S_J1_J2}\end{eqnarray}
 where the magnitudes of $\mathbf{M}_{1}$ and $\mathbf{M}_{2}$ are
fixed, and $\zeta$ is a dimensionless constant which is non-zero
because of thermal and/or quantum fluctuations. This constant is small
in a $1/S$ expansion, where $S$ denotes spin, and remains small
numerically even for $S=1/2$ (see, for instance \cite{Sushkov03}).

It is straightforward to make connection between our itinerant model,
Eq. (\ref{action}), and the $J_{1}-J_{2}$ model. First, the relationship
between the real-space order parameters $\mathbf{M}_{1}$ and $\mathbf{M}_{2}$
and the momentum-space order parameters $\boldsymbol{\Delta}_{X}$
and $\boldsymbol{\Delta}_{Y}$ is $\boldsymbol{\Delta}_{X}=\mathbf{M}_{1}+\mathbf{M}_{2}$
and $\boldsymbol{\Delta}_{Y}=\mathbf{M}_{1}-\mathbf{M}_{2}$. The
scalar product $\mathbf{M}_{1}\cdot\mathbf{M}_{2}$ is then the same
as $\boldsymbol{\Delta}_{X}^{2}-\boldsymbol{\Delta}_{Y}^{2}$ and
the nematic coupling is given by $g=\zeta J_{1}^{2}/J_{2}$. Second,
in the itinerant approach, the hard constraint $\mathbf{M}_{1}^{2}=\mathbf{M}_{2}^{2}=1$
is replaced by the quartic terms, which play the role of soft constraints.
We see that the itinerant and the $J_{1}-J_{2}$ models are indeed
quite similar, and in both models the Ising-nematic order results
from the $Z_{2}$ degeneracy of the stripe magnetic ground state.

Whether the two models have identical phase diagrams and show the
same behavior upon doping, pressure, and alkaline-earth substitution
is a more subtle issue. In the $J_{1}-J_{2}$ model, the analog of
$\alpha$ is $J_{2}^{2}/(\zeta J_{1}^{2})$. Taken at face value,
this quantity is large (because $\zeta$ is small) and depends only
weakly on doping and pressure. For instance, the parameters used in
Ref. \cite{Fang08} yield\textbf{ $\alpha\sim100$}, which places
both undoped and doped iron pnictides in the regime of split second-order
transitions, i.e. region III of the phase diagrams of Figs. \ref{fig_phase_diagram_regions}
and \ref{fig_anisotropic_weak}\textbf{.} On the other hand, in the
itinerant model, $\alpha$ is generally of order one and changes with
doping and pressure due to the changes in the Fermi surface.

It is also unclear (chiefly due to the lack of results) whether the
phase diagram of the $J_{1}-J_{2}$ model contains an intermediate
phase in which one transition is first-order and the other is second-order,
as in Fig. \ref{fig_exp_phase_diagram}. The mean-field analysis of
Eq. (\ref{S_J1_J2}) for a quasi two-dimensional system did find split
second-order nematic and magnetic transitions\cite{Fang08} for small
$\zeta$ and a simultaneous first-order transition\cite{Millis10}
for large enough $\zeta$. A similar result was obtained by Monte
Carlo simulations of Eq. (\ref{S_J1_J2}) in Refs. \cite{Antropov11,JiangpingHu11},
which treated $\zeta$ as a phenomenological input parameter. It was
also shown by mean-field \cite{Fang06}, RG \cite{Qi09}, and Monte
Carlo \cite{Batista11} calculations that in three dimensional systems
with anisotropic magnetic dispersion the degree of anisotropy tunes
the system between the regimes of split second-order transitions and
simultaneous first-order transitions. Making the spin interaction
anisotropic in spin space has the same effect, i.e. it gives rise
to a transformation from split second-order transitions to simultaneous
first-order transitions \cite{Applegate11}. However, in all cases,
the intermediate phase either was not discussed, or was assumed to
be absent \cite{Qi09}. If only the regimes of split second-order
transitions and simultaneous first-order transitions occur in the
$J_{1}-J_{2}$ model, it will be difficult to explain the different
characters of the transitions observed experimentally upon doping,
pressure, or alkaline-earth substitution. Also, to the best of our
knowledge, there have been no studies within the $J_{1}-J_{2}$ model
of whether the magnetic correlation length jumps at the nematic transition.
As we showed in Section \ref{sub_pseudogap}, such a jump (or a sharp
increase), obtained with the itinerant approach, is fundamental to
explain the pseudogap behavior above $T_{N}$.

Several research groups, using $J_{1}-J_{2}$ based models, also put
forward various arguments to relate the transformation from simultaneous
to split nematic and magnetic transitions to doping and pressure \cite{Qi09,Lorenzana_nematic,Lorenzana_tricritical}.
Ref. \cite{Qi09} suggested that the change from simultaneous first-order
to split second-order transitions with doping is due to doping-induced
change of the spin dynamics from Landau-overdamped ($z_{dyn}=2$)
to propagating ($z_{dyn}=1$). However, ARPES measurements later showed
that hot spots - and therefore Landau damping - are present even at
optimal doping \cite{Liu10}. The authors of Ref. \cite{Lorenzana_nematic}
assumed that doping increases the disorder concentration and decreases
the coupling to the lattice, what leads to a decrease in $g$. However,
recent data on isovalent doped pnictides \cite{Dhaka11,Thaler10}
show that, when the chemical potential remains unchanged, the character
of the transitions remain the same as in the parent compound, even
for large doping concentrations and large in-plane disorder. Ref.
\cite{Lorenzana_tricritical} suggested that the phase diagram of
the iron pnictides is close to a magnetic quantum tri-critical point,
but did not analyze the character of the structural transition.

There have also been studies of nematic and magnetic transitions within
the itinerant approach to the iron-pnictides. Ref. \cite{Kim11} considered
a model very similar to the one presented here and argued that the
phase diagram of Fig. \ref{fig_exp_phase_diagram}(b) can be obtained
even for quasi-two dimensional systems if there is a strong enough
coupling to anharmonic elastic terms.

A phase diagram similar to our Fig. \ref{fig_exp_phase_diagram}(b),
with the intermediate regime, was also obtained in Ref. \cite{Cano10}.
These authors assumed that the structural and magnetic transitions
occur independent of each other, and that the two order parameters
are linearly coupled, as in Eq. (\ref{sept_2}). The intermediate
regime then emerges in some range of parameters, primarily due to
the presence of a quartic term in the elastic free-energy. The authors
of \cite{Cano10}, however, did not argue why the intrinsic magnetic
and structural transitions would occur at about the same temperature.

One argument of why this may be the case, without invoking nematic
degrees of freedom, was presented in Ref. \cite{zlatko}. The authors
of \cite{zlatko} argued that, under special conditions (which may
or may not be satisfied in the iron pnictides), a SDW instability
and a CDW instability in the form of orbital currents occur at almost
the same temperature, with the orbital-current instability occurring
first. They further argued that orbital-current order and magnetic
order are orthogonal to each other in momentum space and coexist.

As we already mentioned, some elements of the physics that leads to
our phase diagrams in Fig. \ref{fig_exp_phase_diagram} are similar
to the physics of the $J_{1}-J_{2}$ and phenomenological models.
The key elements that distinguish our study from previous works are
(i) that the whole phase diagram is entirely driven by magnetic degrees
of freedom, and (ii) that the parameters of the effective bosonic
model are derived from the original itinerant model of interacting
fermions, such that the evolution of the system behavior with doping,
pressure, and alkaline-earth substitution is fully described within
the model itself.

Finally, we point out that an RG analysis similar to the one presented
in Section \ref{sec:solution} was performed in Ref. \cite{Qi09}
for $N=3$ components of the magnetic order parameter, and in Ref.
\cite{Millis10} for $N=6$, which refers to the case of incommensurate
magnetic order parameter. Interestingly, although both works obtained
runaway flows, the former associated the first-order instability to
the divergence of the coupling constants, while the latter pointed
out that it happens much earlier, when the condition $u_{L}<g_{L}$
is first satisfied. Our reasoning is similar to the one in Ref. \cite{Millis10}.

\section{concluding remarks\label{sec:conclusions}}

We argued in this work that the development of the preemptive nematic
order and its interplay with the stripe magnetic order can be fully
understood within an itinerant magnetic scenario for the iron pnictides.
We obtained (rather than assumed) the dependence of the nematic coupling
upon doping, pressure, and alkaline-earth substitution, what enabled
us to consistently explain the character of the magneto-structural
transitions for a variety of iron pnictides. We also demonstrated
how the nematic transition induces orbital and structural order, and
triggers the pseudogap behavior observed in the paramagnetic phase
of these materials. \\
 \\
 The authors thank E. Abrahams, J. Analytis, E. Bascones, J. van den
Brink, P. Brydon, S. Bud'ko, P. Canfield, P. Chandra, M. Daghofer,
L. Degiorgi, I. Fisher, A. Goldman, A. Kaminski, A. Kreyssig, W. Ku,
J. Lorenzana, S. Maiti, R. McQueeney, Y. Matsuda, A. Millis, R. Moessner,
R. Prozorov, M. Takigawa, M. Tanatar, Z. Tesanovic, and M. Vavilov
for useful discussions. R. M. F. acknowledges the support from ICAM
and NSF-DMR 0645461, as well as the valuable support from the NSF
Partnerships for International Research and Education (PIRE) program.
A.V.C. acknowledges the support from NSF-DMR 0906953. R. M. F. and
A.V.C. thank the hospitality of the Aspen Center for Physics, where
part of this work has been done, and the support from NSF grant number
1066293. I. E. and J.K. acknowledges the financial support from DAAD
(PPP USA No. 50750339).

\appendix

\section{Microscopic calculation of the Ginzburg-Landau coefficients}

In this Appendix, we show explicitly how to calculate the Ginzburg-Landau
coefficients of Eq. (\ref{action}) starting with the interacting
Hamiltonian $\mathcal{H}=\mathcal{H}_{0}+\mathcal{H}_{\mathrm{int}}$
of Eqs. (\ref{H_0}) and (\ref{H_int}). First, we introduce the $6$-dimensional
creation Nambu operator:

\begin{equation}
\Psi_{\mathbf{k}}^{\dagger}=\left(\begin{array}{cccccc}
c_{\Gamma,\mathbf{k}\uparrow}^{\dagger} & c_{\Gamma,\mathbf{k}\downarrow}^{\dagger} & c_{X,\mathbf{k}\uparrow}^{\dagger} & c_{X,\mathbf{k}\downarrow}^{\dagger} & c_{Y,\mathbf{k}\uparrow}^{\dagger} & c_{Y,\mathbf{k}\downarrow}^{\dagger}\end{array}\right)\label{A_nambu}\end{equation}

After introducing the bosonic fields $\boldsymbol{\Delta}_{X}$ and
$\boldsymbol{\Delta}_{Y}$ via the Hubbard-Stratonovich transformation
and evaluating the products of Pauli matrices, we can write the partition
function as:

\begin{equation}
Z=\int d\Delta_{i}d\Psi\mathrm{e}^{-S\left[\Psi,\Delta_{i}\right]}\label{A_Z}\end{equation}
 with the action written in compact form:

\begin{equation}
S\left[\Psi,\Delta_{i}\right]=-\int_{k}\Psi_{k}^{\dagger}\mathcal{G}_{k}^{-1}\Psi_{k}^{}+\frac{2}{u_{\mathrm{spin}}}\int_{x}\left(\Delta_{X}^{2}+\Delta_{Y}^{2}\right)\label{A_S}\end{equation}

Here, $\Delta_{i}=\left|\boldsymbol{\Delta}_{i}\right|$ and the Green's
function $\mathcal{G}_{k}^{-1}$ is given by:

\begin{equation}
\mathcal{G}_{k}^{-1}=\mathcal{G}_{0,k}^{-1}-\mathcal{V}\label{A_G}\end{equation}
 with the bare term:

\begin{equation}
\mathcal{G}_{0,k}=\left(\begin{array}{ccc}
\hat{G}_{\Gamma,k} & 0 & 0\\
0 & \hat{G}_{X,k} & 0\\
0 & 0 & \hat{G}_{Y,k}\end{array}\right)\label{A_G0}\end{equation}
 and the interacting term:

\begin{equation}
\mathcal{V}=\left(\begin{array}{ccc}
0 & -\hat{\Delta}_{X} & -\hat{\Delta}_{Y}\\
-\hat{\Delta}_{X} & 0 & 0\\
-\hat{\Delta}_{Y} & 0 & 0\end{array}\right)\label{A_V}\end{equation}

Here, we defined the $2\times2$ matrices $\hat{G}_{i,k}=G_{i,k}\mathbb{I}$
and $\hat{\Delta}_{i}=\boldsymbol{\Delta}_{i}\cdot\boldsymbol{\sigma}$,
with identity matrix $\mathbb{I}$, Pauli matrices $\sigma^{j}$,
and $G_{i,k}^{-1}=i\omega_{n}-\xi_{i,\mathbf{k}}$ the non-interacting
single-particle Green's functions.

It is now straightforward to integrate out the fermions, since the
action is quadratic in them, and obtain the effective magnetic action:

\begin{equation}
S_{\mathrm{eff}}\left[\boldsymbol{\Delta}_{X},\boldsymbol{\Delta}_{Y}\right]=-\mathrm{Tr}\ln\left(1-\mathcal{G}_{0,k}\mathcal{V}\right)+\frac{2}{u_{\mathrm{spin}}}\int_{x}\left(\Delta_{X}^{2}+\Delta_{Y}^{2}\right)\label{A_aux_effective_S1}\end{equation}

Here, $\mathrm{Tr}\left(\cdots\right)$ refers to sum over momentum,
frequency and Nambu indices. A series expansion in powers of $\Delta_{i}^{2}$
then gives:

\begin{eqnarray}
S_{\mathrm{eff}}\left[\boldsymbol{\Delta}_{X},\boldsymbol{\Delta}_{Y}\right] & = & \frac{1}{2}\mathrm{Tr}\left(\mathcal{G}_{0,k}\mathcal{V}\right)^{2}+\frac{1}{4}\mathrm{Tr}\left(\mathcal{G}_{0,k}\mathcal{V}\right)^{4}\nonumber \\
 &  & \frac{2}{u_{\mathrm{spin}}}\int_{x}\left(\Delta_{X}^{2}+\Delta_{Y}^{2}\right)\label{A_aux_effective_S}\end{eqnarray}

Evaluation of the traces yields:

\begin{equation}
S_{\mathrm{eff}}\left[\boldsymbol{\Delta}_{X},\boldsymbol{\Delta}_{Y}\right]=\sum_{i}r_{0,i}\Delta_{i}^{2}+\sum_{i,j}u_{ij}\Delta_{i}^{2}\Delta_{j}^{2}\label{A_effective_S}\end{equation}
 with the coefficients:

\begin{eqnarray}
r_{0,i} & = & \frac{2}{u_{\mathrm{spin}}}+2\int_{k}G_{\Gamma,k}G_{i,k}\nonumber \\
u_{ij} & = & \int_{k}G_{\Gamma,k}^{2}G_{i,k}G_{j,k}\label{A_GL}\end{eqnarray}

Due to the $\pi/2$ rotation symmetry relating the $X$ and $Y$ bands,
it follows that $r_{0,1}=r_{0,2}\equiv r_{0}$ and $u_{11}=u_{22}$.
After re-arranging the terms, we obtain:

\begin{eqnarray}
S_{\mathrm{eff}}\left[\boldsymbol{\Delta}_{X},\boldsymbol{\Delta}_{Y}\right] & = & r_{0}\left(\Delta_{X}^{2}+\Delta_{Y}^{2}\right)\nonumber \\
 &  & +\left(\frac{u_{12}+u_{11}}{2}\right)\left(\Delta_{X}^{2}+\Delta_{Y}^{2}\right)^{2}\nonumber \\
 &  & -\left(\frac{u_{12}-u_{11}}{2}\right)\left(\Delta_{X}^{2}-\Delta_{Y}^{2}\right)^{2}\label{A_action_final}\end{eqnarray}
 with the coefficients $u=u_{12}+u_{11}$ and $g=u_{12}-u_{11}$ given
in Eq. (\ref{aux_action}).

\section{Effects of the anisotropic momentum dispersion}

In this Appendix we briefly show that the inclusion of an anisotropic
momentum dispersion in the bare dynamic susceptibilities $\chi_{i,q}$
do not change our main results. After denoting the renormalized susceptibilities
by $\tilde{\chi}_{X,q}^{-1}=\chi_{X,q}^{-1}+\psi-\phi$ and $\tilde{\chi}_{X,q}^{-1}=\chi_{Y,q}^{-1}+\psi+\phi$,
with $\chi_{i,q}^{-1}=r_{0}+f_{i,\mathbf{q}}+\gamma\left|\nu_{n}\right|,$
we can rewrite the mean-field equations (\ref{self_cons}) as:

\begin{eqnarray}
\psi & = & \frac{u}{2}\int_{q}\left(\tilde{\chi}_{X,q}+\tilde{\chi}_{Y,q}\right)\nonumber \\
\phi & = & \frac{g}{2}\int_{q}\left(\tilde{\chi}_{X,q}-\tilde{\chi}_{Y,q}\right)\label{A2_self_cons1}\end{eqnarray}

In the main text, we considered the case of an isotropic momentum
dispersion $f_{i,\mathbf{q}}=q^{2}$. Most generally, $f_{i,\mathbf{q}}$
will have an anisotropic form preserving the tetragonal symmetry of
the system:

\begin{equation}
f_{i,\mathbf{q}}=q_{x}^{2}\left(1\pm\eta\right)+q_{y}^{2}\left(1\mp\eta\right)\label{A2_anisotropic}\end{equation}
 where $-1<\eta<1$ and the upper (lower) sign refers to band $X$
($Y$). Indeed, inelastic neutron scattering measurements find this
form for the dynamic susceptibility in several iron pnictide compounds
\cite{Diallo10,Li10}.

Substituting it in the self-consistent equations (\ref{A2_self_cons1}),
we can re-scale the momentum by $\tilde{q}_{x}=q_{x}\sqrt{1\pm\eta}$
and $\tilde{q}_{y}=q_{y}\sqrt{1\mp\eta}$ depending on whether the
integral involves $\tilde{\chi}_{X,q}$ (upper signs) or $\tilde{\chi}_{Y,q}$
(lower signs). In either case, the Jacobian of the transformation
is the same, yielding:

\begin{eqnarray}
\psi & = & \frac{u}{\sqrt{1-\eta^{2}}}\int_{\tilde{q}}\frac{r_{0}+\psi+\tilde{q}^{2}+\gamma|\nu_{m}|}{\left(r_{0}+\psi+\tilde{q}^{2}+\gamma|\nu_{m}|\right)^{2}-\phi^{2}}\nonumber \\
\phi & = & \frac{g}{\sqrt{1-\eta^{2}}}\int_{\tilde{q}}\frac{\phi}{\left(r_{0}+\psi+\tilde{q}^{2}+\gamma|\nu_{m}|\right)^{2}-\phi^{2}}\label{A2_self_cons}\end{eqnarray}

Thus, comparing the previous equations with the original mean-field
expressions (\ref{self_cons}), we conclude that the only effect of
the anisotropic dispersion is to renormalize the coupling constants
$u$ and $g$ in the same way, yielding:

\begin{eqnarray}
\bar{\bar{u}} & = & \frac{\bar{u}}{\sqrt{1-\eta^{2}}}\nonumber \\
\bar{\bar{g}} & = & \frac{\bar{g}}{\sqrt{1-\eta^{2}}}\label{A2_u_g}\end{eqnarray}

This does not change the value of the ratio $\alpha=u/g=\bar{\bar{u}}/\bar{\bar{g}}$,
implying that the phase diagrams discussed in the main text remain
valid, with the same values for the tricritical points $\alpha_{c1}$
and $\alpha_{c2}$. The only modification will be in the absolute
value of the temperature, since $\bar{r}_{0}$ is proportional to
\textbf{$g$} (see Eq. \ref{r0_alpha}).

\section{Anisotropic 3D model with quadratic dispersion \label{append_quadr}}

In this Appendix we consider the behavior of the magnetic and nematic
tri-critical points in the anisotropic 3D model with the bosonic susceptibility
\begin{eqnarray}
\chi_{i,q}=r_{0}+q_{||}^{2}+\beta^{2}q_{z}^{2}\end{eqnarray}
 where $0\leq\beta\leq1$ and the same momentum cutoff $\Lambda$
is taken for all three momentum components. This model naturally interpolates
between the 2D case ($\beta=0$) and the isotropic 3D case ($\beta=1$).
We show that the behavior of $\alpha_{c1}$ and $\alpha_{c2}$ as
function of $\beta$ is very similar, although not identical, to the
behavior of the two tri-critical points as functions on $\eta_{z}$
displayed in Fig.\ref{ilya_eta}.

The self-consistent equations for $\phi$ and $r$ become:

\begin{eqnarray}
r & = & r_{0}+\bar{u}\int_{0}^{\Lambda_{||}}\frac{dq_{||}}{2\pi}q_{||}\int_{0}^{\Lambda_{z}}dq_{z}\times\nonumber \\
 &  & \left(\frac{1}{r+q_{||}^{2}+\beta^{2}q_{z}^{2}-\phi}+\frac{1}{r+q_{||}^{2}+\beta^{2}q_{z}^{2}+\phi}\right)\nonumber \\
\phi & = & \bar{g}\int_{0}^{\Lambda_{||}}\frac{dq_{||}}{2\pi}q_{||}\int_{0}^{\Lambda_{z}}dq_{z}\times\nonumber \\
 &  & \left(\frac{1}{r+q_{||}^{2}+\beta^{2}q_{z}^{2}-\phi}-\frac{1}{r+q_{||}^{2}+\beta^{2}q_{z}^{2}+\phi}\right)\label{SelfCon}\end{eqnarray}

For simplicity, we set $\bar{g}=1$ below, but the results can be
easily generalized for arbitrary $\bar{g}$ after rescaling $\beta^{2}\rightarrow\beta^{2}/\bar{g}$.
We first do the 2D integral over $q_{||}$ and then evaluate the one-dimensional
integral in the $q_{z}$ direction. We define \begin{eqnarray}
I(c)=\int_{0}^{\Lambda}\frac{dq_{||}}{2\pi}q_{||}\int_{0}^{\Lambda}dq_{z}\frac{1}{q_{||}^{2}+\beta^{2}q_{z}^{2}+c}\end{eqnarray}
 with $c=r\pm\phi$. Evaluating the momentum integrals we obtain \begin{eqnarray}
 &  & 4\pi I(c)=\frac{2\sqrt{\Lambda^{2}+c}}{\beta}\arctan\left(\frac{\beta\Lambda}{\sqrt{\Lambda^{2}+c}}\right)\nonumber \\
 &  & +\Lambda\log\left(\Lambda^{2}+\beta^{2}\Lambda^{2}+c\right)-\frac{2\sqrt{c}}{\beta}\arctan\left(\frac{\beta\Lambda}{\sqrt{c}}\right)\nonumber \\
 &  & -\Lambda\log\left(\beta^{2}\Lambda^{2}+c\right)\end{eqnarray}

As before, we assume that $\Lambda$ is large compared to both $r$
and $\phi$. Then the first line does not depend on $c$ and can be
absorbed into the renormalizion of $r_{0}$, i.e. into $\bar{r}_{0}$.
The 2D and 3D results are indeed reproduced: in the 2D limit $\beta\rightarrow0$,
and we reproduce the logarithmic behavior and Eqs. (\ref{sa_2});
in the 3D limit $\beta\rightarrow1$ and $\Lambda\rightarrow\infty$
such that $\arctan\rightarrow\pi/2$ and we reproduce the characteristic
square root behavior as well as Eqs. \ref{sa_d3}.

A simple analysis shows that the crossover from 2D to 3D behavior
occurs at a rather small $\beta^{2}\sim1/\Lambda$. If $\alpha_{c1}$
and $\alpha_{c2}$ cross, they must cross in this regime. We rescale
$\beta^{2}$ by $\Lambda$ ($\beta=\frac{\tilde{\beta}}{\sqrt{\Lambda}}$)
and also rescale $r$ and $\phi$ by $\Lambda$ ($r=\Lambda\tilde{r},\,\phi=\Lambda\tilde{\phi}$),
obtaining a cut-off independent equation in terms of $\tilde{\beta}$
\begin{eqnarray}
\frac{4\pi I(\tilde{c})}{\Lambda}=-\frac{2\sqrt{\tilde{c}}}{\tilde{\beta}}\arctan\left(\frac{\tilde{\beta}}{\sqrt{\tilde{c}}}\right)-\log\left(\tilde{\beta}^{2}+\tilde{c}\right)\end{eqnarray}

The integral can be plugged into Eqs.(\ref{SelfCon}) to obtain the
self-consistent equations for $\tilde{r}$ and $\tilde{\phi}$. In
order to obtain $\alpha_{c2}$ corresponding to the nematic tri-critical
point, we expand the self-consistent equations to second order in
$\tilde{\phi}$ and look for the value of $\alpha$ when the coefficient
of the $\phi^{2}$-term changes sign. A straightforward analysis yields:
\begin{widetext} \begin{equation}
\alpha_{c2}=\frac{\tilde{\beta}\sqrt{\tilde{r}_{c2}}\left[\tilde{\beta}\sqrt{\tilde{r}_{c2}}\left(3\tilde{\beta}^{2}+5\tilde{r}_{c2}\right)+3\left(\tilde{\beta}^{2}+\tilde{r}_{c2}\right)^{2}\arctan\left(\frac{\tilde{\beta}}{\sqrt{\tilde{r}_{c2}}}\right)\right]}{\tilde{\beta}^{2}\tilde{r}_{c2}\left(6-3\tilde{\beta}^{2}-5\tilde{r}_{c2}\right)+3\left(\tilde{\beta}^{2}+\tilde{r}_{c2}\right)\arctan\left(\frac{\tilde{\beta}}{\sqrt{\tilde{r}_{c2}}}\right)\left[\tilde{\beta}\sqrt{\tilde{r}_{c2}}\left(4-\tilde{\beta}^{2}-\tilde{r}_{c2}\right)+2\left(\tilde{\beta}^{2}+\tilde{r}_{c2}\right)\arctan\left(\frac{\tilde{\beta}}{\sqrt{\tilde{r}_{c2}}}\right)\right]}\label{AlphaC2}\end{equation}
 \end{widetext} where $\tilde{r}_{c2}$ is the critical value for
the onset of Ising-nematic order: \begin{eqnarray}
1=\frac{2\arctan\left(\frac{\tilde{\beta}}{\sqrt{\tilde{r}_{c2}}}\right)}{\tilde{\beta}\sqrt{\tilde{r}_{c2}}}\label{Rc2}\end{eqnarray}

\begin{figure}
\begin{centering}
\includegraphics[width=0.85\columnwidth]{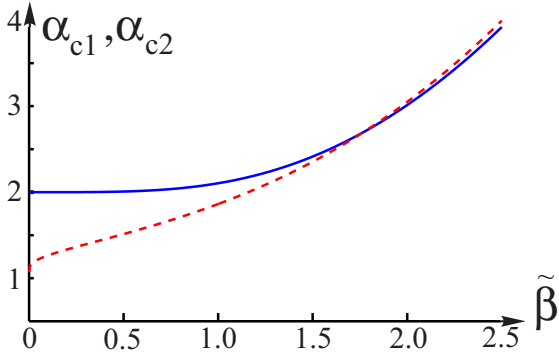} 
\par\end{centering}

\caption{(Color online) The behavior of the magnetic and nematic tri-critical
points $\alpha_{c1}$ (dashed red/light gray line) and $\alpha_{c2}$
(solid blue/dark gray line) as functions of the rescaled $\tilde{\beta}$
in the 3D anisotropic model with the bosonic susceptibility given
by Eq. (\ref{an_suscept_1}). Similar to the other anisotropic 3D
model, $\alpha_{c1}$ and $\alpha_{c2}$ cross at $\tilde{\beta}\approx1.7$,
and at larger $\tilde{\beta}$, $\alpha_{c2}$ becomes smaller than
$\alpha_{c1}$. \label{ilya_beta}}

\end{figure}

To calculate $\alpha_{c1}$ corresponding to the magnetic tri-critical
point, we extend the self-consistent equations to include $M$ in
the standard manner and set $r=\phi$. The equation for the critical
$\tilde{\phi}_{c1}$ at which $M=0$ is \begin{eqnarray}
 &  & \tilde{\phi}_{c1}=\frac{2\sqrt{2}\arctan\left(\frac{\tilde{\beta}}{\sqrt{2\tilde{\phi}_{c1}}}\right)}{\tilde{\beta}}\nonumber \\
 &  & -\log\left(\tilde{\beta}^{2}\right)+\log\left(\tilde{\beta}^{2}+2\tilde{\phi}_{c1}\right).\label{PhiC1}\end{eqnarray}

Expanding the self-consistent equations in powers of $M$, it is straightforward
to obtain the value of $\alpha=\alpha_{c1}$ for which the coefficient
of the quadratic term $M^{2}$ changes sign: \begin{equation}
\alpha_{c1}=\frac{\tilde{\beta}}{\tilde{\beta}-2\sqrt{2}\arctan\left(\frac{\tilde{\beta}}{\sqrt{2\tilde{\phi}_{c1}}}\right)/\sqrt{\tilde{\phi}_{c1}}}\label{AlphaC1}\end{equation}

We plot $\alpha_{c1}$ and $\alpha_{c2}$ as function of $\tilde{\beta}$
in Fig. \ref{ilya_beta}. We see that the two tri-critical points
cross at a certain $\tilde{\beta}$, beyond which $\alpha_{c2}$ becomes
smaller than $\alpha_{c1}$. As expected, as $\tilde{\beta}$ increases,
both $\alpha_{c1}$ and $\alpha_{c2}$ increase, and their ratio approaches
1 from below. The difference between $\alpha_{c1}$ and $\alpha_{c2}$,
however, stays finite.

We also solved numerically the self-consistent equations (\ref{SelfCon})
for $r$ and $\phi$ and obtained the phase diagram in the variables
$\bar{r}_{0}$ and $\alpha$ for various $\beta$. We found the same
four phase diagrams as in Fig. \ref{fig_ilya}. Namely, for small
$\tilde{\beta}$ we recover the behavior of Fig. \ref{fig_ilya} (a),
while for $\tilde{\beta}\gg1$ (but with $\beta\leq1$), we recover
the behavior of Fig. \ref{fig_ilya} (d). The two phase diagrams at
intermediate $\tilde{\beta}$ are also the same as those in Figs.
\ref{fig_ilya} (b), (c).

\section{Linear relationship between the nematic and magnetic order parameters}

In this Appendix, we show how near a simultaneous first-order magneto-structural
transition, the nematic and magnetic order parameters obey an approximately
linear relationship. We start with the self-consistent equations (\ref{aux_r0_m_sc})
in the magnetically ordered phase for arbitrary dimension $2<d\leq3$.
A straightforward manipulation leads to the equivalent equations:

\begin{eqnarray}
\bar{r}_{0} & = & \frac{\bar{u}}{4}\left(2\phi\right)^{\frac{d-2}{2}}-\phi\left(\alpha-1\right)\nonumber \\
\phi & = & \frac{\bar{g}}{4}\left(2\phi\right)^{\frac{d-2}{2}}+\bar{g}\bar{M}^{2}\label{A3_self_m}\end{eqnarray}
 with $\phi=r$. The condition $d\bar{r}_{0}/d\phi=0$ gives the value
of $\Delta\phi$ for which $\bar{r}_{0}$ is maximum and, therefore,
the first instability of the system. Substitution of $\Delta\phi$
in the second equation then gives $\Delta\bar{M}$. Evaluating the
algebraic equations yields: \begin{eqnarray}
\Delta\phi & = & \frac{\left(g/2\right)^{\frac{2}{4-d}}}{2}\left[\frac{\alpha\left(d-2\right)}{\alpha-1}\right]^{\frac{2}{4-d}}\label{A3_jumps}\\
\Delta\bar{M} & = & 2\left(g/2\right)^{\frac{d-2}{2\left(4-d\right)}}\left[\frac{\alpha\left(d-2\right)}{\alpha-1}\right]^{\frac{d-2}{2\left(4-d\right)}}\sqrt{\frac{1-\alpha\left(3-d\right)}{\alpha-1}}\nonumber \end{eqnarray}

These are the values of the nematic and magnetic jumps for $\alpha<\alpha_{c1}=1/\left(d-3\right)$.
Notice that $\Delta\bar{M}\rightarrow0$ for $\alpha\rightarrow\alpha_{c1}$
and that the $d=3$ result of Eq. (\ref{aux_r0_m}) is reobtained
from the second equation.

Using the second equation of (\ref{A3_self_m}), we can expand it
for $\phi$ close to $\Delta\phi$ and $\bar{M}$ close to $\Delta\bar{M}$,
obtaining:

\begin{eqnarray}
\frac{\phi-\Delta\phi}{\Delta\phi} & = & \frac{4\left[1-\alpha\left(3-d\right)\right]}{\left(d-2\right)\left(\alpha+1\right)}\times\nonumber \\
 &  & \left[\left(\frac{\bar{M}-\Delta\bar{M}}{\Delta\bar{M}}\right)+\frac{a_{2}}{a_{1}}\left(\frac{\bar{M}-\Delta\bar{M}}{\Delta\bar{M}}\right)^{2}\right]\nonumber \\
 &  & +\left(\alpha-1\right)\mathcal{O}\left[\left(\frac{M-\Delta M}{\Delta M}\right)^{3}\right]\label{A3_series}\end{eqnarray}
 with:

\begin{equation}
\frac{a_{2}}{a_{1}}=\frac{\left(6-d\right)\left(1-2\alpha\left(3-d\right)\right)+\alpha^{2}\left(22-13d+2d^{2}\right)}{2\left(d-2\right)\left(\alpha+1\right)^{2}}\label{A3_a1_a2}\end{equation}

Thus, the linear relationship between $\phi$ and $M$ is always present
for small enough deviations from the jump. For $\alpha\rightarrow1$,
this linear relationship dominates and extends to larger values of
$N$. Indeed, as $\alpha\rightarrow1$, the jumps $\Delta\phi$ and
$\Delta\bar{M}$ of Eq. (\ref{A3_jumps}) become larger and, consequently,
$\left(\frac{\bar{M}-\Delta\bar{M}}{\Delta\bar{M}}\right)$ becomes
smaller for a fixed $\bar{M}$. Furthermore, all the coefficients
of the series expansion (\ref{A3_series}) of order higher than quadratic
go to zero, and the ratio $a_{1}/a_{2}$ (\ref{A3_a1_a2}) between
the linear and the quadratic coefficients tends to $2$ for any dimension.

\end{document}